\documentclass[aps,prx,showpacs,amssymb,nofootinbib,superscriptaddress,twocolumn]{revtex4-1}
\usepackage{tikz}
\usetikzlibrary{shapes,snakes,backgrounds,fit,decorations.pathreplacing}
\usepackage[ansinew]{inputenc}
\usepackage{bbm}
\usepackage{bm}
\usepackage{amsbsy}
\usepackage{amsthm}
\usepackage{amssymb}
\usepackage{amsfonts}
\usepackage{amsmath}
\usepackage{dsfont} % for symbols like the identity: \mathds{1}
\usepackage{graphicx} % for graphics
\usepackage{epsfig}
\usepackage{epstopdf}
\usepackage{dsfont}
\usepackage{multibib}
\usepackage{color}

\usepackage[colorlinks]{hyperref}
\makeatletter
\newcommand\org@hypertarget{}
\let\org@hypertarget\hypertarget
\renewcommand\hypertarget[2]{%
  \Hy@raisedlink{\org@hypertarget{#1}{}}#2%
  }
\makeatother
\usepackage[figure,table]{hypcap}
\usepackage{MnSymbol}
\usepackage{enumerate}%allows different styles of enumerate environment
\usepackage{float}
\hypersetup{
	bookmarksnumbered,
	pdfstartview={FitH},
	citecolor={darkgreen},
	linkcolor={darkred},
	urlcolor={darkblue},
	pdfpagemode={UseOutlines}}
\definecolor{darkgreen}{RGB}{50,190,50}
\definecolor{darkblue}{RGB}{0,0,190}
\definecolor{darkred}{RGB}{238,0,0}
\usepackage{soul}
\newcommand{\pr}{^{\prime}}

\newcommand{\ket}[1]{\ensuremath{\left|\right.\!{#1}\!\left.\right\rangle}}

\newcommand{\bra}[1]{\ensuremath{\left\langle\right.\!{#1}\!\left.\right|}}

\newcommand{\ketbra}[2]{\ensuremath{|{#1}\rangle\langle{#2}|}}

\newcommand{\scpr}[2]{\ensuremath{\left\langle\right.\hspace*{-1pt} #1 \hspace*{-1pt}\left|\right.\hspace*{-1pt} #2 \hspace*{-1pt}\left.\right\rangle}}

\newcommand{\nl}{\ensuremath{\hspace*{-0.5pt}}}
\newcommand{\nr}{\ensuremath{\hspace*{0.5pt}}}
\newcommand{\sub}[3]{\ensuremath{_{\hspace{#1 pt}\protect\raisebox{#2 pt}{\scriptsize{$ #3$}}}}}
\newcommand{\subtiny}[3]{\ensuremath{_{\hspace{#1 pt}\protect\raisebox{#2 pt}{\tiny{$ #3$}}}}}
\newcommand{\suptiny}[3]{\ensuremath{^{\hspace{#1 pt}\protect\raisebox{#2 pt}{\tiny{$ #3$}}}}}

\newcommand{\expval}[1]{\ensuremath{\left\langle\right.\hspace*{-1pt} #1 \hspace*{-1pt}\left.\right\rangle}}
\newcommand{\comm}[2]{\ensuremath{\left[\right.\! #1 \,, #2 \!\left.\right]}}

\newcommand{\psitheta}{\ensuremath{\psi_{\hspace*{0.0pt}\protect\raisebox{-1.5pt}{\scriptsize{$\theta$}}}}}
\newcommand{\dotpsitheta}{\ensuremath{\dot{\psi}_{\hspace*{0.0pt}\protect\raisebox{-1.5pt}{\scriptsize{$\theta$}}}}}

\newcommand{\tr}{\textnormal{Tr}}
\newcommand{\djj}{d\kern-0.4em\char"16\kern-0.1em}

\renewcommand{\thesubsection}{\Roman{section}.\Alph{subsection}}
\renewcommand{\thesubsubsection}{\Roman{section}.\Alph{subsection}.\arabic{subsubsection}}
\makeatletter
\renewcommand{\p@subsection}{}
\renewcommand{\p@subsubsection}{}
\makeatother

\begin{document}

\title{Flexible resources for quantum metrology}
\author{Nicolai Friis}
\email{nicolai.friis@univie.ac.at}
\affiliation{Institute for Quantum Optics and Quantum Information, Austrian Academy of Sciences, Boltzmanngasse 3, 1090 Vienna, Austria}
\affiliation{Institute for Theoretical Physics, University of Innsbruck, Technikerstra{\ss}e 21a, A-6020 Innsbruck, Austria}
\author{Davide Orsucci}
%\email{davide.orsucci@uibk.ac.at}
\affiliation{Institute for Theoretical Physics, University of Innsbruck, Technikerstra{\ss}e 21a, A-6020 Innsbruck, Austria}
\author{Michalis Skotiniotis}
%\email{michail.skoteiniotis@uab.cat}
\affiliation{F\'{\i}sica Te\`{o}rica: Informaci\'{o} i Fen\`{o}mens Qu\`{a}ntics, Departament de F\'{\i}sica,
Universitat Aut\`{o}noma de Barcelona, 08193 Bellaterra, Spain}
\affiliation{Institute for Theoretical Physics, University of Innsbruck, Technikerstra{\ss}e 21a, A-6020 Innsbruck, Austria}
\author{Pavel Sekatski}
%\email{pavel.sekatski@uibk.ac.at}
\affiliation{Institute for Theoretical Physics, University of Innsbruck, Technikerstra{\ss}e 21a, A-6020 Innsbruck, Austria}
\author{Vedran Dunjko}
%\email{vedran.dunjko@uibk.ac.at}
\affiliation{Institute for Theoretical Physics, University of Innsbruck, Technikerstra{\ss}e 21a, A-6020 Innsbruck, Austria}
\author{Hans J.~Briegel}
%\email{hans.briegel@uibk.ac.at}
\affiliation{Institute for Theoretical Physics, University of Innsbruck, Technikerstra{\ss}e 21a, A-6020 Innsbruck, Austria}
\affiliation{Department of Philosophy, University of Konstanz, 78457 Konstanz, Germany}
\author{Wolfgang D{\"u}r}
%\email{wolfgang.duer@uibk.ac.at}
\affiliation{Institute for Theoretical Physics, University of Innsbruck, Technikerstra{\ss}e 21a, A-6020 Innsbruck, Austria}

\date{\today}
\begin{abstract}
Quantum metrology offers a quadratic advantage over classical approaches to parameter estimation problems by utilizing entanglement and nonclassicality. However, the hurdle of actually implementing the necessary quantum probe states and measurements, which vary drastically for different metrological scenarios, is usually not taken into account. We show that \mbox{for a wide range of} tasks in metrology, 2D cluster states (a particular family of states useful for measurement-based quantum computation) can serve as flexible resources that allow one to efficiently prepare any required state for sensing, and perform appropriate (entangled) measurements using only single qubit operations. Crucially, the overhead in the number of qubits is less than quadratic, thus preserving the quantum scaling advantage. This is ensured by using a compression to a logarithmically sized space that contains all relevant information for sensing. We specifically demonstrate how our method can be used to obtain optimal scaling for phase and frequency estimation in local estimation problems, as well as for the Bayesian equivalents with Gaussian priors of varying widths. Furthermore, we show that in the paradigmatic case of local phase estimation 1D cluster states are sufficient for optimal state preparation and measurement.
\end{abstract}
\maketitle
\vspace*{-12mm}
%\tableofcontents
\section{Introduction}\label{sec:introduction}
\vspace*{-1mm}
Quantum metrology is positioned at the forefront of modern quantum sciences, spearheading the development of future quantum technologies. By utilizing the power of quantum mechanics to gain advantages over previously known techniques in practical tasks such as parameter estimation~\cite{Paris2009, GiovanettiLloydMaccone2011, TothApellaniz2014, DemkowiczDobrzanskiJarzynaKolodynski2015}, state discrimination~\cite{AudenaertCalsamigliaMasanesMunozTapiaAcinBaganVerstraete2007}, or hypothesis testing~\cite{AudenaertNussbaumSzkolaVerstraete2008}, quantum-enhanced measurement procedures have already led to breakthrough discoveries~\cite{LigoVirgoGravWaveDiscovery2016, TsangNairLu2016}. Moreover, nonclassical effects can be harnessed to enhance the precision of determining quantities of interest, including magnetic fields~\cite{WasilevskiJensenKrauterRenemaBalabasPolzik2010, AielloHiroseCappellaro2013}, forces~\cite{ManciniTombesi2003, FermaniManciniTombesi2004}, phases~\cite{MitchellLundeenSteinberg2004, PanChenLuWeinfurterZeilingerZukowski2012}, or frequencies~\cite{WinelandBollingerItanoMooreHeinzen1992, BollingerItanoWinelandHeinzen1996, ChwallaKimMonzSchindlerRiebeRoosBlatt2007}. For many different applications, the quantum advantage manifests as a quadratic scaling gap in terms of the relevant resources~\cite{GiovanettiLloydMaccone2004, BerryWiseman2000, BaganBaigMunozTapia2004, ChiribellaDArianoPerinottiSacchi2004a}, e.g., the number of sensing systems, with respect to the best classical approaches. However, to achieve this so-called Heisenberg scaling, different tasks require different resource states as well as different (potentially non-local) measurements, which have to be separately determined for any specific case, rendering the design of a universally applicable, optimal sensing device difficult. Moreover, this still leaves open the important (and often ignored) question of how the desired states and measurements can be implemented efficiently.

Here we report on the design of a flexible device that allows one to obtain a quantum scaling advantage for a large class of different metrological problems by using only a specific entangled state and single-qubit operations. We show that a 2D cluster state~\cite{BriegelRaussendorf2001, WunderlichWunderlichSingerSchmidtKaler2009} \textemdash\ a particular entangled state associated with a rectangular lattice that can be prepared by commuting, nearest-neighbour interactions among qubits on the lattice \textemdash\ allows achieving Heisenberg scaling for an important group of paradigmatic metrology problems. This includes the sensing of local observables such as magnetic fields~\cite{WasilevskiJensenKrauterRenemaBalabasPolzik2010, AielloHiroseCappellaro2013}, as well as the estimation of phases~\cite{GiovanettiLloydMaccone2004, BerryWiseman2000}, frequencies~\cite{WinelandBollingerItanoMooreHeinzen1992, BollingerItanoWinelandHeinzen1996, ChwallaKimMonzSchindlerRiebeRoosBlatt2007}, and certain interaction strengths~\cite{SkotioniotisSekatskiDuer2015}. Crucially, we show that this can be done both in the local (frequentist) approach with arbitrarily many repetitions, and in the (single-shot) Bayesian approach for arbitrary cost functions and priors (see, e.g., Ref.~\cite{JarzynaDemkowiczDobrzanski15}), including flat~\cite{BerryWiseman2000, BaganBaigMunozTapia2004, ChiribellaDArianoPerinottiSacchi2004a} and Gaussian priors with varying width~\cite{MacieszczakFraasDemkowiczDobrzanski2014, DemkowiczDobrzanski2011}. The key difference between these estimation problems lies in the incorporation of a priori available knowledge about the estimated parameter. In local estimation, no quantification of prior knowledge is required in principle, but it is often assumed that fluctuations around a well-known value of the parameter are being estimated in order to make use of the quantum Fisher information (QFI) as a relevant figure of merit. In Bayesian estimation, the initial information is encoded in a prior probability distribution that is updated according to Bayes' law after each individual measurement.

%\vspace*{-0.5mm}
The optimal probe state for these different problems vary strongly, ranging from Greenberger-Horne-Zeilinger (GHZ) states in the case of local phase estimation, to
\clearpage
\newpage
\begin{figure}[ht!]
\includegraphics[keepaspectratio,width=0.44\textwidth]{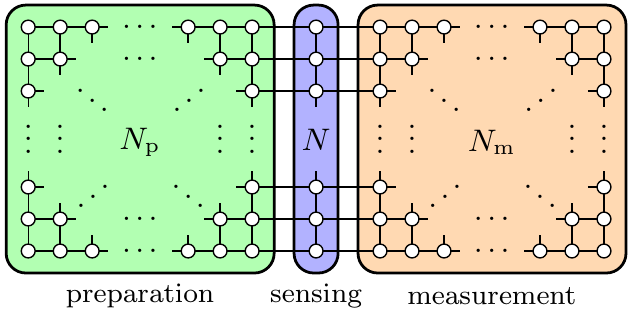}
\vspace*{-2mm}
\caption{\textbf{Cluster state for Metrology}. 2D cluster states can be efficiently used for quantum metrology if the numbers of qubits used for state preparation ($N_{\mathrm{p}}$) and measurement ($N_{\mathrm{m}}$) increase less than quadratically with $N$, the number of qubits used for sensing.
\label{fig:cluster state for metrology}}
\end{figure}

\noindent
certain superpositions of states with different Hamming weights (e.g., with sine-shaped profiles for the coefficients~\cite{BerryWiseman2000}) for Bayesian phase estimation (for flat priors). Moreover, also the corresponding optimal measurements are vastly different, including simple local measurements for GHZ states, but also complicated, entangled measurements on all qubits~\cite{ChiribellaDArianoPerinottiSacchi2004b, ChiribellaDArianoSacchi2005}, e.g., discrete Fourier basis measurements for Bayesian estimation with flat priors~\cite{BerryWiseman2000}. In particular, some states and measurements may be significantly more difficult to realize than others.

The 2D cluster state allows one to deal with all of these problems. On the one hand, the fact that it is a universal resource for measurement-based quantum computation (MBQC)~\cite{RaussendorfBriegel2001, VanDenNestMiyakeDuerBriegel2006} trivially enables arbitrary state preparation and measurements on a subset of the qubits in the cluster, provided the latter is large enough. On the other hand, MBQC provides a simple, unifying framework in which state preparation and measurements can be assigned an unambiguous resource cost in terms of the overall number of qubits in the cluster\footnote{Note that, in a different context, cluster states have previously also been used for specific metrology tasks directly (i.e., using all qubits for sensing)~\cite{RosenkranzJaksch2009}.}, as illustrated in Fig.~\ref{fig:cluster state for metrology}. To guarantee a quantum scaling advantage for metrological applications, the probe preparation and measurements must be efficiently executable. That is, any metrological scaling advantage is lost if the size of the cluster required for a given estimation strategy with an $N$-qubit probe grows as $N^{2}$ or stronger, in which case it becomes favourable to use all qubits in the cluster as individual, classical probes instead.

We show that the preparation of optimal probe states and corresponding suitable measurements for local as well as Bayesian phase and frequency estimation can indeed be carried out efficiently using 2D cluster states. For the local scenario, we explicitly construct the preparation and measurement strategy achieving optimality. For the Bayesian scenario, we present a construction that can generate all optimal probe states with a linear overhead in $N$. We then introduce a compression procedure that can be implemented on a 2D cluster with $O(N\log^{2}\!N)$ qubits, which enables one to efficiently perform measurements even when the circuit descriptions of the corresponding unitaries are of exponential size in the number of qubits of the compressed space. These constructions allow achieving Heisenberg scaling for phase and frequency estimation scenarios using the 2D cluster in a flexible manner. Crucially, this flexibility holds the potential for yielding (nearly) optimal scaling performance for a variety of estimation problems, and hence goes beyond the capabilities of architectures dedicated to specific individual tasks~\cite{BerryHigginsBartlettMitchellPrydeWiseman2009}. To further illustrate these general results, we discuss a particular choice of probe states and measurements that can be efficiently implemented in our framework, for which Heisenberg scaling can be achieved for Gaussian priors of varying widths.

This paper is structured as follows. In Section~\ref{sec:2D Cluster States as Universal Resources for Phase and Frequency Estimation} we first discuss the basic structure of parameter estimation problems and the general form of all optimal probe states. We then argue that 2D cluster states provide flexible resources to achieve Heisenberg scaling in phase and frequency estimation problems by using an efficient compression to the subspace of the optimal probes. In Section~\ref{sec:quantum advantage Bayesian estimation main text} we then show how Heisenberg scaling can be achieved in Bayesian phase (and frequency) estimation, before demonstrating in Section~\ref{sec:Efficient State Preparation} how the necessary probe states can be prepared in a measurement-based architecture consisting of $O(N)$ qubits. Finally, we introduce the explicit construction of the efficient compression algorithm required for the measurements in Section~\ref{sec:Efficient Unary-to-Binary Compression}. At last, we discuss our findings and their implications in Section~\ref{sec:conclusion}, including generalization to the estimation of quantities other than phases and frequencies.

\section{2D Cluster States as Universal Resources for Phase and Frequency Estimation}\label{sec:2D Cluster States as Universal Resources for Phase and Frequency Estimation}

\subsection{Parameter Estimation Problems}\label{sec:Parameter Estimation Problems}

In typical parameter estimation procedures, one wishes to determine an unknown parameter $\theta$ that is not directly measurable. To this end, a probe state described by a density operator $\rho_{o}$ is prepared, which undergoes a dynamical evolution governed by $\theta$, encoding the parameter in the resulting state $\rho(\theta)$. The evolution can in principle be an arbitrary quantum channel but we are here mainly interested in pure states $\rho_{o}=\ket{\psi}\!\!\bra{\psi}$ and unitary channels, where
\begin{align}
    \rho(\theta) &=\,U_{\theta}\ket{\psi}\!\bra{\psi}U^{\dagger}_{\theta}\,,
    \label{eq:unitary channel pure state}
\end{align}
for a unitary $U_{\theta}=\exp(-i\theta H)$ generated by the Hamiltonian\footnote{We work in units where $\hbar=1$. In addition, we adopt the usual convention of Hamiltonian estimation where the eigenvalues of $H$ (and hence $\theta$) are taken to be dimensionless. For example, for frequency estimation one then has $\theta=\omega t$, where the time $t$ is assumed to be known precisely.} $H=H^{\dagger}$. For example, in phase (and frequency) estimation, one considers a local Hamiltonian for $N$ qubits, i.e.,
\begin{align}
    H   &=\,\sum\limits_{i=1}^{N}H_{i}
\end{align}
and $H_{i}$ acts nontrivially only on the $i$th qubit. Typically, one has
\begin{align}
    H_{i}   &\equiv\, \tfrac{1}{2}Z\ \forall i\,,
\end{align}
where $Z$ is the usual Pauli operator, but other local Hamiltonians can be brought to this form by local unitaries. After the encoding, a measurement of the probe state $\rho(\theta)$ is performed, which can be represented by a positive-operator valued measure (POVM), i.e., a set $\{E_{m}\}$ of positive semi-definite operators $E_{m}\geq0$ satisfying $\sum_{m}E_{m}=\mathds{1}$, where $\mathds{1}$ is the identity operator. For an introduction to POVM measurements see, e.g.,~\cite[pp.~90]{NielsenChuang2000} or~\cite{Paris2012}.

From the measurement outcomes, labelled $m$, an estimate of the parameter in question can be obtained. The precise nature of the estimator depends on the type of estimation scenario, distinguishing, for example, between local and Bayesian estimation mentioned previously. All these scenarios have in common that the precision of the estimation [as quantified by some figure of merit, e.g., the mean-square error (MSE)] improves with the number $N$ of probe systems. For classical strategies based on product states, this increase is at most linear in $N$, which is referred to as the standard quantum limit (or shot noise scaling). However, using approaches based on the optimal quantum mechanical probes the improvement in this figure of merit can be quadratic in $N$, i.e., achieving (optimal) Heisenberg scaling. For reviews of parameter estimation techniques and quantum metrology we direct the reader to, e.g., Refs.~\cite{Paris2009, TothApellaniz2014, DemkowiczDobrzanskiJarzynaKolodynski2015} or the \hyperlink{sec:appendix}{Appendix}.

In local phase (and frequency) estimation one typically considers many repetitions of the same measurement that provide an estimate, whose variance one is interested in minimizing using the available resources. In this scenario, the optimal $N$-qubit probe state is a GHZ state
\begin{align}
    \ket{\psi\subtiny{0}{0}{\mathrm{GHZ}}}  &=\,\tfrac{1}{\sqrt{2}}\Bigl(\,\ket{0}^{\otimes N}+\ket{1}^{\otimes N}\Bigr)\,,
    \label{eq:GHZ state}
\end{align}
and the accompanying optimal measurements are local $X$ measurements. This can be determined via the QFI, the relevant figure of merit for local estimation, as we explain in more detail in Appendix~\ref{sec:local estimation}. In Bayesian parameter estimation (see, e.g., Refs.~\cite{Personick1971,DemkowiczDobrzanskiJarzynaKolodynski2015} or Appendix~\ref{sec:Bayesian estimation}), the situation is somewhat different. Here one quantifies the initial knowledge (or belief) about the parameter by a prior probability distribution that is updated after each single measurement. In this case, a figure of merit is the average variance of the updated distribution. In the Bayesian estimation scenario, the optimal probes and measurements depend on the shape of the prior and the cost function used. For instance, for phase estimation with flat priors (i.e., no prior knowledge), the optimal probe state achieving Heisenberg scaling is given by
\begin{align}
    \ket{\psi\subtiny{0}{0}{\mathrm{opt}}}  &=\,\sum\limits_{n=0}^{N}\,\psi\sub{0}{-1}{n}\,\ket{n}\,,
    \label{eq:probe_state 2}
\end{align}
where $\ket{n}$ are eigenstates of $H$ corresponding to its $N+1$ different eigenvalues, and the coefficients $\psi\sub{0}{-1}{n}$ have a sinusoidal profile (see, e.g., Ref.~\cite{BerryWiseman2000}), i.e.,
\begin{align}
    \psi\sub{0}{-1}{n}  &=\,\sqrt{\frac{2}{N+2}}\,\sin\left(\frac{(n+1)\pi}{N+2}\right)\,.
    \label{eq:Berry wiseman state}
\end{align}
Although different from the optimal measurement, we find that for the state in Eq.~(\ref{eq:probe_state 2}) a projective measurement in the basis obtained via the quantum Fourier transform (QFT) of the basis $\{\ket{n}\}$ allows for Heisenberg scaling for Bayesian phase and frequency estimation with Gaussian priors of varying widths, as we discuss in Section~\ref{sec:quantum advantage Bayesian estimation main text}, as well as in Appendices~\ref{sec:Quantum Advantage in Bayesian Estimation} and~\ref{sec:Bayesian frequency estimation in MBQC}.

The crucial observation required to extend the applicability of this approach to arbitrary priors (and cost functions) lies in noticing that in $N$-qubit phase (and frequency) estimation scenarios of any kind, $H$ only has $N+1$ different eigenvalues. For each of these values, only one representative eigenstate needs to be selected. Moreover, within the subspaces corresponding to fixed eigenvalues one may choose those eigenstates that can be prepared most efficiently. Instead of the typical Dicke states that are symmetric with respect to the exchange of the qubits, we therefore employ eigenstates corresponding to a unary encoding of $n$, i.e.,
\begin{align}
    \ket{n}\subtiny{-1}{0}{\mathrm{un}} &=\,\ket{1}^{\otimes n}\ket{0}^{\otimes N-n}.
\end{align}
All optimal probe states can hence be chosen to be of the form of Eq.~(\ref{eq:probe_state 2}) with $\ket{n}\equiv\ket{n}\subtiny{-1}{0}{\mathrm{un}}$ for some choice of the coefficients $\psi\sub{0}{-1}{n}$. Most importantly, all of these probe states have support in an $(N+1)$-dimensional subspace of the $2^{N}$-dimensional overall Hilbert space.

Therefore, the problem of optimal state preparation and measurements for $N$ qubits can be translated to that of $\lambda:=\lceil\log(N\nl+\nl1)\rceil$ qubits (where the logarithm is understood
to be to base $2$), provided that one can efficiently and coherently convert the unary encoding $\ket{n}\subtiny{-1}{0}{\mathrm{un}}$ to a binary encoding in $\lambda$ qubits. More precisely, one can initially prepare a state of $\lambda$ qubits and convert it (efficiently) to the desired $N$-qubit state for sensing (using at least $N-\lambda$ auxiliary qubits). After the parameter has been encoded, one performs the reverse procedure before carrying out the final measurement on $\lambda$ qubits. In Section~\ref{sec:Efficient Unary-to-Binary Compression} we present a quantum circuit of size $O(N\log^{2}\!N)$ (and its MBQC representation) achieving exactly such a unary-to-binary compression. On the logarithmically small space of these $\lambda$ qubits the probe state preparation and measurement can then be carried out even with exponential overhead in $\lambda$ while maintaining Heisenberg scaling.

\subsection{Parameter Estimation in MBQC Architectures}\label{sec:Parameter Estimation in MBQC Architectures}

\begin{figure*}[ht!]
\includegraphics[width=0.88\textwidth]{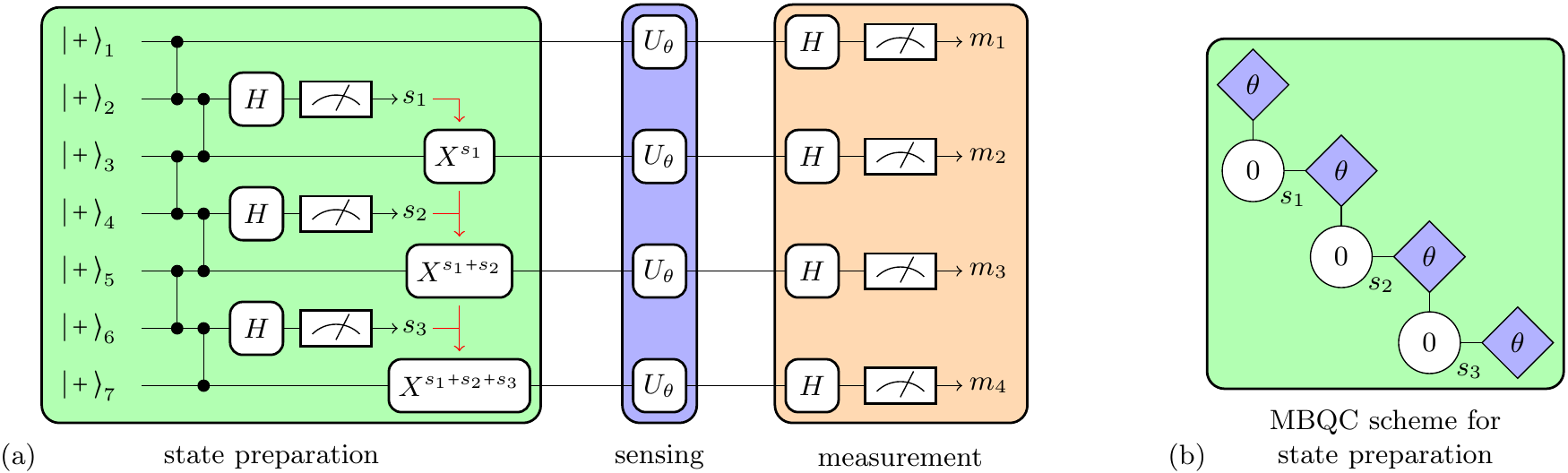}
\caption{\label{fig:local phase estimation in MBQC}%(Color online)
\textbf{Local phase estimation in MBQC}. (a) The circuit representation of an efficient local phase estimation procedure based on a seven-qubit architecture for MBQC is shown. The vertical lines
(\nr\protect\raisebox{-3pt}{\protect\begin{tikzpicture}
\protect\draw[fill=black,line width=1pt] (0,0.135) circle (1.35pt);
\protect\draw[black,line width=0.3pt] (0,0.135) -- (0,-0.135);
\protect\draw[fill=black,line width=1pt] (0,-0.135) circle (1.35pt);
\end{tikzpicture}}\nr) represent controlled phase gates $C\nl Z_{ij}$ applied to the respective qubit pairs $(i,j)$. Here, the rounded boxes correspond to applications of single-qubit gates, where $H=(X+Z)/\sqrt{2}$ is the Hadamard gate, and the symbol~\protect\raisebox{-3pt}{\protect\begin{tikzpicture}[Measurement/.style={rectangle,draw=black,fill=white,thick,minimum height=4mm, minimum width=8mm}]
    \protect\node at (4.0,0) [Measurement]{};
    \protect\draw[color=black] (4.27,-0.05) arc (45:135:0.38);
    \protect\draw[color=black] (4,-0.1) -- (4.15,0.15);
\end{tikzpicture}}
indicates a measurement in the computational basis $\{\ket{s}|s=0,1\}$ with outcome $s$. In the preparation stage (green), the resource state, a four-qubit GHZ state, is created by measurements of the three qubits of a 1D seven-qubit cluster state. Given the measurement outcome $s_{n}$ of the qubit labelled $2n$, the qubit $(2n+1)$ is corrected locally by a Pauli-$X$ operation if $\sum_{i=1}^{n}s_{n}$ is odd. After the local corrections, the encoding transformation $U_{\theta}$ is applied, imprinting the parameter that is to be estimated. In the final measurement stage (orange), the remaining qubits in the cluster are locally measured. In (b), the preparation and sensing stages are illustrated as MBQC measurement patterns in a graphical notation (see, e.g., Ref.~\protect\cite{BrowneBriegel2006}). Measured qubits are represented by circles inscribed with the corresponding measurement angle in the $x-y$ plane (here $\varphi=0$), while output qubits are indicated by diamonds~(\protect\raisebox{-3pt}{\protect\begin{tikzpicture}[Diamond/.style={diamond,draw=black,thick,fill=white,minimum size=3mm}]
    \protect\node at (4.0,0) [Diamond]{};
\end{tikzpicture}}). The connecting lines between qubits indicate the initial application of $C\nl Z$ gates, and all qubits are assumed to have been initialized in the state $\ket{+}$.
}
\end{figure*}

The premise for taking advantage of the quadratic scaling gap in resources (here, the number of qubits) between the quantum strategy described in the previous section and the best classical strategy is that the required probe states and measurements can be implemented efficiently. Here, we will take efficiency to mean that the overhead in the number of qubits used for the execution of the quantum strategy, including preparation and measurement, must grow less strongly than $N^{2}$. To illustrate this requirement, consider a situation where an array of qubits is provided and one is given the task of using the array most efficiently for the estimation of a parameter. For instance, an array of spins (which may otherwise be used for quantum computation or quantum simulation) could be exposed to a magnetic field with fixed direction but unknown strength for this purpose. If one has the ability to prepare arbitrary quantum states of these (spin) qubits, then one may initialize a GHZ state for local phase estimation, or the corresponding optimal state for Bayesian phase estimation (or any other estimation problem for that matter). However, as we have seen in the previous section, states and measurements that offer advantages for different metrological problems are in general quite distinct, and the conversion from one to the other may involve arbitrarily long sequences of entangling operations. The preparation and measurement hence comes at a cost that we wish to quantify.

An approach that allows for preparing arbitrary quantum states and performing any measurements on them, while naturally including a resource count for these tasks is MBQC. In this paradigm, introduced in Ref.~\cite{RaussendorfBriegel2001}, an array of qubits is initialized in a particular (entangled) quantum state, typically a so-called cluster state~\cite{BriegelRaussendorf2001}. A cluster state is a type of graph state, i.e., it can be represented by a graph (a set of vertices $v_{i}$ and edges $e_{ij}$ connecting the vertices). Each vertex represents a qubit initialized in the state $\ket{+}$, and controlled phase gates $C\nl Z$, given by
\begin{align}
    C\nl Z_{ij}  &=\,\ket{0}\!\!\bra{0}_{i}\otimes\mathds{1}_{j}\,+\,\ket{1}\!\!\bra{1}_{i}\otimes Z_{j}\,=\,C\nl Z_{ji}\,,
    \label{eq:controlled phase gate}
\end{align}
are applied to each pair of qubits connected by an edge. For simplicity, we will here only consider 2D cluster states where the underlying graph is a regular, rectangular lattice, but in principle, also other graph states~\cite{HeinDuerEisertRaussendorfVanDenNestBriegel2006} could be considered for our purposes. By applying only single-qubit gates and carrying out local measurements on a subset of all qubits in a 2D cluster, arbitrary unitary operations can be implemented on the remaining qubits~\cite{VanDenNestMiyakeDuerBriegel2006}. Performing a unitary transformation in the circuit model of quantum computation hence translates to a sequence of measurement angles for single-qubit measurements in the cluster. For a more detailed introduction to MBQC see Refs.~\cite{BrowneBriegel2006,BriegelBrowneDuerRaussendorfVanDenNest2009}, or Appendix~\ref{sec:Measurement-Based Quantum Computation}.

In other words, a number of the initial qubits can be sacrificed to obtain a probe state of fewer qubits, which is more suitable for a given metrological task at hand. Note that using the unmodified cluster state as a probe state itself does not provide a scaling advantage with respect to classical strategies, i.e., its QFI is $O(N)$. Similarly, additional qubits can be used to implement arbitrary measurements by performing appropriate unitaries followed by computational basis measurements. Here, one needs to ensure that only the part of the cluster used to prepare the probe state is subjected to the transformation encoding the parameter. This can be achieved, e.g., by appropriately timed Pauli-$X$ operations on the qubits used for the measurement at the middle and at the end of the interaction period. For spins this corresponds to the general practice of refocusing of the magnetisation, i.e., a spin echo.

Crucially, the overall number of qubits required for the preparation and measurement of this $N$-qubit probe state must grow less than quadratically with $N$ to maintain a potential metrological scaling advantage. This is possible, for instance, for local phase estimation, where the optimal measurement strategy can be carried out with $2N-1$ qubits in a 1D cluster state as shown in Fig.~\ref{fig:local phase estimation in MBQC}. As we will show in the following, such efficient constructions also exist for Bayesian phase (and frequency) estimation problems. In Section~\ref{sec:Efficient State Preparation}, we demonstrate that all probe states (including the optimal ones) of the form of Eq.~(\ref{eq:probe_state 2}) can be efficiently prepared from a 2D cluster state using only local operations. In Section~\ref{sec:Efficient Unary-to-Binary Compression} we then present the unary-to-binary compression requiring $O(N\log^{2}\!N)$ qubits of the cluster to reduce the problem of implementing optimal measurements to the subspace of $\lambda:=\lceil\log(N\nl+\nl1)\rceil$ qubits. On this subspace, projective measurements in any basis can be carried out efficiently, provided that the unitary transformation relating it to a computational-basis measurement requires no more than $O(2^{\lambda})$ (nearest neighbour) gates. This is the case, for instance, for the QFT measurement, which performs optimally for flat priors~\cite{BerryWiseman2000} and achieves Heisenberg scaling for Gaussian priors of varying widths as we will show next.

\section{Quantum Advantage in Bayesian Estimation}\label{sec:quantum advantage Bayesian estimation main text}

We now briefly discuss the Bayesian phase estimation scenario, more details on which can be found in Appendix~\ref{sec:Bayesian estimation}, and show that the combination of sine states and QFT measurements can achieve Heisenberg scaling. In Bayesian parameter estimation, the initial knowledge about the parameter is encoded in a prior probability distribution $p(\theta)$. When a measurement with POVM elements $\{E_{m}\}$ is performed on the parameter-encoded state $\rho(\theta)$, the conditional probability of obtaining the outcome labelled $m$ is
\begin{align}
    p(m|\nr\theta\nr)   &=\,\tr\bigl(E_{m}\nr\rho(\theta)\bigr)\,.
\end{align}
To obtain the unconditional probability for the same outcome, these values are weighed according to one's prior belief, i.e.,
\begin{align}
    p(m)    &=\,\int\!\!d\theta\,p(\theta)\,p(m|\nr\theta\nr).
\end{align}
The information obtained in a measurement with outcome $m$ is then used to update this belief via Bayes' law, obtaining the posterior distribution $p(\theta|m)$ given by
\begin{align}
    p(\theta|m) &=\,\frac{p(m|\nr\theta\nr)\,p(\theta)}{p(m)}.
\end{align}
In turn, the posterior distribution provides an estimate $\hat{\theta}(m)$ for the parameter via
\begin{align}
    \hat{\theta}(m) &=\,\int\!\!d\theta\,p(\theta|m)\,\theta\,.
\end{align}
As a figure of merit for this estimation procedure one then quantifies the width of the posterior by a suitable measure $V_{\mathrm{post}}\suptiny{1}{-1}{(m)}$ and averages over all possible outcomes, such that
\begin{align}
    \overline{V}_{\!\mathrm{post}}  &=\,\sum\limits_{m}p(m)\,V_{\mathrm{post}}\suptiny{1}{-1}{(m)}\,.
\end{align}
For instance, when the parameter in question has support over all of $\mathbb{R}$ (e.g., for frequency estimation, see Appendix~\ref{sec:Bayesian frequency estimation in MBQC}), one may use the MSE
\begin{align}
    V_{\mathrm{post}}\suptiny{1}{-1}{(m)}    &=\,V[p(\theta|m)]\,=\,\int\!d\theta\,p(\theta|m)\,\bigl(\theta-\hat{\theta}(m)\bigr)^{2}.
\end{align}
Here, we want to focus on phase estimation, i.e., the case where the parameter has support on the interval $[-\pi,\pi]\subset\mathbb{R}$. When the prior is appropriately narrow, one may still use the MSE, which allows the use of some simple techniques (e.g., a Bayesian version of the Cram{\'e}r-Rao inequality, see the Appendix~\ref{sec:bayesian cramer rao bound} and Ref.~\cite{JarzynaDemkowiczDobrzanski15}) for the comparison with classical strategies. Nonetheless, wrapped distributions and covariant measures of their width are in general more suitable for phase estimation. As an example, one can consider the wrapped Gaussian distribution of the form
\begin{align}
    p(\theta)   &=\,\frac{1}{\sqrt{2\pi}\nr\sigma}\,\sum\limits_{q=-\infty}^{\infty}\,e^{-\nr\frac{(\theta-\theta_{o}+2\pi q)^{2}}{2\sigma^{2}}}\,,
\end{align}
where $q\in\mathbb{Z}$, and the mean angle is
\begin{align}
    \arg\bigl(\expval{e^{i\theta}}_{p(\theta)}\bigr)  &=\,\int\limits_{-\pi}^{\pi}\!\!d\theta\,p(\theta)\,=\,\theta_{o}.
\end{align}
The non-negative parameter $\sigma$ can be identified with the circular standard deviation
\begin{align}
    S   &=\,\sqrt{\ln(1/|\expval{e^{i\theta}}_{p(\theta)}|^{2})}\,=\,\sigma\,,
\end{align}
corresponding to the width of the underlying Gaussian distribution. However, for our purposes, it is more useful

\begin{figure}[ht!]
%%%trim={<left> <lower> <right> <upper>}
\includegraphics[width=0.48\textwidth,trim={0cm 0.1cm 0cm 0cm},clip]{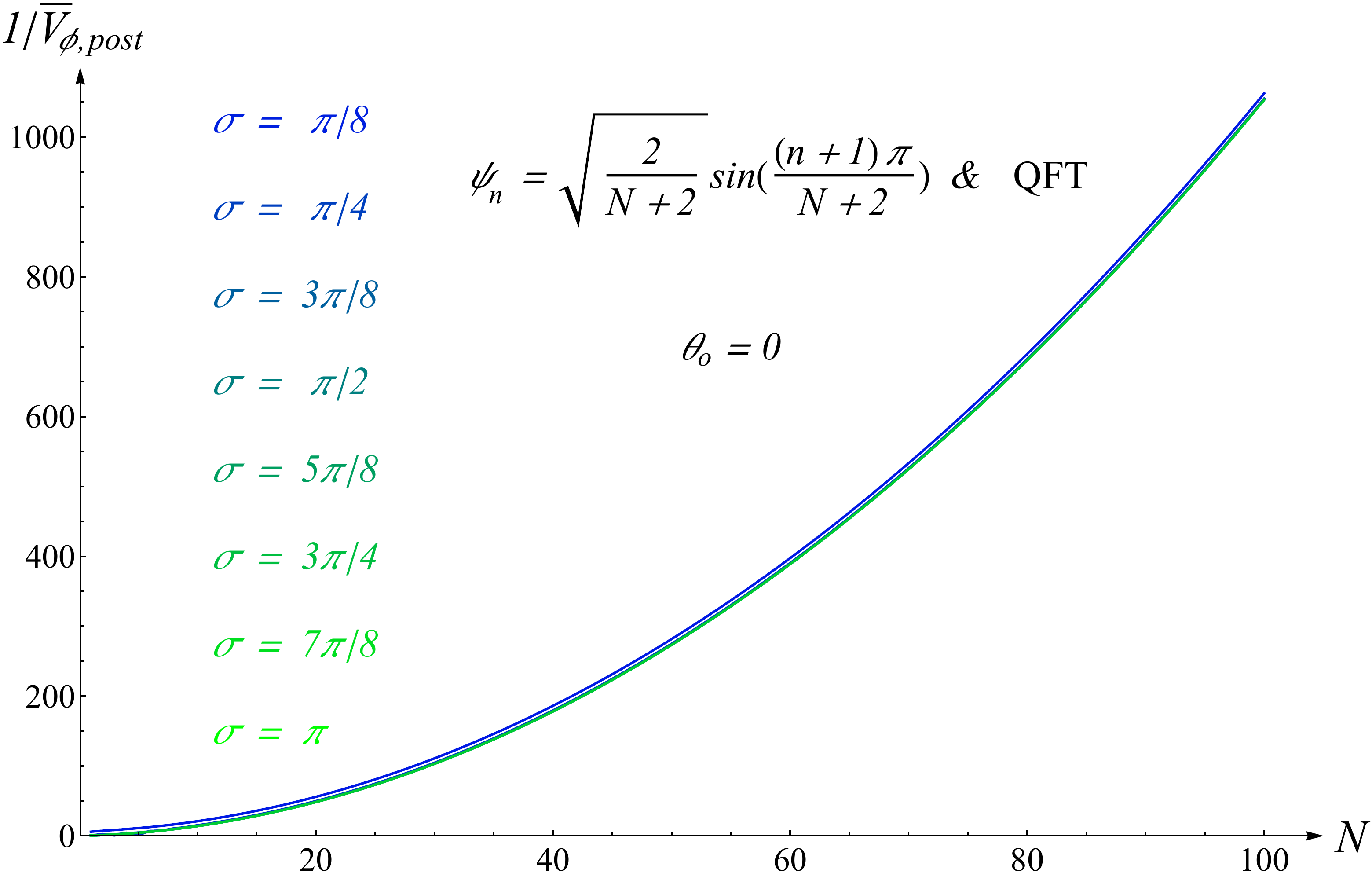}\\
%\hspace*{-8cm}(b)
\caption{\label{fig:Berry-Wiseman main text}%(Color online)
\textbf{Example for quantum strategy:}. The inverse of the average phase variance $\overline{V}_{\!\phi,\mathrm{post}}$ of the posterior is shown for up to $N=100$ qubits for the measurement strategy using probe states with coefficients as in Eq.~(\ref{eq:Berry wiseman state}) and QFT measurements. Although $N$ is an integer with $N\geq1$, the curves have been plotted for continuous values of $N$ for the purpose of illustration. The prior is chosen to be a wrapped Gaussian with $\theta_{o}=0$ and curves are shown for values of $\sigma$ from $\tfrac{\pi}{8}$ (blue) to $\pi$ (green) in steps of $\tfrac{\pi}{8}$. The curves, although difficult to tell apart visually, are distinct. Additional analysis of this measurement strategy using the MSE and comparisons with classical strategies can be found in Fig.~\ref{fig:Berry-Wiseman comparison}, whereas similar results for Bayesian frequency estimation are presented in Appendix~\ref{sec:Bayesian frequency estimation in MBQC}.}
\end{figure}
\noindent
to quantify the width of this wrapped distribution by the Holevo phase variance~\cite{Holevo1984} $V_{\!\phi}$, given by
\begin{align}
    V_{\!\phi}  &=\,|\expval{e^{i\theta}}_{p(\theta)}|^{-2}\,-\,1\,=\,e^{\sigma^{2}}\,-\,1\,.
    \label{eq:holevo phase var}
\end{align}
Likewise, we will quantify the width of the posterior by
\begin{align}
    V_{\!\phi,\mathrm{post}}\suptiny{1}{-1}{(m)}  &=\,|\expval{e^{i\theta}}_{p(\theta|m)}|^{-2}\,-\,1\,.
\end{align}
For the probe states of Eq.~(\ref{eq:probe_state 2}) with the sinusoidal profile of Eq.~(\ref{eq:Berry wiseman state}), and the QFT measurement represented by the basis $\{\ket{e_{k}}\}$, where
\begin{align}
    \ket{e_{k}} &=\,\frac{1}{\sqrt{N+1}}\sum\limits_{n=0}^{N}e^{i\nr n\nr\tfrac{2\pi k}{N+1}}\,\ket{n}\subtiny{-1}{0}{\mathrm{un}}\,,
\end{align}
we then calculate the average phase variance $\overline{V}_{\!\phi,\mathrm{post}}=\sum_{m}p(m)\,V_{\!\phi,\mathrm{post}}\suptiny{1}{-1}{(m)}$. The results for various values of $\sigma$ and for up to $100$ qubits are shown in Fig.~\ref{fig:Berry-Wiseman main text}. The numerical results indicate that for all widths of the priors the example quantum strategy exhibits Heisenberg scaling. In Appendix~\ref{sec:Quantum Advantage in Bayesian Estimation} we discuss the performance of this measurement strategy in more detail and give a comparison with the performance of classical strategies, which can be shown to exhibit shot noise scaling.

%%%%%%%%%%%%%%%%%%%%%%%%%%%%%%%%%%%%%%%%%%%%%%%%%%%%%%%%%%%%%%%%%%%%%%%%%%%%%%%%%%%%%%%%%%%%%%%%%%%%%%%%%%%%%%%%%%%%%%%%%%%%

\section{Efficient Preparation of Probe States}\label{sec:Efficient State Preparation}

In this section we present a method that allows for the efficient preparation of the probe state of Eq.~(\ref{eq:probe_state 2}), which immediately generalizes to any state in the subspace of optimal probes spanned by $\{\ket{n}\subtiny{-1}{0}{\mathrm{un}}\}_{n=0,\ldots,N}$. This method relies on the simple observation that in the bit-string $(u_{1}u_{2}u_{3}\ldots u_{N})$ representing the state
\begin{align}
    \ket{n}\subtiny{-1}{0}{\mathrm{un}} &=\,\ket{1}^{\otimes n}\ket{0}^{\otimes N-n}\,=\,\ket{u_{1}}\ket{u_{2}}\ldots \ket{u_{N}},
\end{align}
i.e., where $u_{k}\in\{0,1\}$ and $n=\sum_{k}u_{k}$, the $n$ entries $u_{1},u_{2},\ldots,u_{n}=1$ are always to the left of the entries $u_{n+1},\ldots,u_{N}=0$. In other words, the $k$-th qubit can only be in the state $|1\rangle$, if all of the $k-1$ qubits before are also in the state $|1\rangle$.

\begin{figure}[hb!]
\includegraphics[width=0.4\textwidth]{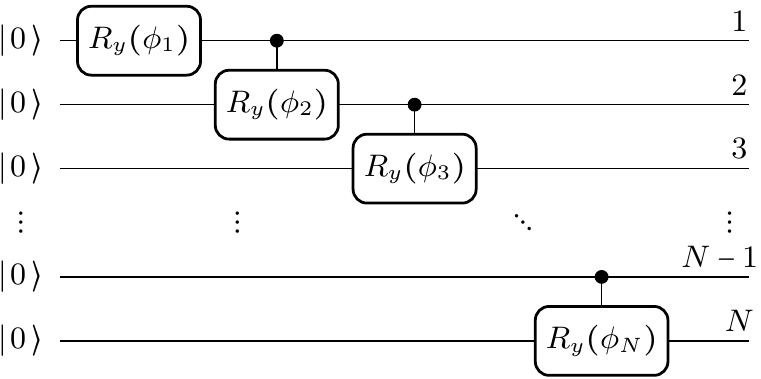}
\vspace*{-2mm}
\caption{\label{fig:circuit for BW}%(Color online)
\textbf{Circuit preparing the sine state}. After an initial single-qubit $Y$-rotation on the first qubit, a sequence of controlled $Y$-rotations, where the black dots ($\bullet$) indicate the control qubits, creates a state of the form of Eq.~(\ref{eq:probe_state 2}).}
\end{figure}

Focussing on the sine state of Eq.~(\ref{eq:Berry wiseman state}) as an example, note that the coefficients are all real and positive. Initializing all qubits in the state $\ket{0}$, the circuit preparing the sine state must hence be a cascade of $N$ (controlled) single-qubit $Y$-rotations $C\nl R_{y}(\phi_{i})$, whose angles $\{\phi_{i}\}_{i=1,\ldots,N}$ determine the weights $\psi\sub{0}{-1}{n}$, see Fig.~\ref{fig:circuit for BW}. This becomes apparent when inspecting the single-qubit Pauli-$Y$ rotations
\begin{align}
    R_{y}(\phi) &=\,\exp\bigl(i\nr \tfrac{\phi}{2}Y\bigr)\,=\,
    \begin{pmatrix}
        \ \,\cos\bigl(\tfrac{\phi}{2}\bigr)   &  \sin\bigl(\tfrac{\phi}{2}\bigr)\,\\[1mm]
        -\sin\bigl(\tfrac{\phi}{2}\bigr)   &  \cos\bigl(\tfrac{\phi}{2}\bigr)
    \end{pmatrix}\,.
\end{align}
The action of the circuit in Fig.~\ref{fig:circuit for BW} then transforms the $k$-th qubit to the state $\cos\bigl(\tfrac{\phi_{k}}{2}\bigr)\ket{0}+\sin\bigl(\tfrac{\phi_{k}}{2}\bigr)\ket{1}$ if the $(k\nr-\nr1)$-th qubit is in the state $\ket{1}$. All together, these $N$ rotations are parametrized by angles $\phi_{n}\in[0,\pi/4)$, such that both the sine and the cosine in the above expression are non-negative. It is straightforward to verify that the output of the circuit is the state of Eq.~(\ref{eq:probe_state 2}) with amplitudes
\begin{align}
  \psi\sub{0}{-1}{n}  &=
  \begin{cases}
   \displaystyle\cos\bigl(\tfrac{\phi_{n+1}}{2}\bigr)\prod_{k=1}^{n} \sin\bigl(\tfrac{\phi_{k}}{2}\bigr) &  \ \ \forall n \in \{0,1,\ldots,N-1\}\\
   \displaystyle\prod_{k=1}^{N} \sin\bigl(\tfrac{\phi_{k}}{2}\bigr) &  \ \ \mbox{for}\ $n\,=\,N$
  \end{cases}\,.
  \label{eq:BW weights vs angles}
\end{align}
%\newpage
\vspace*{-3mm}

\newpage
\noindent
Note that $\psi\sub{0}{-1}{0}$ uniquely determines $\phi_{1}$ and that each of the $\psi\sub{0}{-1}{n}$ depends only on $\{\phi_{k}\}_{k=1}^{n+1}$. This allows inverting Eq.~(\ref{eq:BW weights vs angles}) and expressing the angles $\phi_{n}$ as
\vspace*{-1mm}
\begin{align}
  \phi_{n}  &=
  \begin{cases}
        \displaystyle2\arccos(\psi\sub{0}{-1}{0})   \phantom{\frac{\psi}{\sqrt{\sum \psi}}} &   \ \mbox{for}\ n=1\\
        \displaystyle2\arccos\left(\frac{\psi_{n-1}}{\sqrt{1-\sum_{k=0}^{n-2} \psi_{k}^{2}}}\right)    &\ \mbox{for}\ n\in\{2,3,\ldots, N\}
  \end{cases}\;,
\end{align}
which allows reconstructing the rotation angles for any real, non-negative choice of $\{\psi\sub{0}{-1}{n}\}$.

\begin{figure}[hb!]
\includegraphics[width=0.38\textwidth]{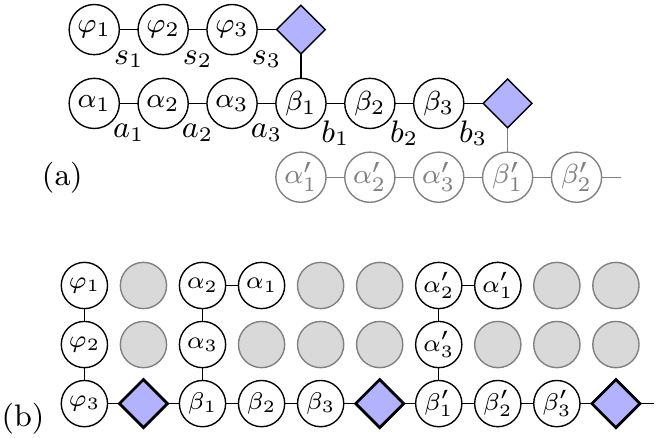}
\vspace*{-1.5mm}
\caption{\label{fig:MBQC_pattern for BW state}%(Color online)
\textbf{MBQC pattern for sine state}. In (a) the measurement pattern for the preparation of the sine state of Eq.~(\ref{eq:probe_state 2}) is shown (in part). The measurement angles $\varphi_{i}$ $(i=1,2,3)$ determine the angle $\phi_{1}$ of the first rotation $R_{y}(\phi_{1})$ in Fig.~\ref{fig:circuit for BW}, while the angles $\alpha_{i}$ and $\beta_{i}$ are chosen to realize $R_{y}(-\phi_{2}/2)$ and $R_{y}(\phi_{2}/2)$, respectively, which combine with the $C\nl Z$ gate of the cluster to realize the first controlled operation in Fig.~\ref{fig:circuit for BW}. The initial Hadamard gates to switch the qubits initialized in $\ket{+}$ to $\ket{0}$ are also included in this measurement pattern. (b) shows the pattern as part of an initial 2D cluster. Assuming that each qubit in the cluster is initially connected to its nearest neighbours, the qubits indicated by isolated gray disks have to be disconnected from the remaining cluster by $Z$-measurements. The qubits indicated by (blue) diamonds represent the probe state qubits, which are subsequently exposed to the transformation $U_{\theta}$.}
\end{figure}

Having found the circuit shown in Fig.~\ref{fig:circuit for BW}, the only difficulty is to arrange the required measurements such that the overall preparation procedure can be embedded efficiently in a rectangular 2D structure, which is shown in Appendix~\ref{sec:Probe State Preparation in MBQC}. We hence arrive at the MBQC measurement pattern depicted in Fig.~\ref{fig:MBQC_pattern for BW state}, which generates the sine state of Eq.~(\ref{eq:probe_state 2}) with weights as in Eq.~(\ref{eq:Berry wiseman state}). It requires a square 2D cluster of (at most) $3\times(4N-2)$ qubits to prepare an $N$-qubit probe state. Crucially, the number of qubits in the cluster increases only linearly with the size of the probe. Moreover, any other probe state in the subspace spanned by the vectors $\{\ket{n}\subtiny{-1}{0}{\mathrm{un}}\}_{n=0,\ldots,N}$ can be prepared with the same efficiency in a similar way by replacing the Pauli-$Y$ rotations by other single-qubit unitaries.

Next, we will show in Section~\ref{sec:Efficient Unary-to-Binary Compression} how a large class of useful measurements of the encoded probe states (including the QFT measurement) can be carried out efficiently.

%%%%%%%%%%%%%%%%%%%%%%%%%%%%%%%%%%%%%%%%%%%%%%%%%%%%%%%%%%%%%%%%%%%%%%%%%%%%%%%%%%%%%%%%%%%%%%%%%%%%%%%%%%%%%%%%%%%%%%%%%%%%

\section{Efficient Unary-to-Binary Compression}\label{sec:Efficient Unary-to-Binary Compression}

Finally, we turn to the implementation of the measurements required to achieve Heisenberg scaling. In principle, the optimal measurement for a given prior and cost function may be an arbitrarily complicated measurement in an entangled basis of $N$-qubit states, for example, a projective measurement in the QFT basis (see, e.g., Ref.~\cite[Chapter~5]{NielsenChuang2000} or~\cite{CleveWatrous2000, TakahashiKunihiroOhta2007}).

\begin{figure*}[ht!]
\begin{center}
(a)\hspace*{3mm}\includegraphics[width=0.83\textwidth]{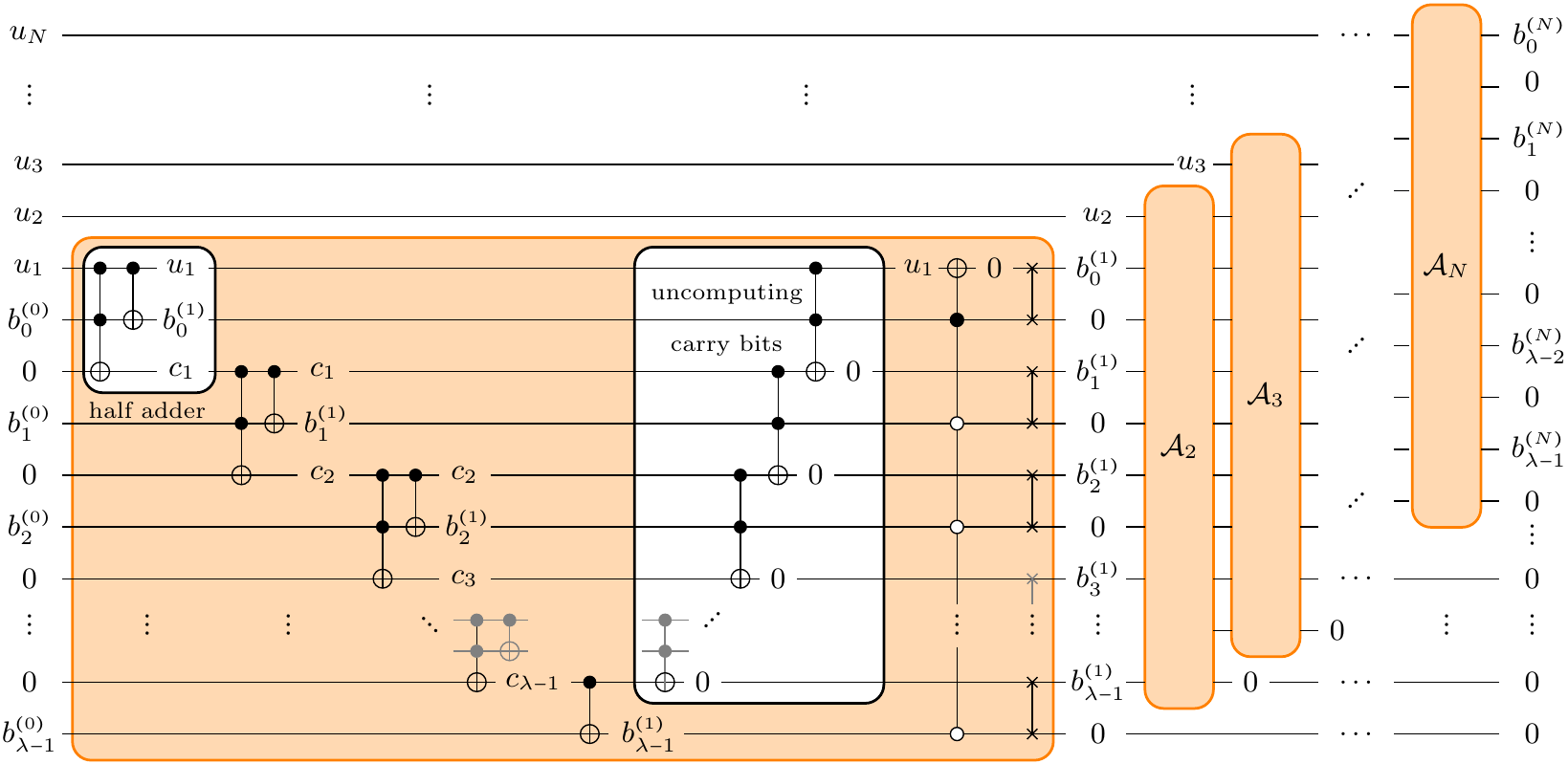}\\[2mm]
(b)\hspace*{2mm}\includegraphics[width=0.83\textwidth]{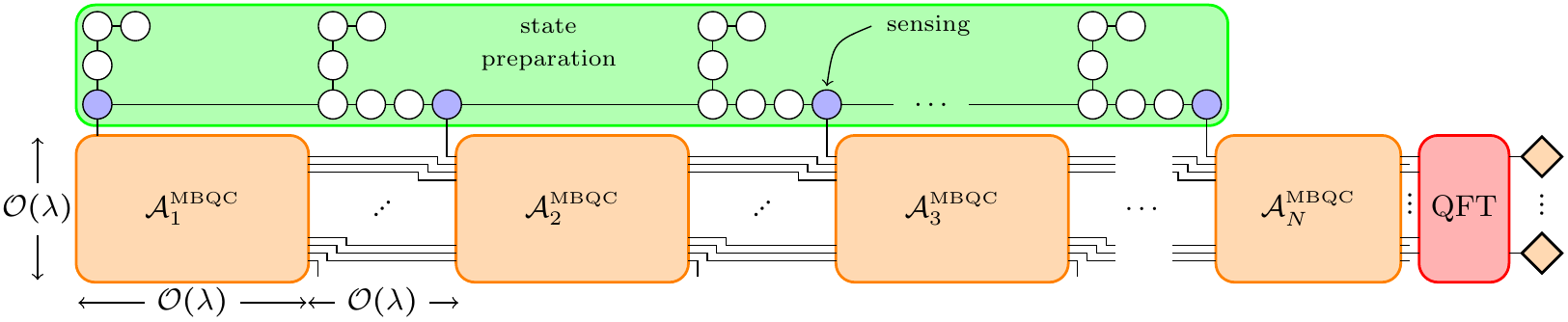}
\end{center}
\vspace*{-3mm}
\caption{\label{fig:Unary to binary compression}%(Color online)
\textbf{Unary-to-binary compression \& Bayesian estimation in MBQC}. The circuit depicted in (a) coherently compresses the $N$-qubit unary encoding $\ket{n}\subtiny{-1}{0}{\mathrm{un}}=\ket{u_{1},u_{2},\ldots,u_{N}}$ of the number $n$ (for $0\leq n\leq N$) to the binary representation $\ket{b_{0}\suptiny{0}{0}{(N)},b_{1}\suptiny{0}{0}{(N)},\ldots,b_{\lambda-1}\suptiny{0}{0}{(N)}}$ on $\lambda=\lceil\log(N\nl+\nl1)\rceil$ qubits. In each of the $N$ steps $\mathcal{A}_{k}$, one of the unary bits is added to the bits of the binary representation by way of $\lambda$ half adder circuits. Each of the latter consists of a Toffoli gate writing the carry bit on one of the $\lambda-1$ auxiliary qubits initialized in the state $\ket{0}$, and a CNOT gate carrying out the modulo-$2$ addition. The final half adder does not require its own auxiliary qubit or Toffoli gate, since the last carry bit always takes the value $0$. After the binary addition, the carry bits and the respective unary register are uncomputed, i.e., coherently erased. For the carry bits this is achieved by Toffoli gates, while the register carrying the value $u_{k}$ is switched to $0$ by a generalized Toffoli (a CNOT gate with multiple controls) conditioned on the binary encoding of the number $k$ (shown in $\mathcal{A}_{1}$ for $k=1$, where $\circ$ and $\bullet$, indicate conditioning on the states $\ket{0}$ and $\ket{1}$, respectively). A final parallel application of nearest neighbour swap gates
(\nr\protect\raisebox{-3pt}{\protect\begin{tikzpicture}
\protect\draw[black,line width=0.3pt] (0-0.05,0.13-0.05) -- (0+0.05,0.13+0.05);
\protect\draw[black,line width=0.3pt] (0-0.05,0.13+0.05) -- (0+0.05,0.13-0.05);
\protect\draw[black,line width=0.3pt] (0,0.13) -- (0,-0.13);
\protect\draw[black,line width=0.3pt] (0-0.05,-0.13-0.05) -- (0+0.05,-0.13+0.05);
\protect\draw[black,line width=0.3pt] (0-0.05,-0.13+0.05) -- (0+0.05,-0.13-0.05);
\end{tikzpicture}}\nr) arranges the auxiliary and binary register qubits appropriately for the application of the next step $\mathcal{A}_{k+1}$. The circuit depth and size of each $\mathcal{A}_{k}$ is $O(\lambda)$. In (b), the complete measurement pattern for Bayesian phase estimation in MBQC is shown, incorporating the preparation scheme (green) of Fig.~\ref{fig:MBQC_pattern for BW state} into the same 2D cluster as the measurement procedure. Note that for the parameter encoding, only the preparation part (green) should be exposed to the transformation, while the remaining cluster must be shielded or dynamically decoupled (see, e.g., Ref.~\protect\cite{ViolaKnillLloyd1998}). Each of the $\mathcal{A}_{k}$ circuits from (a) can be translated to a measurement pattern $\mathcal{A}_{k}\suptiny{0}{0}{\mathrm{MBQC}}$ on $O(\lambda^{2})$ qubits of the cluster, which are connected to the $k$-th output qubit of the preparation phase (blue disks). Black lines indicate ``teleportation wires" of length $O(\lambda)$, i.e., that additional qubits have to be introduced to connect the correct input qubits (blue) to the corresponding parts $\mathcal{A}_{k}\suptiny{0}{0}{\mathrm{MBQC}}$ of the cluster. After the unary-to-binary compression, measurements (e.g., the QFT) can be carried out efficiently on the logarithmically small subspace even if their MBQC implementation requires $O(2^{\lambda})$ qubits.
\vspace*{-2mm}
}
\end{figure*}

Fortunately, closer inspection reveals that we do not require arbitrary measurements on $N$ but only on $\lambda:=\lceil\log(N\nl+\nl1)\rceil$ qubits, where the logarithm is understood to be to base $2$. This is the case because all encoded information about the phase is stored within the $(N\nl+\nl1)$-dimensional subspace spanned by the vectors $\{\ket{n}\subtiny{-1}{0}{\mathrm{un}}\}_{n=0,\ldots,N}$. All optimal measurements can hence be restricted to this subspace. To exploit this observation, we will now present an efficient algorithm that coherently compresses the information encoded in the probe state on the $2^{N}$-dimensional Hilbert space of $N$ qubits to the exponentially smaller space of $\lambda:=\lceil\log(N\nl+\nl1)\rceil$ qubits.

The principle of operation of this $N$-step compression algorithm, shown in Fig.~\ref{fig:Unary to binary compression}~(a), is to switch from the unary encoding of the number $n$ in the state
\begin{align}
    \ket{n}\subtiny{-1}{0}{\mathrm{un}} &=\,\ket{1}^{\otimes n}\ket{0}^{\otimes N-n}\,=\,\ket{u_{1},u_{2},\ldots,u_{N}}
\end{align}
to a binary encoding of the same number via a unitary transformation and extend the result to superpositions of different states $\ket{n}\subtiny{-1}{0}{\mathrm{un}}$ by linearity. The unary-to-binary conversion is achieved by successive binary addition of each bit in the string $(u_{1},u_{2},\ldots,u_{N})$ to the bit string of an auxiliary register of length $\lambda$ initially representing the number $0$. The corresponding qubits are initialized in the state
\begin{align}
    \ket{b_{\lambda-1}\suptiny{0}{0}{(0)},b_{\lambda-2}\suptiny{0}{0}{(0)},\ldots,b_{1}\suptiny{0}{0}{(0)},b_{0}\suptiny{0}{0}{(0)}}   &=\,\ket{0}^{\otimes\lambda}\,.
\end{align}
In the $k$-th step of the procedure, the bit $u_{k}$ is added to the binary representation $(b_{\lambda-1}\suptiny{0}{0}{(k-1)},\ldots,b_{0}\suptiny{0}{0}{(k-1)})$ of the number $n\suptiny{0}{0}{(k-1)}=\sum_{i=0}^{\lambda-1}b_{i}\suptiny{0}{0}{(k-1)}2^{i}$, where $n\suptiny{0}{0}{(k)}=k$ for $0 \leq k \leq n$ and $n\suptiny{0}{0}{(k)}=n$ for $n< k\leq N$.

The binary addition of $u_{k}$ to the least significant digit $b_{0}\suptiny{0}{0}{(k-1)}$ of $n\suptiny{0}{0}{(k-1)}$ is performed by a half adder circuit, see Fig.~\ref{fig:Unary to binary compression}~(a). It, in turn consists of a  CNOT gate for the modulo-$2$ addition, producing the original value $u_{k}$ and the new binary digit $b_{0}\suptiny{0}{0}{(k)}=u_{k}\oplus b_{0}\suptiny{0}{0}{(k-1)}$, where $\oplus$ denotes addition modulo $2$. The CNOT is preceded by a Toffoli gate whose target is an additional auxiliary qubit which stores the carry bit (see, e.g., Ref.~\cite{VedralBarencoEkert1996, CuccaroDraperKutinMoulton2004, Pius:PhD2015} for quantum arithmetic operations). This carry bit is then added to the next binary digit $b_{1}\suptiny{0}{0}{(k-1)}$ by another half adder. The procedure carries on until reaching the final binary digit $b_{\lambda-1}\suptiny{0}{0}{(k-1)}$, where the half adder can be replaced by a simple CNOT gate, since the register size was chosen such that the final carry bit is always $0$.

Subsequently, the qubits corresponding to the carry bits and $u_{k}$ have to be disentangled from the qubits carrying the binary encoding. For the carry bits, this is achieved by another cascade of Toffoli gates [see Fig.~\ref{fig:Unary to binary compression}~(a)], since the carry bit can only have the value $1$, if both of the previously added bits have the value $1$ as well. To coherently erase $u_{k}$, note that the binary string $(b_{\lambda-1}\suptiny{0}{0}{(k)},\ldots,b_{0}\suptiny{0}{0}{(k)})$ encodes the number $k$ only if $u_{k}=1$. We can hence flip the corresponding qubit conditioned on the binary encoding of $k$ using a generalized Toffoli gate. Using the already existing ancillas (which have previoulsy been returned to the state $\ket{0}$), this multi-controlled CNOT gate can be realized in a standard construction using $\lambda-1$ nearest-neighbour (NN) SWAP gates, preceding and following an array of $2(\lambda-1)$ Toffolis on three adjacent qubits along with a single CNOT~\cite[p.~184]{NielsenChuang2000}. Conditioning on states $\ket{0}$ rather than $\ket{1}$ requires at most $2\lambda$ additional single-qubit $X$ gates. Having disentangled all other qubits from the $\lambda$ qubits storing the binary encoding, we perform another $\lambda$ NN SWAPS in anticipation of inputting the next unary digit $u_{k+1}$.

Taking into account that each Toffoli or NN SWAP gate can be realized with a constant overhead in NN CNOT and single-qubit gates, we find that the circuit for $\mathcal{A}_{k}$ requires at most $O(\lambda)$ NN CNOT and single-qubit gates. The entire unary-to-binary compression algorithm consists of $N$ such elements, resulting in a circuit size of $O(N \log N)$ on an input of length $O(\log N)$, which can hence be realized with at most $O(N\log^{2}\!N)$ qubits in MBQC, see Fig.~\ref{fig:Unary to binary compression}~(b).

On the logarithmically sized (in $N$) output, any measurement can then be performed efficiently as long as the corresponding unitary on $\lambda:=\lceil\log(N\nl+\nl1)\rceil$ qubits requires no more than $2^{\lambda}$ NN gates. While this does not cover all possible unitaries (e.g., the construction discussed in Ref.~\cite[p.~193]{NielsenChuang2000} requires $O(\lambda^{2}2^{2\lambda})$ two-qubit and single-qubit gates), some particularly useful unitaries may be much less costly. For instance, an implementation of the QFT on a $\lambda$-qubit linear nearest-neighbour architecture presented in Ref.~\cite{TakahashiKunihiroOhta2007} has circuit size $O(\lambda\log\lambda)$ and depth $O(\lambda)$, meaning an overhead of only $O(\lambda^{2})$ qubits (depth times input length) in a measurement-based setting.

\vspace*{-4mm}
\section{Discussion}\label{sec:conclusion}
\vspace*{-2mm}

In summary, we have shown that 2D architectures for MBQC provide flexible resources for quantum-enhanced metrology tasks. That is, an initial array of qubits prepared in a 2D cluster state and local operations are used to achieve Heisenberg scaling for phase and frequency estimation in both the local (frequentist) and the Bayesian approach to parameter estimation.
In the Bayesian scenario, the preparation procedure presented can be applied to execute strategies with optimal states for arbitrary priors and cost functions. This flexibility allows outperforming other approaches where a fixed probe state (e.g., an array of differently sized GHZ states) is used for different task without adaption to the specific problem at hand. The efficient compression algorithm further allows to perform measurements with up to exponential circuit sizes. This includes the QFT measurement that is optimal for flat priors, provides Heisenberg scaling for Gaussian priors of varying widths, and is expected to perform similarly well also for other priors under certain regularity conditions.

In principle, our results can be generalized also to scenarios beyond phase and frequency estimation. For all local Hamiltonians that are not proportional to $Z$, appropriate local corrections can be applied on the sensing qubits before and after the encoding such that the overall transformation commutes with the controlled phase gates used to create the cluster. For instance, when $H=\tfrac{1}{2}X$, Hadamard gates before and after $U_{\theta}$ produce an encoding transformation that commutes with $C\nl Z$ and can hence be applied after the entire cluster for sensing and measurements has been prepared. Moreover, when the corresponding states and measurements giving Heisenberg scaling are known, a similar method can also be employed for nonlocal interaction Hamiltonians, provided that they are proportional to a product of Pauli operators, or linear combinations of products of only one type of Pauli operators. For example, for parameter estimation with Ising-type couplings of the form $H=\sum_{i,j}c_{ij}X_{i}\otimes X_{j}$, GHZ states and local measurements achieve Heisenberg scaling~\cite{SkotioniotisSekatskiDuer2015}, which can hence be efficiently implemented in our scheme. Nonetheless, many interesting questions regarding the applicability to general dynamics and scaling beyond the Heisenberg limit~\cite{BoixoFlammiaCavesGeremia2007,RoyBraunstein2008, NapolitanoKoschorreckDubostBehboodSewellMitchell2011, Luis2010, RivasLuis2010} remain.
%in this context.

Our results are of practical significance since they suggest that a single platform, 2D cluster states, can be flexible enough for a plethora of precision-enhanced parameter estimation tasks. In addition, this platform could in principle also be part of an integrated device, where a parameter estimation strategy is used to learn about, e.g., stray fields or the particular form of noise processes. For this purpose, part of the 2D cluster state can be used for sensing, while the remaining qubits are used to perform MBQC. The gathered information from the parameter estimation can then be used to improve the performance of the computation: By learning stray fields, one can compensate for systematic errors. By learning the particular shape of a noise process, one can adapt to an optimized error correction code, thereby reducing the overhead for fault-tolerant implementations.

At the same time, this connection between computational and metrological resources provides interesting insights. The advantage in metrology is provided by the entanglement of the cluster state, i.e., the $C\nl Z$ gates applied to neighbouring pairs of qubits, which ensures the improved performance with respect to an array of unentangled, individual qubits. At the same time, it is known that metrological advantages can, but need not arise solely from entanglement~\cite{BoixoDattaDavisFlammiaShajiCaves2008, SahotaQuesada2015, KokDunninghamRalph2017}. For example, nonclassicality in terms of squeezing can lead to Heisenberg scaling in precision~\cite{GaibaParis2009, FriisSkotiniotisFuentesDuer2015} without any entanglement when the average energy is considered as the resource. This work hence also contributes to the discussion of the required physical resources for parameter estimation~\cite{SahotaQuesadaJames2016}, and the relationship between computational power and metrology~\cite{DemkowiczDobrzanskiMarkiewicz2015}.

Finally, open questions remain regarding the role of noise~\cite{EscherDeMatosFilhoDavidovich2011, DemkowiczDobrzanskiKolodinskiGuta2012, KnyshChenDurkin2014}, especially in connection with adaptive approaches to computation and error-correction involving metrology~\cite{DuerSkotiniotisFroewisKraus2014, SekatskiSkotiniotisDuer2016, SekatskiSkotiniotisKolodynskiDuer2016, CombesFerrieCesareTierschMilburnBriegelCaves2014, OrsucciTierschBriegel2016, PirandolaLupo2017}. Although noise is known to be problematic in the limit of infinitely many qubits since it is known to restrict to a linear scaling of precision, i.e., $\mathcal{I}\leq\kappa N$ for some constant $\kappa$, the approach presented here holds the promise of significantly outperforming classical strategies for finite system sizes. Indeed, this follows from the observation that the constant $\kappa$ strongly depends on the strength and type of the noise~\cite{DemkowiczDobrzanskiKolodinskiGuta2012, SekatskiSkotiniotisKolodynskiDuer2016} and can be arbitrarily large if the noise is weak enough. Meanwhile, the overhead needed for preparation and measurement of the optimal state does not depend on the noise, leaving room for an arbitrarily large advantage of our scheme over classical strategies for any fixed $N$. In addition, techniques that deal with errors and maintain a metrological advantage are known (see, e.g.,~\cite{DuerSkotiniotisFroewisKraus2014, SekatskiSkotiniotisDuer2016, SekatskiSkotiniotisKolodynskiDuer2016}) and may be applicable here. We leave such extensions for future work, along with the explicit determination of optimal~\cite{SandersMilburnZhang1997, FroewisSkotiniotisKrausDuer2014} and ``pretty good" states~\cite{SkotioniotisFroewisDuerKraus2015} for specific metrological tasks in our framework, where recent algorithmic approaches~\cite{Knott2016} may prove to be useful.

\vspace*{-1mm}
\begin{acknowledgements}
\vspace*{-1mm}
We are grateful to Jan Ko{\l}ody{\'n}ski and Markus Tiersch for valuable discussions and comments. This work was supported by the Austrian Science Fund (FWF) through Grants No.~SFB FoQuS F4012, No.~P28000-N27 and the START  project Y879-N27, as well as DK ALM:W1259, the Templeton World Charity Foundation Grant No.~TWCF0078/AB46, the Swiss National Science Foundation Grant No.~P300P2\_167749, the Spanish MINECO through Grant No.~FIS2013-40627-P, the Generalitat de Catalunya CIRIT contract 2014-SGR966 and by TherMiQ (Grant Agreement 618074).
\end{acknowledgements}

%%%%%%%%%%%%%%%%%%%%%%%%%%%%%%%%%%%%%%%%%%%%%%%%%%%%%%%%%%%%%%%%%%%%%%%%%%%%%%%%%%%%%%%%%%%%%%%%%%%%%%%%%%%%%%%%%%%
%%%%%%%%%%%%%%%%%%%%%%%%%%%%%%%%%%%%%%%%%%%%%%%%%%%%%%%%%%%%%%%%%%%%%%%%%%%%%%%%%%%%%%%%%%%%%%%%%%%%%%%%%%%%%%%%%%%

%\newpage
\hypertarget{sec:appendix}
\appendix
\section*{Appendix}
%\addcontentsline{Appendix}
\renewcommand{\thesubsubsection}{A.\Roman{subsection}.\arabic{subsubsection}}
\renewcommand{\thesubsection}{A.\Roman{subsection}}
\renewcommand{\thesection}{A}
\setcounter{equation}{0}
\numberwithin{equation}{section}
\setcounter{figure}{0}
\renewcommand{\thefigure}{A.\arabic{figure}}

\begin{center}
%\begin{large}
\textbf{Table of Contents}\\[3.5mm]
%\end{large}
\hspace*{-2mm}
\begin{tabular}{l p{7.4cm} l}
  \ref{sec:local estimation}    &   Local Parameter Estimation $.\ldots\ldots\ldots\ldots\ldots\ldots.$    &   \pageref{sec:local estimation}\\[1mm]
  \ref{sec:Bayesian estimation} &   Bayesian Parameter Estimation $.\ldots\ldots\ldots\ldots\ldots.$    &   \pageref{sec:Bayesian estimation}\\[1mm]
  \ref{sec:Optimal classical Bayesian estimation strategy} &   Classical Bayesian Estimation Strategies $.\ldots\ldots.$    &   \pageref{sec:Optimal classical Bayesian estimation strategy}\\[1mm]
  \ref{sec:Quantum Advantage in Bayesian Estimation} &   Quantum Advantage in Bayesian Estimation $\ldots$    &   \nr\pageref{sec:Quantum Advantage in Bayesian Estimation}\\[1mm]
  \ref{sec:Bayesian frequency estimation in MBQC} &   Bayesian Frequency Estimation $.\ldots\ldots\ldots\ldots\ldots.$    &   \nr\pageref{sec:Bayesian frequency estimation in MBQC}\\[1mm]
  \ref{sec:Measurement-Based Quantum Computation} &   Measurement-Based Quantum Computation $\ldots.$    &   \nr\pageref{sec:Measurement-Based Quantum Computation}
\end{tabular}
\end{center}

\subsection{Local Parameter Estimation}\label{sec:local estimation}

%%%%%%%%%%%%%%%%%%%%%%%%%%%%%%%%%%%%%%%%%%%%%%%%%%%%%%%%%%%%%%%%%%%%%%%%%%%%%%%%%%%%%%%%%%%%%%%%%%%%%%%%%%%%%%%%%%%%%%%%%%%%

In this appendix, we give a detailed description of the local parameter estimation scenario and show how Heisenberg scaling can be achieved using a GHZ state and local measurements.

\subsubsection{The Local Estimation Scenario}\label{sec:The local parameter estimation scenario}

We consider a typical parameter estimation scenario, where $\theta$, the quantity of interest, is encoded in a density operator $\rho(\theta)$ by a dynamical (unitary) transformation $U_{\theta}=e^{-i\nr\theta\nr H}$, i.e.,
\begin{align}
    \rho(\theta)    &=\,U_{\theta}\,\rho(0)\,U_{\theta}^{\dagger}\,.
    \label{eq:unitary encoding appendix}
\end{align}
We then perform a measurement with POVM elements $\{E_{m}\}$ which yields an outcome $m$. The (conditional) probability of obtaining the measurement outcome $m$ (given that the parameter has the value $\theta$) is then
\begin{align}
    p(m|\nr\theta\nr) &=\,\tr\bigl(E_{m}\nr\rho(\theta)\bigr).
    \label{eq:conditional m given theta}
\end{align}
To each measurement outcome $m$, an estimator $\hat{\theta}(m)$ assigns a corresponding estimate for the value of $\theta$. The estimator is called \emph{unbiased} if it assigns the value $\theta$ on average, that is, if the expected value of the estimator satisfies
\begin{align}
    \expval{\hat{\theta}(m)} &=\,\sum\limits_{m}p(m|\nr\theta\nr)\,\hat{\theta}(m)    \,=\,\theta\,.
    \label{eq:unbiasdness condition}
\end{align}
This requirement ensures the \emph{accuracy} of the measurement procedure, but not its \emph{precision}, which is determined by the variance $V[\hat{\theta}(m)]$ of the estimator. We use the mean-square error (MSE) given by
\begin{align}
    V[\hat{\theta}(m)]   &=\,\sum\limits_{m}p(m|\nr\theta\nr)\,\bigl(\hat{\theta}(m)-\theta\bigr)^{2}\,,
    \label{eq:MSE variance local estimation}
\end{align}
and $\sigma=\sqrt{V[\hat{\theta}(m)]}$ is the associated standard deviation. Unfortunately, it is often the case that a given estimator offers high precision only within a small range of the parameter $\theta$, but not globally, as we shall discuss for a simple example in Appendix~\ref{sec:Local Metrology using GHZ states}. Such estimators are hence useful \emph{locally}, i.e., for estimating small fluctuations of the parameter around some known value. In such local estimation scenarios, accuracy is guaranteed even when unbiasedness as specified in Eq.~(\ref{eq:unbiasdness condition}) is required to hold only in the vicinity of this value.

To increase the precision, the procedure consisting of preparation, encoding, and measurement may be repeated a number of times, say $\nu$, providing estimates $\theta^{(i)}$ $(i=1,\ldots,\nu)$, from which the mean value
\begin{align}
\bar{\theta}_{\nu}    &=\,\frac{1}{\nu}\sum\limits_{i=1}^{\nu}\theta^{(i)}
\end{align}
and the associated MSE
\begin{align}
V_{\!\nu}[\theta^{(i)}]    &=\,\frac{1}{\nu}\sum\limits_{i=1}^{\nu}\bigl(\theta^{(i)}-\bar{\theta}_{\nu}\bigr)^{2}
\end{align}
can be calculated. As $\nu$ increases, the mean and variance computed from the measurement data converge to the expected value $\expval{\hat{\theta}(m)}$ of the estimates and the expected value of the corresponding variance, $V[\hat{\theta}(m)]$, respectively. Trusting that the results of the individual runs are independent and identically distributed (i.i.d.), the variance of the distribution of mean values with $\nu$ samples decreases linearly with $\nu$. The overall expected precision associated to the result $\bar{\theta}_{\nu}$ is hence quantified by the standard error of the mean, given by $\sigma_{\nu}=\sigma/\sqrt{\nu}=\sqrt{V[\hat{\theta}(m)]/\nu}$. In other words, the precision increases with the number of runs, but the options for choosing a probe state, measurement, and estimator still leave room for improvement.

It is here that measurement strategies using genuine quantum features such as entanglement and nonclassicality can provide advantages with respect to classical strategies. To determine the potential gain and to allow comparisons with the best classical protocol it is useful to eliminate the choice of estimator, and consider the important Cram{\'e}r-Rao bound, before discussing an example estimation scenario in Appendix~\ref{sec:Local Metrology using GHZ states}.

%%%%%%%%%%%%%%%%%%%%%%%%%%%%%%%%%%%%%%%%%%%%%%%%%%%%%%%%%%%%%%%%%%%%%%%%%%%%%%%%%%%%%%%%%%%%%%%%%%%%%%%%%%%%%%%%%%%%%%%%%%%%

\subsubsection{The Cram{\'e}r-Rao bound}\label{sec:cramer rao bound}

For any unbiased estimator the variance $V[\hat{\theta}(m)]$ can be shown (see, e.g., Refs.~\cite{Cramer:Methods1946, VanTrees1968, Frieden2004} or Appendix~\ref{sec:cramer rao bound proof}) to satisfy the \emph{Cram{\'e}r-Rao} (CR) inequality
\begin{align}
    V[\hat{\theta}(m)]   &\geq\,\frac{1}{I\bigl(\rho(\theta)\bigr)}\,,
    \label{eq:Cramer Rao bound}
\end{align}
where $I\bigl(\rho(\theta)\bigr)$ is the \emph{Fisher information} (FI) given by
\begin{align}
    I\bigl(\rho(\theta)\bigr)   &=\,\sum\limits_{m}p(m|\nr\theta\nr)\,\Bigl(\frac{\partial}{\partial\theta}\log p(m|\nr\theta\nr)\Bigr)^{2}\nonumber\\[1mm]
    &=\,\sum\limits_{m}\frac{\bigl(\tr[E_{m}\nr\dot{\rho}(\theta)]\bigr)^{2}}{\tr\bigl(E_{m}\nr\rho(\theta)\bigr)}\,.
    \label{eq:Fisher information}
\end{align}
Here it is noteworthy that, on the one hand, the FI does not depend on the choice of the estimator (as long as it is unbiased), and one can hence determine a lower bound for the variance based solely on the initial state and the chosen measurement. On the other hand, the FI typically depends on the value of the parameter and an unbiased estimator for which the CR inequality globally becomes an equality may not exist for all values. However, estimators can be found for which the bound is tight locally, and globally in the asymptotic limit of $\nu\rightarrow\infty$, see, e.g., Ref.~\cite{Kay1993}.

One may then further ask, what the optimal measurement strategy is for a given probe state and parameter encoding. The maximization of the FI over all possible POVMs then yields (see, e.g.,~\cite{BraunsteinCaves1994}) the \emph{quantum Fisher information} (QFI) $\mathcal{I}\bigl(\rho(\theta)\bigr)$, given by
\begin{align}
    \mathcal{I}\bigl(\rho(\theta)\bigr) &=\,2\,\tr\bigl(\hat{S}_{\theta}\,\dot{\rho}(\theta)\bigr)\,,
    \label{eq:quantum fisher information}
\end{align}
where the operator $\hat{S}_{\theta}\equiv \hat{S}[\rho(\theta)]$, called the \emph{symmetric logarithmic derivative} (SLD) is implicitly given by the relation
\begin{align}
    \hat{S}_{\theta}\nr\rho(\theta)\,+\,\rho(\theta)\nr\hat{S}_{\theta} &=\,\dot{\rho}(\theta)\,,
    \label{eq:SLD}
\end{align}
and where the dot indicates the partial derivative with respect to $\theta$, i.e., $\dot{\rho}=\tfrac{\partial}{\partial\theta}\rho$. The corresponding quantum \emph{Cram{\'e}r-Rao} bound is hence simply $V[\hat{\theta}(m)]\geq1/\mathcal{I}\bigl(\rho(\theta)\bigr)$. The optimal measurement for which the FI and the QFI coincide is a projective measurement in the eigenbasis of the SLD $\hat{S}_{\theta}$~\cite{BraunsteinCaves1994}.

For unitary encodings that we consider here, see Eq.~(\ref{eq:unitary encoding appendix}), the QFI is independent of the value of the parameter, $\mathcal{I}\bigl(\rho(\theta)\bigr)=\mathcal{I}\bigl(\rho(\theta\pr)\bigr)\ \forall \theta,\theta\pr$. To see this, simply note that in such a case $\rho(\theta\pr)=U^{\ }_{\!\theta\pr-\theta}\nr\rho(\theta)U_{\theta\pr-\theta}^{\dagger}$ and
\begin{align}
    \dot{\rho}(\theta)  &=\,i\comm{\rho(\theta)}{H}\,.
\end{align}
Therefore the derivative appearing in the QFI in Eq.~(\ref{eq:quantum fisher information}) is just $\dot{\rho}(\theta\pr)=U_{\!\theta\pr-\theta}^{\ }\nr\dot{\rho}(\theta)U_{\theta\pr-\theta}^{\dagger}$. Using Eq.~(\ref{eq:SLD}) one then finds that the SLDs are related in the same way, i.e., $\hat{S}_{\theta\pr}=U^{\ }_{\!\theta\pr-\theta}\nr\hat{S}_{\theta}U_{\theta\pr-\theta}^{\dagger}$. Cyclically permuting the unitary operators in the trace then gives the result, $\mathcal{I}\bigl(\rho(\theta)\bigr)=\mathcal{I}\bigl(\rho(\theta\pr)\bigr)$.

If we additionally restrict to pure probe states $\ket{\psi}$ as before, the QFI takes the simple form (see, e.g., Ref.~\cite{DemkowiczDobrzanskiJarzynaKolodynski2015})
\begin{align}
    \mathcal{I}(\ket{\psi}) &   =\,4\Bigl(\,\scpr{\dotpsitheta}{\dotpsitheta}\,-\,|\!\scpr{\dotpsitheta}{\psitheta}\!|^{2}\Bigr)\,,
    \label{eq:pure state unitary encoding Fisher}
\end{align}
where $\ket{\psitheta}=U_{\theta}\ket{\psi}$ is the encoded state and the dot indicates a partial derivative with respect to $\theta$. Since $U_{\theta}=e^{-i\theta H}$, a simple computation then reveals that the QFI for such scenarios is proportional to the variance of the Hamiltonian generating the dynamics, i.e.,
\begin{align}
    \mathcal{I}(\ket{\psi}) &=\,4\,\bigl(\expval{H^{2}}_{\psitheta}\,-\,\expval{H}_{\psitheta}^{2}\bigr)\,=\,4(\Delta H)^{2}\,
    \label{eq:pure state unitary encoding Fisher H}
\end{align}
and the SLD coincides with $\dot{\rho}(\theta)$. The QFI is hence maximal for pure states that maximize the variance of $H$, see, e.g., Refs.~\cite{DemkowiczDobrzanskiJarzynaKolodynski2015,LiuJingZhongWang2014}.

Let us now consider an estimation scenario where a probe state of $N$ qubits is subject to a local transformation, i.e., where the Hamiltonian is of the form $H=\sum_{i=1,\ldots,N}H_{i}$ and $H_{i}$ acts nontrivially only on the $i$th qubit. For simplicity, we assume that each qubit undergoes the same local transformation, $H_{i}=H_{j}\equiv H_{1}\forall i,j$, and that the local Hamiltonian has eigenvalues\footnote{Any deviation from this assumption enters the problem as a constant factor multiplying the parameter $\theta$, and can hence be absorbed into $\theta$.} $\pm\tfrac{1}{2}$ with the corresponding eigenstates denoted by $\ket{0}$ and $\ket{1}$. We may further align our reference frame such that $H_{1}=S_{z}=\tfrac{1}{2}Z$, where $S_{i}$ is the spin-$\tfrac{1}{2}$ angular momentum operator for direction $i=x,y,z$ and $X,Y,Z$ denote the usual Pauli operators. A comment on estimation scenarios for other Hamiltonians can be found in Section~\ref{sec:conclusion} of the main text, but here we are restricting our discussion to phase estimation scenarios where $U_{\theta}=\bigotimes_{n=1}^{N}U_{\theta}\suptiny{0}{0}{(n)}$, such that $U_{\theta}\suptiny{0}{0}{(n)}=\exp(-i\theta S_{z}\suptiny{0}{0}{(n)})$ acts only on the $n$th qubit. For ease of notation, we will drop the superscript $(n)$ in the following when referring to single-qubit operations and there is no risk of confusion.

If the probe state is classical, i.e., a product state of the form $\ket{\psi_{1}}\otimes\ket{\psi_{2}}\otimes\ldots\otimes\ket{\psi_{N}}$, then the QFI becomes maximal when the local single-qubit probe states are all chosen to be $\ket{+}=\bigl(\ket{0}+\ket{1}\bigr)/\sqrt{2}$, maximizing the variance of $H_{1}=S_{z}$. From Eq.~(\ref{eq:pure state unitary encoding Fisher H}) it then follows immediately that the largest possible value of the QFI for a classical probe of $N$ qubits is
\begin{align}
    \mathcal{I}\bigl(\ket{+}^{\otimes N}\bigr) &=\,4\nr N\,\bigl(\expval{H_{1}^{2}}\,-\,\expval{H_{1}}^{2}\bigr)\,=\,N\,.
    \label{eq:pure state unitary encoding Fisher product state}
\end{align}
The corresponding SLD is easily found to be $\hat{S}_{\theta}=\bigl(\cos\theta\nr S_{y}-\sin\theta\nr S_{x}\bigr)^{\otimes N}$, i.e., the optimal measurement is realized by single-qubit projective measurements in the basis $U_{\theta+\pi/2}\ket{\pm}$, where $\ket{\pm}$ are the eigenstates of $S_{x}=\tfrac{1}{2}X$. It hence becomes obvious that classical measurement strategies can (at most) decrease the variance linearly with the number of qubits. This scaling behaviour is referred to as the \emph{standard quantum limit}. As we shall discuss next, a different scaling behaviour can be achieved for quantum probes.

%%%%%%%%%%%%%%%%%%%%%%%%%%%%%%%%%%%%%%%%%%%%%%%%%%%%%%%%%%%%%%%%%%%%%%%%%%%%%%%%%%%%%%%%%%%%%%%%%%%%%%%%%%%%%%%%%%%%%%%%%%%%

\subsubsection{A Proof of the Cram{\'e}r-Rao Bound}\label{sec:cramer rao bound proof}

In this section we present a proof of the Cram{\'e}r-Rao bound of Eq.~(\ref{eq:Cramer Rao bound}) following Refs.~\cite{VanTrees1968,Frieden2004}. For an unbiased estimator $\hat{\theta}(m)$, we can write the unbiasedness condition of Eq.~(\ref{eq:unbiasdness condition}) as
\begin{align}
    \sum\limits_{m}p(m|\nr\theta\nr)\,\bigl(\hat{\theta}(m)-\theta\bigr) &=\,0\,,
    \label{eq:unbiasdness condition appendix}
\end{align}
where we have used that the conditional probability is normalized, i.e., $\sum_{m}p(m|\nr\theta\nr)=1$. Differentiating the condition of Eq.~(\ref{eq:unbiasdness condition appendix}) with respect to the parameter we have
\begin{align}
    \sum\limits_{m}\frac{\partial p(m|\nr\theta\nr)}{\partial\theta}\,\bigl(\hat{\theta}(m)-\theta\bigr)\,-\,\sum\limits_{m}p(m|\nr\theta\nr)    &=\,0\,,
\end{align}
which we can rewrite as
\begin{align}
    \sum\limits_{m}\bigl(\hat{\theta}(m)-\theta\bigr)\,p(m|\nr\theta\nr)\,\frac{\partial}{\partial\theta}\,\log p(m|\nr\theta\nr)    &=\,1\,.
\end{align}
Then, we define the quantities $x:=\sqrt{p(m|\nr\theta\nr)}\bigl(\hat{\theta}(m)-\theta\bigr)$ and
\begin{align}
    y   &:=\,\sqrt{p(m|\nr\theta\nr)}\,\frac{\partial}{\partial\theta}\,\log p(m|\nr\theta\nr)\,,
\end{align}
and use the Cauchy-Schwarz inequality
\begin{align}
    \left|\sum\hspace*{-5mm}\int x\,y\right|^{2}    &\leq\,\sum\hspace*{-5mm}\int \,\left|x\right|^{2}\,\sum\hspace*{-5mm}\int \,\left|y\right|^{2}
    \label{eq:Cauchy Schwarz ineq}
\end{align}
to arrive at
\begin{align}
    \sum\limits_{m}p(m|\nr\theta\nr)\,\bigl(\hat{\theta}(m)-\theta\bigr)^{2}\times
    \sum\limits_{n}p(n|\nr\theta\nr)\,\Bigl(\frac{\partial}{\partial\theta}\,\log p(n|\nr\theta\nr)\Bigr)^{2}
    &\geq1.
    \label{eq:cramer rao proof}
\end{align}
The first factor on the left-hand side of Eq.~(\ref{eq:cramer rao proof}) is just the variance
\begin{align}
    V[\hat{\theta}(m)]   &=\,\sum\limits_{m}p(m|\nr\theta\nr)\,\bigl(\hat{\theta}(m)-\theta\bigr)^{2}\,.
    \label{eq:local variance appendix}
\end{align}
Dividing by the second factor, which we identify with the Fisher information
\begin{align}
    I\bigl(\rho(\theta)\bigr)   &=\,\sum\limits_{m}p(m|\nr\theta\nr)\,\Bigl(\frac{\partial}{\partial\theta}\log p(m|\nr\theta\nr)\Bigr)^{2}\,,
    \label{eq:Fisher information appendix}
\end{align}
we finally obtain the \emph{Cram{\'e}r-Rao} inequality
\begin{align}
    V[\hat{\theta}(m)]   &\geq\,\frac{1}{I\bigl(\rho(\theta)\bigr)}\,.
    \label{eq:Cramer Rao bound appendix}
\end{align}

%%%%%%%%%%%%%%%%%%%%%%%%%%%%%%%%%%%%%%%%%%%%%%%%%%%%%%%%%%%%%%%%%%%%%%%%%%%%%%%%%%%%%%%%%%%%%%%%%%%%%%%%%%%%%%%%%%%%%%%%%%%%

\subsubsection{Heisenberg Scaling in Local Metrology}\label{sec:Local Metrology using GHZ states}

Let us now revisit the local phase estimation scenario for an entangled state, for instance, the $N$-qubit GHZ state, given by
\begin{align}
    \ket{\psi\subtiny{0}{0}{\mathrm{GHZ}}}  &=\,\tfrac{1}{\sqrt{2}}\Bigl(\,\ket{0}^{\otimes N}+\ket{1}^{\otimes N}\Bigr)\,.
    \label{eq:GHZ state}
\end{align}
A quick calculation of the QFI of Eq.~(\ref{eq:pure state unitary encoding Fisher}) for this state provides the result $\mathcal{I}(\ket{\psi\subtiny{0}{0}{\mathrm{GHZ}}})=N^{2}$. The precision may hence quadratically increase with the number of qubits. This optimal scaling behaviour is usually called the \emph{Heisenberg limit}. To see how one can practically achieve Heisenberg scaling, let us consider a simple parity measurement, that is, a projective measurement with outcomes $m=+1$ (even) and $m=-1$ (odd), and associated POVM elements
\begin{align}
    E_{\mathrm{even}}&=\,\sum\limits_{n\ \mathrm{even}}\,E_{n}\,\ \ \mbox{and}\ \ E_{\mathrm{odd}}\,=\,\sum\limits_{n\ \mathrm{odd}}\,E_{n}\,,
\end{align}
where $E_{n}$ projects into the subspace where $n$ qubits are in the state $\ket{-}$ and $(N-n)$ qubits are in the state $\ket{+}$. Denoting the single-qubit projectors as $P_{\pm}=\ket{\pm}\!\!\bra{\pm}$, we can write
\begin{align}
    E_{n}   &=\,\sum\limits_{i}\pi_{i}\bigl(P_{-}^{\otimes n}P_{+}^{\otimes N-n}\bigr)\,,
    \label{eq:single qubit n minus outcome POVM}
\end{align}
where the sum is over all $\tbinom{N}{n}$ permutations $\pi_{i}$. One then straightforwardly finds
\begin{align}
    \tr\bigl(E_{n}\nr\rho(\theta)\bigr) &=\,\frac{1}{2^{N}}\binom{N}{n}\,\bigl(1+(-1)^{n}\cos(\nl N\theta)\bigr)\,,
\end{align}
which in turn yields the conditional probabilities
\begin{subequations}
\begin{align}
    p(+|\nr\theta\nr) &=\,\tr\bigl(E_{\mathrm{even}}\nr\rho(\theta)\bigr)\,=\,\cos^{2}\!\bigl(\tfrac{N\theta}{2}\bigr)\,,\\[1mm]
    p(-|\nr\theta\nr) &=\,\tr\bigl(E_{\mathrm{odd}}\nr\rho(\theta)\bigr)\,=\,\sin^{2}\!\bigl(\tfrac{N\theta}{2}\bigr)\,.
\end{align}
\end{subequations}
Using the definition in Eq.~(\ref{eq:Fisher information}) one can then verify that this measurement is optimal, i.e., the FI and QFI coincide, $I(\ket{\psi\subtiny{0}{0}{\mathrm{GHZ}}})=\mathcal{I}(\ket{\psi\subtiny{0}{0}{\mathrm{GHZ}}})=N^{2}$.

We then only need to find a suitable estimator. We can construct such an estimator from the expected value of the associated observable $M$, which has the spectral decomposition $M=E_{\mathrm{even}}-E_{\mathrm{odd}}=X^{\otimes N}$, such that $\expval{M}=\cos(N\theta)$. Crucially, note that the required measurements are just local $X$-measurements, the results of which are multiplied to obtain the overall measurement result in each run, i.e., $m=m_{1}m_{2}\ldots m_{N}$. For $\theta\in[0,\tfrac{\pi}{N}]$ we then assign the estimator
\begin{align}
    \hat{\theta}(m)  &=\,\tfrac{1}{N}\arccos(m)\,=\,\tfrac{1-m}{2N}\pi\,=\,\begin{cases} 0   &   \mbox{if}\ \ m=+1 \\ \tfrac{\pi}{N}   &   \mbox{if}\ \ m=-1 \end{cases}\,.
\end{align}
Computing the mean and variance for this estimator one finds,
\begin{subequations}
\begin{align}
    \expval{\hat{\theta}(m)} &=\,\tfrac{\pi}{N}\,\sin^{2}\!\bigl(\tfrac{N\theta}{2}\bigr)\,,\\[1mm]
    V[\hat{\theta}(m)]   &=\,\theta^{2}\,+\,\sin^{2}\!\bigl(\tfrac{N\theta}{2}\bigr)\,\bigl(\tfrac{\pi^{2}}{N^{2}}-\tfrac{2\pi\theta}{N}\bigr)\,.
\end{align}
\end{subequations}
The estimator is only unbiased for $\theta=\tfrac{\pi}{2N}$, but in this case the variance admits Heisenberg scaling and takes the value $\tfrac{\pi^{2}}{4N^{2}}$. However, one can do better than this by averaging over the outcomes before assigning an estimate, rather than averaging the individual estimates. Practically speaking, one can view this as estimating $m(\theta)=\cos(N\theta)$ followed by a simple reparametrization using $\theta(m)=\arccos(m)/N$. This estimator is unbiased by definition, since $m(\theta)=\expval{M}$ and one finds the variance
\begin{align}
    V[m(\theta)]    &=\,\sum\limits_{m}p(m|\nr m(\theta)\nr)\bigl(m-\expval{M}\bigr)^{2}\,=\,\sin^{2}\!(\nl N\theta)\,.
\end{align}
Propagating the error through the reparameterization then yields
\begin{align}
    V[\theta(m)]    &=\,V[m(\theta)]\,\Bigl(\frac{\partial \theta}{\partial m}\Bigr)^{2}\,=\,\frac{1}{N^{2}}\,.
\end{align}
One can hence get a quadratic scaling advantage for local phase estimation using an $N$-qubit GHZ state and local measurements. By extension via error propagation, Heisenberg scaling is also maintained for frequency estimation by reparameterizing $\theta=\omega t$ for any fixed interrogation time $t$. As shown in Fig.~\ref{fig:local phase estimation in MBQC} in the main text, the preparation of an $N$-qubit GHZ state can be realized using a $(2N-1)$-qubit 1D cluster state, which hence constitutes a resource for local phase and frequency estimation at the Heisenberg limit.

%%%%%%%%%%%%%%%%%%%%%%%%%%%%%%%%%%%%%%%%%%%%%%%%%%%%%%%%%%%%%%%%%%%%%%%%%%%%%%%%%%%%%%%%%%%%%%%%%%%%%%%%%%%%%%%%%%%%%%%%%%%%%%%%%%%%
%%%%%%%%%%%%%%%%%%%%%%%%%%%%%%%%%%%%%%%%%%%%%%%%%%%%%%%%%%%%%%%%%%%%%%%%%%%%%%%%%%%%%%%%%%%%%%%%%%%%%%%%%%%%%%%%%%%%%%%%%%%%%%%%%%%%
%%%%%%%%%%%%%%%%%%%%%%%%%%%%%%%%%%%%%%%%%%%%%%%%%%%%%%%%%%%%%%%%%%%%%%%%%%%%%%%%%%%%%%%%%%%%%%%%%%%%%%%%%%%%%%%%%%%%%%%%%%%%%%%%%%%%

\subsection{Bayesian Parameter Estimation}\label{sec:Bayesian estimation}

In Appendix~\ref{sec:The Bayesian Estimation Scenario}, we first review the basic structure of Bayesian parameter estimation problems. We then discuss an inequality that serves as a Bayesian analogue of the Cram{\'e}r-Rao bound in Appendix~\ref{sec:bayesian cramer rao bound} and present a simple proof in Appendix~\ref{sec:bayesian cramer rao bound proof}, before highlighting an interesting connecting between Bayesian estimation and noisy local estimation in Appendix~\ref{sec:connection of Gaussian Bayesian to noisy local}. Finally, we investigate the limitations of the MSE cost function for Bayesian estimation in Appendix~\ref{sec:limitations of the MSE approach}.

\subsubsection{The Bayesian Estimation Scenario}\label{sec:The Bayesian Estimation Scenario}

Much like in the local estimation scenario discussed in Appendix~\ref{sec:local estimation}, the Bayesian scenario considers the estimation of a parameter $\theta$ that has been encoded onto a quantum state $\rho(\theta)$ by performing a measurement given by some POVM $\{E_{m}\}$. As before, the conditional probability to obtain the outcome $m$ given that the parameter has the value $\theta$ is
\begin{align}
    p(m|\nr\theta\nr) &=\,\tr\bigl(E_{m}\nr\rho(\theta)\bigr).
    \label{eq:conditional m given theta 2}
\end{align}
However, where the local estimation scenario requires only that the parameter be close to values for which an unbiased estimator is available, the Bayesian estimation scenario captures all previously held belief about $\theta$ in a probability distribution referred to as the \emph{prior} $p(\theta)$. Performing a single measurement, the probability to obtain the outcome $m$ is then simply
\begin{align}
    p(m)    &=\,\int\!d\theta\,p(\theta)\,p(m|\nr\theta\nr)\,=\,\tr\bigl(E_{m}\nr\Gamma\bigr)\,,
    \label{eq:prob for m unconditional}
\end{align}
where we have defined the quantity
\begin{align}
    \Gamma &=\,\int\!d\theta\,p(\theta)\,\rho(\theta)\,,
    \label{eq:Gamma}
\end{align}
following the notation of Ref.~\cite{Personick1971}). Given some outcome $m$, we then want to provide an estimate $\hat{\theta}(m)$ for the value of the parameter. To this end, note that Bayes' law lets us determine $p(\theta|m)$, the probability that the parameter had the value $\theta$ given the outcome $m$, as
\begin{align}
    p(\theta|m) &=\,\frac{p(m|\nr\theta\nr)\,p(\theta)}{p(m)}\,.
    \label{eq:Bayes law}
\end{align}
As an estimate we then simply average the possible values of $\theta$ weighted with the corresponding probabilities $p(\theta|m)$, i.e.,
\begin{align}
    \hat{\theta}(m)  &=\,\int\!d\theta\,p(\theta|m)\,\theta\,=\,\int\!d\theta\,\frac{p(m|\nr\theta\nr)\,p(\theta)}{p(m)}\,\theta\,=\,\frac{\tr\bigl(E_{m}\nr\eta\bigr)}{\tr\bigl(E_{m}\nr\Gamma\bigr)}\,,
    \label{eq:estimator}
\end{align}
where we have inserted from Eqs.~(\ref{eq:Bayes law}) and~(\ref{eq:prob for m unconditional}), and defined the new quantity~\cite{Personick1971}
\begin{align}
    \eta &=\,\int\!d\theta\,p(\theta)\,\rho(\theta)\,\theta\,.
    \label{eq:eta}
\end{align}
Thus, the estimate for $\theta$ given that the outcome $m$ was observed, depends on the prior $p(\theta)$ and the encoding of the parameter in the state $\rho(\theta)$ via the quantities $\Gamma$ and $\eta$ from Eqs.~(\ref{eq:Gamma}) and~(\ref{eq:eta}), respectively, and on the chosen POVM $\{E_{m}\}$. Note that the estimator used for the Bayesian estimation scenario need not be unbiased in the sense of Eq.~(\ref{eq:unbiasdness condition}). Instead, on average, we now expect the estimator to assign the same mean value as the prior, i.e.,
\begin{align}
    \bar{\theta}    &=\sum\limits_{m}p(m)\,\hat{\theta}(m)=\sum\limits_{m}\tr\bigl(E_{m}\eta\bigr)=\tr(\eta)=\int\!d\theta\,p(\theta)\,\theta\,.
\end{align}

As a figure of merit for the precision of the estimate, we then wish to quantify how close $\hat{\theta}(m)$ is to $\theta$ according to our updated belief. We are hence interested in the variance $V_{\mathrm{post}}\suptiny{1}{-1}{(m)}$ of the \emph{posterior} $p(\theta|m)$ given the outcome $m$. Using the MSE approach as in Eq.~(\ref{eq:MSE variance local estimation}), but now with the posterior instead of the conditional probability $p(m|\nr\theta\nr)$, we write
\begin{align}
    &V_{\mathrm{post}}\suptiny{1}{-1}{(m)}    \,=\,V[p(\theta|m)]\,=\,\int\!d\theta\,p(\theta|m)\,\bigl(\theta-\hat{\theta}(m)\bigr)^{2}
    \label{eq:final variance given m}\\[1mm]
    &=\,\frac{1}{p(m)}\Bigl[\tr\bigl(E_{m}\!\int\!d\theta\,p(\theta)\,\rho(\theta)\,\theta^{2}\bigr)\,-\,\frac{\bigl(\tr(E_{m}\nr\eta)\bigr)^{2}}{\tr(E_{m}\nr\Gamma)}\Bigr]\,,
    \nonumber
\end{align}
where we have used~(\ref{eq:Bayes law}) and~(\ref{eq:estimator}). In general, the width of the posterior may decrease or increase with respect to the width of the prior, depending on the measurement outcome. It is therefore more useful to average over different outcomes and define
\begin{align}
    \overline{V}_{\!\mathrm{post}}    &=\sum\limits_{m}p(m)\,V_{\mathrm{post}}\suptiny{1}{-1}{(m)}=\int\!d\theta\,p(\theta)\,\theta^{2}
    -\sum\limits_{m}\frac{\bigl(\tr(E_{m}\nr\eta)\bigr)^{2}}{\tr(E_{m}\nr\Gamma)}.
    \label{eq:average final variance}
\end{align}
Here, a comment on the choice of $\overline{V}_{\!\mathrm{post}}$ as a figure of merit for the average increase in the knowledge is in order. For parameters (and priors) that have support on the entirety of $\mathbb{R}$, the MSE is certainly a useful choice. However, when estimating parameters with bounded support other quantifiers of the width of the posterior may be more appropriate. For instance, for phase estimation one may consider the Holevo phase variance as discussed in Section~\ref{sec:quantum advantage Bayesian estimation main text}. We will nonetheless consider the MSE in the following. This has several reasons. First, the MSE can still be useful for phase estimation when the priors are suitably narrow (see Appendix~\ref{sec:limitations of the MSE approach}) and it allows to establish some simple bounds (see Appendix~\ref{sec:bayesian cramer rao bound}) for the optimal classical estimation strategies, as we shall explain in Appendix~\ref{sec:bound for multiqubit measurements}. Second, the MSE is of course useful for frequency estimation problems (see Appendix~\ref{sec:Bayesian frequency estimation in MBQC}), where the parameter range is not bounded. We hence allow the parameter to take values $\theta\in[-\infty,\infty]$ for the remainder of this work.

As a simple example, consider a Gaussian prior of width $\sigma>0$ centered at $\theta=\theta_{o}$, that is,
\begin{align}
    p(\theta)   &=\,\frac{1}{\sqrt{2\pi}\nr\sigma}\,e^{-\nr\frac{(\theta-\theta_{o})^{2}}{2\sigma^{2}}}\,,
    \label{eq:Gaussian prior}
\end{align}
with $\bar{\theta}=\theta_{o}$ and $V[p(\theta)]=\int\!d\theta\,p(\theta)\,(\theta-\theta_{o})^{2}=\sigma^{2}$. The first term on the right-hand side of Eq.~(\ref{eq:average final variance}) then evaluates to
\begin{align}
    \int\!d\theta\,p(\theta)\,\theta^{2}    &=\,
    \sigma^{2}\,+\,\theta_{o}^{2}\,,
    \label{eq:Gaussian theta squared exp value}
\end{align}
while the remaining term
\begin{align}
    \theta_{o}^{2}   &\,\leq\,\sum\limits_{m}\frac{\bigl(\tr(E_{m}\nr\eta)\bigr)^{2}}{\tr(E_{m}\nr\Gamma)}
    \,<\,\sigma^{2}\,+\,\theta_{o}^{2}
\end{align}
determines the average decrease in width of the posterior with respect to the prior.

%%%%%%%%%%%%%%%%%%%%%%%%%%%%%%%%%%%%%%%%%%%%%%%%%%%%%%%%%%%%%%%%%%%%%%%%%%%%%%%%%%%%%%%%%%%%%%%%%%%%%%%%%%%%%%%%%%%%%%%%%%%%%

\subsubsection{A Bayesian Cram{\'e}r-Rao bound}\label{sec:bayesian cramer rao bound}

The average variance of the posterior can be bounded from below using the \emph{van~Trees inquality} (see, e.g., Ref.~\cite{VanTrees1968,GillLevit1995} or Appendix~\ref{sec:bayesian cramer rao bound proof})
\begin{align}
    \overline{V}_{\!\mathrm{post}}    &\geq\,\frac{1}{I\bigl(p(\theta)\bigr)\,+\,\bar{I}\bigl(\rho(\theta)\bigr)}\,,
    \label{eq:Bayesian cramer rao}
\end{align}
which can be viewed as a type of Cram{\'e}r-Rao bound for Bayesian estimation, where
\begin{align}
    I\bigl(p(\theta)\bigr)  &=\,\int\!d\theta\,p(\theta)\,\Bigl(\frac{\partial}{\partial\theta}\log p(\theta)\Bigr)^{2}\,,
    \label{eq:classical Fisher prior}
\end{align}
is the classical Fisher information of the prior and
\begin{align}
    \bar{I}\bigl(\rho(\theta)\bigr)   &=\,\int\!d\theta\,p(\theta)\,I\bigl(\rho(\theta)\bigr)\,
    \label{eq:Fisher posterior}\\[1mm]
    &=\,\int\!d\theta\,p(\theta)\,\sum\limits_{m}p(m|\nr\theta\nr)\,\Bigl(\frac{\partial}{\partial\theta}\log p(m|\nr\theta\nr)\Bigr)^{2}\,.
    \nonumber
\end{align}
is the averaged (over the unknown parameter~$\theta$) FI associated to the state $\rho(\theta)$ and the POVM $\{E_{m}\}$ as specified in Eq.~(\ref{eq:Fisher information}).

Since the QFI $\mathcal{I}\bigl(\rho(\theta)\bigr)$ arises as a maximization of the FI $I\bigl(\rho(\theta)\bigr)$ over all POVMs, we have $\mathcal{I}\bigl(\rho(\theta)\bigr)\geq I\bigl(\rho(\theta)\bigr)$. If, as before for the local case, we consider the parameter to be encoded by a unitary transformation of the form of Eq.~(\ref{eq:unitary encoding appendix}), the QFI is independent of $\theta$, as we have shown in Appendix~\ref{sec:cramer rao bound}. This allows us to bound the average FI by the QFI, i.e.,
\begin{align}
    \bar{I}\bigl(\rho(\theta)\bigr)   &\leq\,\int\!d\theta\,p(\theta)\,\mathcal{I}\bigl(\rho(\theta)\bigr)\,=\,\mathcal{I}\bigl(\rho(\theta)\bigr)\,,
\end{align}
and consequently we can modify the van~Trees inequality to
\begin{align}
    \overline{V}_{\!\mathrm{post}}    &\geq\,\frac{1}{I\bigl(p(\theta)\bigr)\,+\,\mathcal{I}\bigl(\rho(\theta)\bigr)}\,.
    \label{eq:Bayesian cramer rao with QFI}
\end{align}
In contrast to the (quantum) Cram{\'e}r-Rao inequality~(\ref{eq:Cramer Rao bound}), the bounds in~(\ref{eq:Bayesian cramer rao}) and~(\ref{eq:Bayesian cramer rao with QFI}) are generally not tight, so they do not allow us to conclude that a measurement strategy exists such that $1/\overline{V}_{\!\mathrm{post}}$ grows quadratically with $N$. And while it can indeed be shown~\cite{JarzynaDemkowiczDobrzanski15} that Heisenberg scaling is asymptotically achievable for arbitrary priors in the Bayesian regime we require an explicit description of the involved states and measurements to determine whether these can be efficiently implemented.

Nonetheless, a simple consequence of the van Trees inequality pertains to the classical scaling behaviour. Recall from Eq.~(\ref{eq:pure state unitary encoding Fisher product state}) that the maximal value of the QFI for product states is proportional to $N$. This implies that $\overline{V}_{\!\mathrm{post}}$ decreases at most linearly with $N$  for classical strategies, i.e., $1/\overline{V}_{\!\mathrm{post}}\leq N+I\bigl(p(\theta)\bigr)$, where $I\bigl(p(\theta)\bigr)$ is a constant independent of $N$. For instance, for the Gaussian prior of Eq.~(\ref{eq:Gaussian prior}), which we want to focus on in the following, we have $I\bigl(p(\theta)\bigr)=1/\sigma^{2}$.

At this point, two comments on the choice of Gaussian priors are in order. First, note that there exists an interesting connection between Bayesian estimation with Gaussian priors and local estimation subject to parallel, Gaussian noise~\cite{MacieszczakFraasDemkowiczDobrzanski2014}. As is outlined in Appendix~\ref{sec:connection of Gaussian Bayesian to noisy local}, this connection provides an alternative way of computing the variance $\overline{V}_{\!\mathrm{post}}$ via the (quantum) Fisher information of the probe state after a noisy channel. Here, we do not explicitly consider the problem of noisy metrology in more detail, but we refer the interested reader to Refs.~\cite{EscherDeMatosFilhoDavidovich2011,DemkowiczDobrzanskiKolodinskiGuta2012}.

The second comment concerns the fact that the probability distribution of Eq.~(\ref{eq:Gaussian prior}) has support on the entire real line, whereas for phase estimation, $\theta$ only takes values in an interval of length $2\pi$. In addition, the use of the MSE means that differences between estimates and parameter values larger than $\pi$ are disproportionately penalized. Intuitively it is clear that this becomes an issue when the width of the Gaussian prior becomes comparable with (half of) the length of the interval for $\theta$. In Appendix~\ref{sec:limitations of the MSE approach} this problem is discussed in more detail.

For sufficiently narrow priors the MSE is hence still a useful cost function for the variance and (non-wrapped) Gaussians can be employed instead of the more complicated wrapped Gaussians to simplify calculations. Moreover, the use of the MSE (rather than some circular statistics equivalent or covariant cost function, cf. Ref.~\cite{DemkowiczDobrzanskiJarzynaKolodynski2015}) as a measure for the precision of the estimate allows us to remain within the framework of Ref.~\cite{Personick1971}. It also permits us to apply the Bayesian Cram{\'e}r-Rao bound of Ineq.~(\ref{eq:Bayesian cramer rao with QFI}), which provides a straightforward comparison with classical strategies, as we shall discuss in Appendix~\ref{sec:bound for multiqubit measurements}. Finally, note that these considerations arise for the phase estimation problem discussed in this section, but are no cause for concern in the frequency estimation paradigm, which is presented in Appendix~\ref{sec:Bayesian frequency estimation in MBQC}.

%%%%%%%%%%%%%%%%%%%%%%%%%%%%%%%%%%%%%%%%%%%%%%%%%%%%%%%%%%%%%%%%%%%%%%%%%%%%%%%%%%%%%%%%%%%%%%%%%%%%%%%%%%%%%%%%%%%%

\subsubsection{A Proof of the Bayesian Cram{\'e}r-Rao Bound}\label{sec:bayesian cramer rao bound proof}

We now want to present an explicit proof that the average variance $\overline{V}_{\mathrm{post}}$ of the posterior $p(\theta|m)$ can be bounded from below by the \emph{van Trees inequality}~\cite{GillLevit1995}, which is the Bayesian equivalent of the Cram{\'e}r-Rao bound, given by
\begin{align}
    \overline{V}_{\!\mathrm{post}}    &\geq\,\frac{1}{I\bigl(p(\theta)\bigr)\,+\,\bar{I}\bigl(\rho(\theta)\bigr)}\,,
    \label{eq:Bayesian cramer rao appendix}
\end{align}
where $I\bigl(p(\theta)\bigr)$ is the classical Fisher information of the prior, given by
\begin{align}
    I\bigl(p(\theta)\bigr)  &=\,\int\!d\theta\,p(\theta)\,\Bigl(\frac{\partial}{\partial\theta}\log p(\theta)\Bigr)^{2}\,,
    \label{eq:classical Fisher prior appendix}
\end{align}
and $\bar{I}\bigl(\rho(\theta)\bigr)=\bar{I}\bigl(\rho(\theta),\{E_{m}\}\bigr)$ is the Fisher information associated to the state $\rho(\theta)$ and the POVM $\{E_{m}\}$, averaged over the (unknown) parameter $\theta$. That is, it is given by
\begin{align}
    \bar{I}\bigl(\rho(\theta)\bigr)   &=\,\int\!d\theta\,p(\theta)\,I\bigl(\rho(\theta)\bigr)\,
    \label{eq:Fisher posterior appendix}\\[1mm]
    &=\,\int\!d\theta\,p(\theta)\,\sum\limits_{m}p(m|\nr\theta\nr)\,\Bigl(\frac{\partial}{\partial\theta}\log p(m|\nr\theta\nr)\Bigr)^{2}\,.
    \nonumber
\end{align}
In the frequency estimation scenario, the parameter $\theta$ is typically allowed to take on any value in $\mathbb{R}$, but the prior is assumed to have compact support, such that $p(\pm\infty)=0$. In the phase estimation scenario, on the other hand, the parameter can take values in the interval $[a,a+2\pi]$ for some $a\in\mathbb{R}$ and w.l.o.g. one may pick $a=0$. In this case, one may assume that the probability densities are either wrapped, e.g., the prior satisfies $p(\theta)=p(\theta\!\mod2\pi)$ and $\theta$ is to be understood as $\theta\!\mod2\pi$. Alternatively, one can also treat $\theta$ to be any real number, and require that the prior be sufficiently narrow. In the latter scenario, one can still use the MSE approach for the variance, but care needs to be taken with the initial width of the prior, as discussed in Appendix~\ref{sec:limitations of the MSE approach}. With this in mind, we now discuss a proof of Eq.~(\ref{eq:Bayesian cramer rao appendix}). First, note that
\begin{align}
    \int\limits_{a}^{b}\!d\theta\,\hat{\theta}(m)\,\frac{\partial}{\partial\theta}\bigl(p(\theta)\,p(m|\nr\theta\nr)\bigr)    &=\,
    \hat{\theta}(m)\,\bigl[p(\theta)\,p(m|\nr\theta\nr)\bigr]_{a}^{b}\,=\,0\,,
    \label{eq:Bayesian cramer rao proof 1}
\end{align}
due to the assumptions above for $(a,b)=(0,2\pi)$ or $(a,b)=(-\infty,+\infty)$, respectively. Similarly, integration by parts immediately lets us evaluate the integral
\begin{align}
    \int\limits_{a}^{b}\!d\theta\,\theta\,\frac{\partial}{\partial\theta}\bigl(p(\theta)\,p(m|\nr\theta\nr)\bigr)    &=\,
    -\, \int\limits_{a}^{b}\!d\theta\,p(\theta)\,p(m|\nr\theta\nr)\,,
    \label{eq:Bayesian cramer rao proof 2}
\end{align}
where we have eliminated the boundary term using the previous assumptions. Using Bayes' law [see Eq.~(\ref{eq:Bayes law})] and the preliminary results of Eqs.~(\ref{eq:Bayesian cramer rao proof 1}) and~(\ref{eq:Bayesian cramer rao proof 2}), we can then calculate
\begin{align}
    &\int\!d\theta\,(\hat{\theta}(m)-\theta)\frac{\partial}{\partial\theta}\bigl(p(\theta)\,p(m|\nr\theta\nr)\bigr)  \,=\,
    \int\!d\theta\,p(\theta)\,p(m|\nr\theta\nr)
    \nonumber\\[1mm]
    &=\,p(m)\int\!d\theta\,p(\theta|m)\,=\,p(m)\,,
    \label{eq:Bayesian cramer rao proof 3}
\end{align}
since $p(\theta|m)$ is normalized. When we sum over the possible measurement outcomes, we must hence just get
\begin{align}
    \sum\limits_{m}\int\!d\theta\,(\hat{\theta}(m)-\theta)\frac{\partial}{\partial\theta}\bigl(p(\theta)\,p(m|\nr\theta\nr)\bigr)  &=\,
    \sum\limits_{m}p(m)\,=\,1\,.
    \label{eq:Bayesian cramer rao proof 4}
\end{align}
On the other hand, we can rewrite parts of the integrand as
\begin{align}
    \frac{\partial}{\partial\theta}\bigl(p(\theta)\,p(m|\nr\theta\nr)\bigr) &=\,
    p(\theta)\,p(m|\nr\theta\nr)\frac{\partial}{\partial\theta}\log\bigl(p(\theta)\,p(m|\nr\theta\nr)\bigr)
\end{align}
and use the Cauchy-Schwarz inequality from Eq.~(\ref{eq:Cauchy Schwarz ineq}) with $x=(\hat{\theta}(m)-\theta)\sqrt{p(\theta)\,p(m|\nr\theta\nr)}$ and
\begin{align}
    y   &=\,\sqrt{p(\theta)\,p(m|\nr\theta\nr)}\,\frac{\partial}{\partial\theta}\log\bigl[p(\theta)\,p(m|\nr\theta\nr)\bigr]
\end{align}
to arrive at the inequality
\begin{align}
    1   &\leq\,\sum\limits_{m}\int\!d\theta\,p(\theta)\,p(m|\nr\theta\nr)\,(\hat{\theta}(m)-\theta)^{2}\,\times\,
    \label{eq:Bayesian cramer rao proof 5}\\[1mm]
    &\ \ \times\,\sum\limits_{n}\int\!d\theta\pr\,p(\theta\pr)\,p(n|\nr\theta\pr\nr)\,\Bigl(\frac{\partial}{\partial\theta\pr}\log\bigl[p(\theta\pr)\,p(n|\nr\theta\pr\nr)\bigr]\Bigr)^{2}\,.
    \nonumber
\end{align}
The first factor on the right-hand side of Eq.~(\ref{eq:Bayesian cramer rao proof 5}) is just $\overline{V}_{\!\mathrm{post}}$ from Eq.~(\ref{eq:average final variance}). The second factor can be split into three terms by squaring
\begin{align}
    \frac{\partial}{\partial\theta}\log\bigl[p(\theta)\,p(m|\nr\theta\nr)\bigr]    &=\,
    \frac{\partial}{\partial\theta}\log p(\theta)\,+\,\frac{\partial}{\partial\theta}\log p(m|\nr\theta\nr)\,.
\end{align}
Summing over the normalized conditional probability $p(m|\theta)$, the first term gives the classical Fisher information for the prior, i.e.,
\begin{align}
    \int\!d\theta\,p(\theta)\,\Bigl(\frac{\partial}{\partial\theta}\log p(\theta)\Bigr)^{2}  &=\,I\bigl(p(\theta)\bigr)\,,
    \label{eq:classical Fisher prior appendix 2}
\end{align}
while the term containing the square of $\tfrac{\partial}{\partial\theta}\log p(m|\nr\theta\nr)$ gives $\bar{I}\bigl(\rho(\theta)\bigr)$ as defined in Eq.~(\ref{eq:Fisher posterior appendix}). The remaining cross term is of the form
\begin{align}
    &\sum\limits_{m}\int\!d\theta\,p(\theta)\,p(m|\nr\theta\nr)\,\Bigl(\frac{\partial}{\partial\theta}\log p(\theta)\Bigr)\,\Bigl(\frac{\partial}{\partial\theta}\log p(m|\nr\theta\nr)\Bigr)
    \nonumber\\[1mm]
    &=\,\sum\limits_{m}\int\!d\theta\,\Bigl(\frac{\partial}{\partial\theta}\,p(\theta)\Bigr)\,\Bigl(\frac{\partial}{\partial\theta}\,p(m|\nr\theta\nr)\Bigr)
    \nonumber\\[1mm]
    &=\,\int\!d\theta\,\Bigl(\frac{\partial}{\partial\theta}\,p(\theta)\Bigr)\,\Bigl(\frac{\partial}{\partial\theta}\,\sum\limits_{m}p(m|\nr\theta\nr)\Bigr)\,=\,0\,,
\end{align}
which vanishes since the sum over $p(m|\nr\theta\nr)$ is independent of $\theta$, i.e.,
\begin{align}
    \sum\limits_{m}p(m|\nr\theta\nr)   &=\,\sum\limits_{m}\tr\bigl(E_{m}\nr\rho(\theta)\bigr)\,=\,\tr(\rho)\,=\,1\,.
\end{align}
Dividing both sides of the inequality in~(\ref{eq:Bayesian cramer rao proof 5}) by the sum of the nonzero terms of the second factor, we arrive at the Bayesian Cram{\'e}r-Rao bound
\begin{align}
    \overline{V}_{\!\mathrm{post}}    &\geq\,\frac{1}{I\bigl(p(\theta)\bigr)\,+\,\bar{I}\bigl(\rho(\theta)\bigr)}\,.
    \label{eq:Bayesian cramer rao appendix 2}
\end{align}

%%%%%%%%%%%%%%%%%%%%%%%%%%%%%%%%%%%%%%%%%%%%%%%%%%%%%%%%%%%%%%%%%%%%%%%%%%%%%%%%%%%%%%%%%%%%%%%%%%%%%%%%%%%%%%%%%%%%

\subsubsection{Relating Noisy Local Estimation with Bayesian Estimation for Gaussian Priors}\label{sec:connection of Gaussian Bayesian to noisy local}

In this appendix we discuss an interesting connection between noisy local estimation and Bayesian estimation for Gaussian priors. We hence consider a local estimation scenario as in Section~\ref{sec:The local parameter estimation scenario}, where ``parallel" noise is present on top of the unitary encoding of Eq.~(\ref{eq:unitary encoding appendix}). That is, the noise is generated by the same Hamiltonian as the encoding of the parameter but distributed according to some probability distribution $\tilde{p}(\theta)$. The state encoding the parameter is then given by
\begin{align}
    \tilde{\rho}(\theta)    &=\,U_{\theta}\nr\tilde{\rho}(0)\nr U_{\theta}^{\dagger}\,,
\end{align}
where the noise can be understood as part of preparing the initial state
\begin{align}
    \tilde{\rho}(0) &=\,\int\!\!d\theta\pr\,\tilde{p}(\theta\pr)\,U_{\theta\pr}\nr\ket{\psi}\!\bra{\psi}\nr U_{\theta\pr}^{\dagger}
\end{align}
starting from some pure state $\ket{\psi}$. We further assume that the noise has a Gaussian profile centered around zero, that is, the noise distribution is
\begin{align}
    \tilde{p}(\theta)    &=\,p(\theta+\theta_{o})\,,
\end{align}
where $\theta_{o}$ is the mean of the Gaussian prior $p(\theta)$ of Eq.~(\ref{eq:Gaussian prior}). We can now see how the encoded state of this noisy local scenario corresponds to the quantity $\Gamma$ from Eq.~(\ref{eq:Gamma}) in the Bayesian scenario, i.e.,
\begin{align}
    \Gamma  &=\,\int\!d\theta\,p(\theta)\,\rho(\theta)\,=\,\int\!d\theta\,p(\theta)\,U_{\theta}\nr\ket{\psi}\!\bra{\psi}\nr U_{\theta}^{\dagger}\nonumber\\[1mm]
    &=\,\int\!d\theta\pr\,p(\theta\pr+\theta_{o})\,U_{\theta\pr+\theta_{o}}\nr\ket{\psi}\!\bra{\psi}\nr U_{\theta\pr+\theta_{o}}^{\dagger}\,=\,\tilde{\rho}(\theta_{o})\,,
    \label{eq:Gamma appendix}
\end{align}
where we have substituted $\theta\pr=\theta-\theta_{o}$. To establish a similar connection for $\eta$ from Eq.~(\ref{eq:eta}), we make use of the fact that the prior (and the noise distribution in the local scenario) are Gaussian, such that
\begin{align}
    \frac{d}{d\theta}\,p(\theta)    &=\,\dot{p}(\theta)\,=\,-\,\frac{\theta-\theta_{o}}{\sigma^{2}}\,p(\theta)\,.
\end{align}
With this, we find
\begin{align}
    \eta    &=\!\int\!\!d\theta\,\theta\,p(\theta)\,\rho(\theta)\,=\,\theta_{o}\!\int\!\!d\theta\,p(\theta)\,\rho(\theta)\,-\,\sigma^{2}\!\!\int\!\!d\theta\,\dot{p}(\theta)\,\rho(\theta)
    \nonumber\\[1mm]
    &=\,\theta_{o}\nr\Gamma\,+\,i\sigma^{2}\!\!\int\!\!d\theta\nr p(\theta)\nr\comm{\rho(\theta)}{H}\,=\,\theta_{o}\nr\Gamma\,+\,\sigma^{2}\dot{\tilde{\rho}}(\theta_{o})\,,
    \label{eq:eta appendix}
\end{align}
where the dot indicates a partial derivative w.r.t. $\theta$. Reinserting the expressions for $\Gamma$ and $\eta$ into Eq.~(\ref{eq:average final variance}), the last term gives
\begin{align}
    \sum\limits_{m}\frac{\bigl(\tr(E_{m}\nr\eta)\bigr)^{2}}{\tr(E_{m}\nr\Gamma)}   &=\,
    \sigma^{4}\sum\limits_{m}\frac{\bigl(\tr[E_{m}\nr\dot{\tilde{\rho}}(\theta_{o})]\bigr)^{2}}{\tr\bigl(E_{m}\nr\tilde{\rho}(\theta_{o})\bigr)}
    \,+\,\theta_{o}^{2}\,.
    \label{eq:average final variance second term Gaussian prior}
\end{align}
The first term on the right-hand side of Eq.~(\ref{eq:average final variance second term Gaussian prior}) can easily be recognized as the Fisher information $I\bigl(\tilde{\rho}(\theta_{o})\bigr)$ from Eq.~(\ref{eq:Fisher information}) for the POVM $\{E_{m}\}$ in the local scenario with parallel Gaussian noise. Combining this result with Eqs.~(\ref{eq:average final variance}) and~(\ref{eq:Gaussian theta squared exp value}), we find
\begin{align}
    \overline{V}_{\!\mathrm{post}}    &=\,\sigma^{2}\,-\,\sigma^{4}\,I\bigl(\tilde{\rho}(\theta_{o})\bigr)\,.
    \label{eq:average final variance analogy}
\end{align}
Since the variance of the initial Gaussian prior is just $V[p(\theta)]=\sigma^{2}$, one arrives at the conclusion that the average decrease in variance in the Bayesian scenario with Gaussian prior, $\Delta V=V[p(\theta)]-\overline{V}_{\!\mathrm{post}}$ for any given POVM is proportional to the Fisher information at $\theta=\theta_{o}$ for the same POVM in the local scenario with parallel Gaussian noise, $\Delta V=\sigma^{4}I\bigl(\tilde{\rho}(\theta_{o})\bigr)$. In particular, for the optimal POVM one obtains the QFI, which is independent of the value of the parameter for the unitary encoding with parallel noise, and hence
\begin{align}
    \Delta V_{\mathrm{opt}} &=\,\sigma^{4}\,\mathcal{I}(\tilde{\rho})\,.
\end{align}
This result immediately informs us about an important property of the Bayesian scenario. Since $\mathcal{I}(\tilde{\rho})$ is the QFI in a scenario with parallel noise that can be viewed as dephasing, one cannot expect Heisenberg scaling of $\mathcal{I}(\tilde{\rho})$, i.e., that $\mathcal{I}(\tilde{\rho})$ increases quadratically with $N$, see Refs.~\cite{EscherDeMatosFilhoDavidovich2011,DemkowiczDobrzanskiKolodinskiGuta2012}. Instead, it is clear that $\mathcal{I}(\tilde{\rho})\leq1/\sigma^{2}$ since $\overline{V}_{\!\mathrm{post}}\geq0$. On the other hand, one expects that $\mathcal{I}(\tilde{\rho})$ approaches the bound $1/\sigma^{2}$ from below as $N$ increases. As suggested in Ref.~\cite{KnyshChenDurkin2014}, it is reasonable to assume that
\begin{align}
    \mathcal{I}(\tilde{\rho})   &=\,\frac{1}{\sigma^{2}}\,-\,\frac{K}{N^{\alpha}}
\end{align}
for some positive constant $K$ and some power $\alpha\geq1$, such that $\overline{V}_{\!\mathrm{post}}=K\nr\sigma^{4}/N^{\alpha}$. Therefore, a scaling advantage of a quantum strategy with respect to a classical strategy is obtained if one finds an (efficiently preparable) state and POVM such that $\alpha>1$.

%%%%%%%%%%%%%%%%%%%%%%%%%%%%%%%%%%%%%%%%%%%%%%%%%%%%%%%%%%%%%%%%%%%%%%%%%%%%%%%%%%%%%%%%%%%%%%%%%%%%%%%%%%%%%%%%%%%%%%%%%%%%%

\subsubsection{Limitations of the MSE Approach}\label{sec:limitations of the MSE approach}

Here, we aim to discuss the limitations of applicability of the mean square error (MSE) cost function for Bayesian phase estimation, i.e., for a scenario where the parameter $\theta$ is encoded by a unitary $U_{\theta}=e^{-i\theta H}$, with $H=\tfrac{1}{2}Z$ for each qubit. Since the difference between the two eigenvalues of $H$ is $1$, it is immediately apparent that values of $\theta$ that differ by $2\pi$ cannot be distinguished in such a scenario. This periodicity is not accurately reflected in the use of the MSE, since estimates that differ by integer multiples of $2\pi$ are unduly penalized. In a local estimation scenario where small fluctuations around a fixed value of the parameter are being estimated, this is not an issue. Similarly, this is of no concern for Bayesian estimation when the prior is sufficiently localized, but can become an issue for larger values of $\sigma$ [where we focus on Gausian priors as in Eq.~(\ref{eq:Gaussian prior})]. We are therefore interested in quantifying for which values of $\sigma$ the approach using the MSE cost function becomes problematic.

We will take a pragmatic point of view and consider the MSE approach as useful, if this post-processing of the measurement data provides an increase in knowledge in the sense of an average decrease of the width of the posterior $p(\theta|m)$. We therefore ask, what the minimal MSE of the posterior can be in principle, given a fixed Gaussian prior of width $\sigma$. When obtaining a measurement outcome $m$, the corresponding estimate may in principle only be understood modulo $2\pi$. In other words, if no prior knowledge is available, and one were to trust the estimate of the parameter unconditionally, the posterior would be a ``comb" of Dirac delta functions $\delta(\theta-2\pi k)$ for all values $k$ such that $\theta-2\pi k$ lies within the allowed range of parameters. For an unrestricted range, $\theta\in\mathbb{R}$, we hence have infinitely many side-peaks at distances $2\pi k$ for $k\in\mathbb{Z}$. If we take into account the prior information, some of these peaks are suppressed by its shape, e.g., as $\exp(-\tfrac{\theta^{2}}{2\sigma^{2}})$ for a Gaussian prior. The optimally reachable
\newpage

\begin{figure}[ht!]
\includegraphics[width=0.44\textwidth]{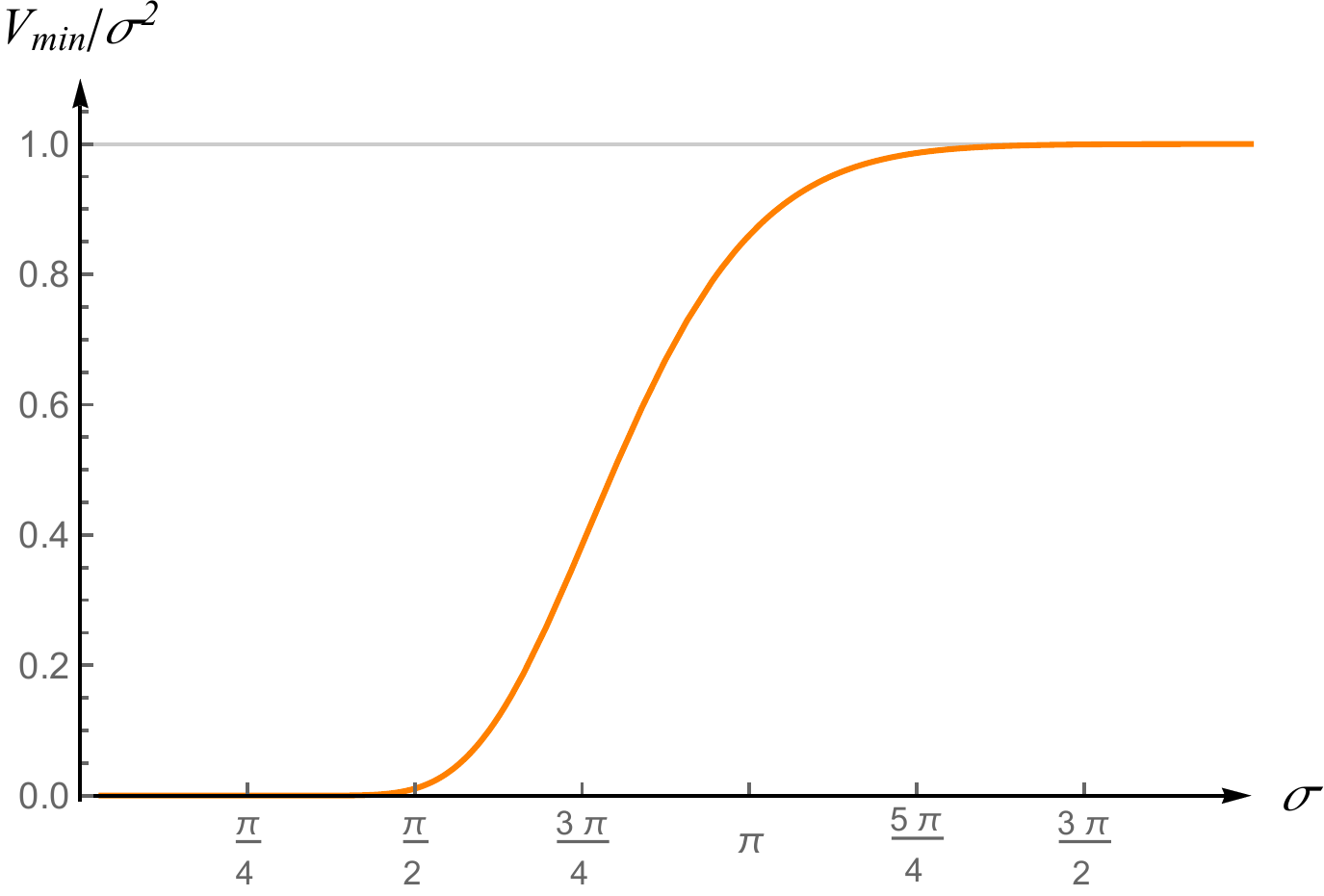}
\caption{\label{fig:MSE limitations}%(Color online)
\textbf{Limitations of the MSE approach}. The minimally achievable MSE $V_{\mathrm{min}}$ of the posterior, normalized by the variance $\sigma^{2}$ of the prior, is plotted against $\sigma$. For small $\theta$, the width of the optimal posterior is close to zero and the MSE appropriately captures the increase in knowledge about $\theta$. However, when the width of the prior reaches a threshold value close to $\pi/2$, the minimal MSE of the posterior drastically increases and quickly saturates at the initial width of the prior. In this regime the MSE does not reflect the increase in knowledge about $\theta$ in a meaningful way.}
\end{figure}
\noindent
posterior is then given by
\begin{align}
    p_{opt}(\theta) &=\,\mathcal{N}\sum\limits_{k\in\mathbb{Z}}e^{-\tfrac{\theta^{2}}{2\sigma^{2}}}\delta(\theta-2\pi k)\,,
\end{align}
where the normalization is given by $\mathcal{N}^{-1}=\sum\limits_{k\in\mathbb{Z}}\exp(-\tfrac{(2\pi k)^{2}}{2\sigma^{2}})$. The MSE of this distribution is
\begin{align}
    V_{\mathrm{min}} &=\,\mathcal{N}\sum\limits_{k\in\mathbb{Z}}e^{-\tfrac{(2\pi k)^{2}}{2\sigma^{2}}}(2\pi k)^{2}\,.
\end{align}
As illustrated in Fig.~\ref{fig:MSE limitations}, the MSE of this optimal posterior strongly increases from around $\sigma\approx\pi/2$, and from around $5\pi/4$ the width stays virtually constant as compared to the MSE of the prior. Of course this does not mean that the measurement process does not provide information about the parameter. Clearly, knowing the value of $\theta$ exactly modulo $2\pi$ is more useful than a uniform prior. However, the MSE simply fails to capture this distinction. We hence have to keep this limited applicability of the approach using (non-wrapped, Gaussian) priors and the MSE cost function in mind. Specifically, we restrict our analysis to Gaussian priors of widths smaller or equal than $1$.

%%%%%%%%%%%%%%%%%%%%%%%%%%%%%%%%%%%%%%%%%%%%%%%%%%%%%%%%%%%%%%%%%%%%%%%%%%%%%%%%%%%%%%%%%%%%%%%%%%%%%%%%%%%%%%%%%%%%%%%%%%%%%
%%%%%%%%%%%%%%%%%%%%%%%%%%%%%%%%%%%%%%%%%%%%%%%%%%%%%%%%%%%%%%%%%%%%%%%%%%%%%%%%%%%%%%%%%%%%%%%%%%%%%%%%%%%%%%%%%%%%%%%%%%%%%

\subsection{Classical Bayesian Estimation Strategies}\label{sec:Optimal classical Bayesian estimation strategy}

After introducing the quantities of interest for Bayesian parameter estimation in the previous appendix, we now want to illustrate these techniques for classical Bayesian estimation. This provides the opportunity to establish a direct comparison with the results obtained for a strategy exploiting quantum features that we will present in Appendix~\ref{sec:Quantum Advantage in Bayesian Estimation}.

We consider a strategy to be classical, if no quantum correlations are used for the state preparation or measurement, which corresponds to the choice of product states for $N$ qubits along with single-qubit measurements. The Bayesian approach allows updating the estimation strategy based on the outcomes of previous measurements. Consequently, a parallel strategy of $N$ individual probes that are prepared and measured in the same way may not be optimal even among the classical measurement schemes. At the same time, the explicit evaluation of a sequential measurement strategy with intermediate updates is computationally extremely demanding. To give a fair representation of the performance of classical strategies we hence consider a bound for the sequential measurement scheme in Appendix~\ref{sec:bound for multiqubit measurements}, and compute the average variance explicitly for the optimal parallel strategy in Appendix~\ref{sec:optimal parallel strategy}. In preparation for these scenarios, we begin with the single-qubit Bayesian estimation problem in Appendix~\ref{sec:Single-qubit measurements}

%%%%%%%%%%%%%%%%%%%%%%%%%%%%%%%%%%%%%%%%%%%%%%%%%%%%%%%%%%%%%%%%%%%%%%%%%%%%%%%%%%%%%%%%%%%%%%%%%%%%%%%%%%%%%%%%%%%%%%%%%%%%%

\subsubsection{Single-Qubit Measurements}\label{sec:Single-qubit measurements}

For the scenario that we consider here, i.e., Gaussian priors as in Eq.~(\ref{eq:Gaussian prior}) and unitary parameter encodings as in Eq.~(\ref{eq:unitary encoding appendix}), the optimal single-qubit measurement strategy for Bayesian estimation is similar to that of the local scenario. That is, the probe state is chosen to be $\ket{+}$, i.e., a uniform superposition of the eigenstates of $H$. The optimal accompanying measurement is a projective measurement with POVM elements
\begin{align}
    \tilde{E}_{\pm} &=\,U_{\theta_{o}+\pi/2}\,\ket{\pm}\!\bra{\pm}\,U^{\dagger}_{\theta_{o}+\pi/2}\,,
    \label{eq:single qubit measurement POVM}
\end{align}
which corresponds to a measurement in a direction on the equatorial plane of the Bloch sphere that is orthogonal to the direction obtained by rotating $\ket{\pm}$ by the expected value $\theta_{o}$ of the prior. This can be seen by noting that probe states and measurement directions can be restricted to the equatorial plane, followed by an optimization over the angle defining their relative orientation. For this combination of state and measurement, the conditional probabilities to obtain the outcomes ``$+$" or ``$-$" are
\begin{align}
    p(\pm|\nr\theta\nr) &=\,\tr\bigl(\tilde{E}_{\pm}\rho(\theta)\bigr)\,=\,\frac{1}{2}\bigl(1\pm\sin(\theta-\theta_{o})\bigr)\,,
    \label{eq:classical single-qubit cond prob}
\end{align}
such that $p(\pm|\theta_{o})=1/2$. We further compute
\vspace*{-1mm}
\begin{subequations}
\begin{align}
    \tr\bigl(\tilde{E}_{\pm}\Gamma\bigr)  &=\,p(m=\pm)\,=\,\tfrac{1}{2}\,,\\[1mm]
    \tr\bigl(\tilde{E}_{\pm}\eta\bigr)  &=\,\tfrac{1}{2}\bigl(\theta_{o}\,\pm\,\sigma^{2}\,e^{-\sigma^{2}/2}\bigr)\,.
\end{align}
\end{subequations}
The corresponding estimates are then easily found by inserting into Eq.~(\ref{eq:estimator}), yielding
\begin{align}
    \theta_{m=\pm}  &=\,\theta_{o}\,\pm\,\sigma^{2}e^{-\sigma^{2}/2}\,.
    \label{eq:estimator single qubit}
\end{align}

\newpage
\begin{figure}[h!]
\includegraphics[width=0.45\textwidth]{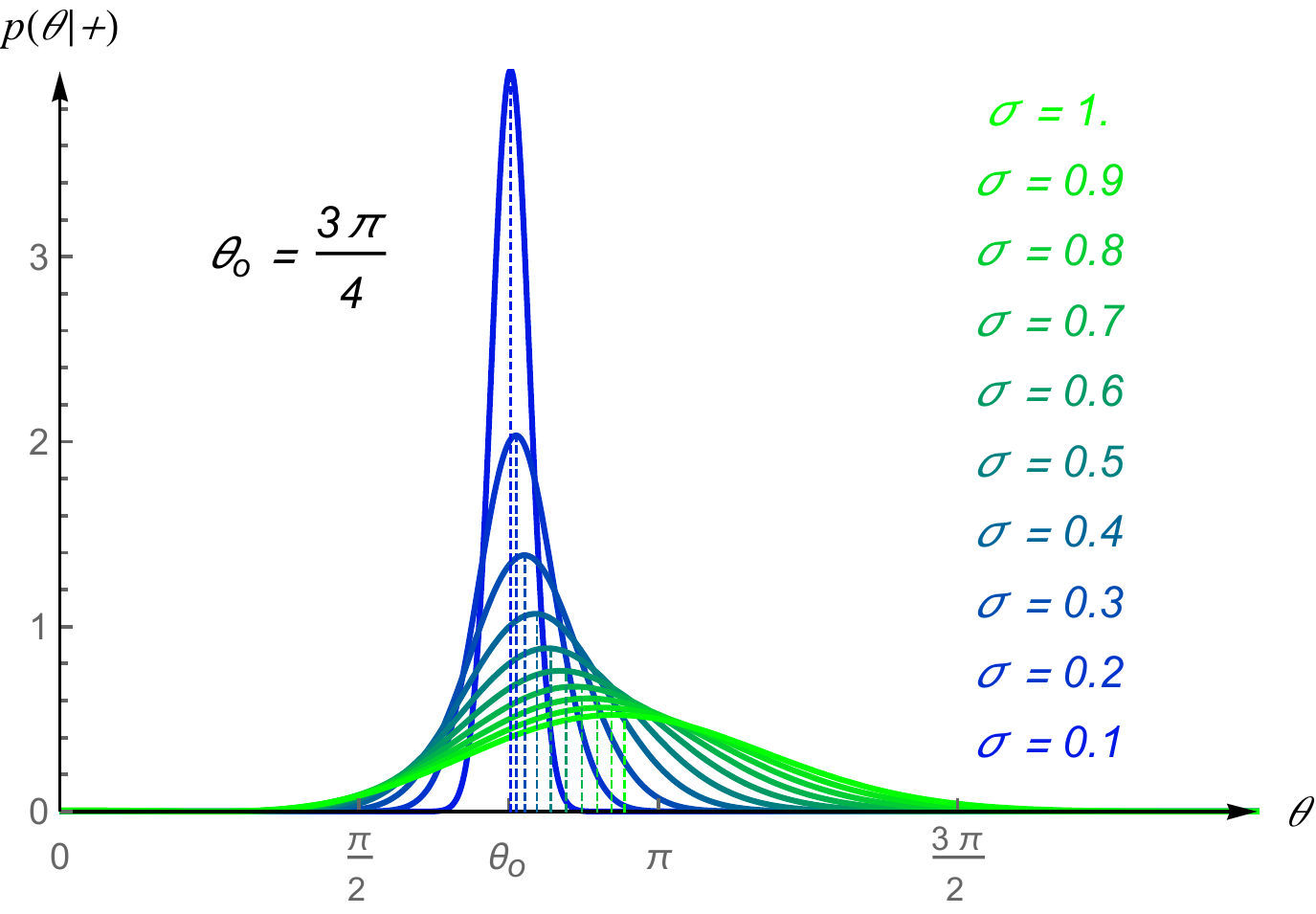}
\vspace*{-1.5mm}
\caption{\label{fig:posterior single qubit}%(Color online)
\textbf{Posterior distributions for single-qubit probe}. The posterior distributions for the outcome $m=+$ for the optimal single-qubit measurement are shown when starting from a Gaussian prior of width $\sigma$ (shown for $\sigma=0.1,\ldots,1$ in steps of $0.1$). The prior is centered at $\theta=\theta_{o}$, while the mean value of the posterior $p(\theta|m=+)$ is just the estimator $\theta_{m=+}$ from Eq.~(\ref{eq:estimator single qubit}), indicated by the vertical dashed lines.}
\end{figure}

Together with Eqs.~(\ref{eq:average final variance}) and~(\ref{eq:Gaussian theta squared exp value}) we then find the average variance of the posterior
\begin{align}
    \overline{V}_{\!\mathrm{post}}    &=\,\sigma^{2}\bigl(1\,-\,\sigma^{2}\nr e^{-\sigma^{2}}\bigr)\,.
    \label{eq:average final variance single qubit}
\end{align}
Since $0<\sigma^{2}\nr e^{-\sigma^{2}}<1$ for finite, nonzero $\sigma$, the variance decreases on average, $\overline{V}_{\!\mathrm{post}}<\sigma^{2}$, but it becomes apparent that the decrease in width quantified by $\Delta V:=\sigma^{2}-\overline{V}_{\!\mathrm{post}}$ has a maximum for $\sigma=2$. This signifies that the MSE approach using Gaussian priors is not useful for priors of large width when considering phase estimation (see Appendix~\ref{sec:limitations of the MSE approach} for a discussion of this issue). It is also interesting to note that the posterior distributions conditional on the outcomes $m=\pm$ are given by
\begin{align}
    p(\theta|m=\pm) &=\,\frac{1}{\sqrt{2\pi}\nr\sigma}\bigl(1\pm\sin(\theta-\theta_{o})\bigr)e^{-\nr\frac{(\theta-\theta_{o})^{2}}{2\sigma^{2}}}\,.
\end{align}
Unlike the prior, the posterior distributions illustrated in Fig.~\ref{fig:posterior single qubit} are no longer Gaussian, and they are not symmetric around their mean values $\theta=\theta_{m=\pm}$.

%%%%%%%%%%%%%%%%%%%%%%%%%%%%%%%%%%%%%%%%%%%%%%%%%%%%%%%%%%%%%%%%%%%%%%%%%%%%%%%%%%%%%%%%%%%%%%%%%%%%%%%%%%%%%%%%%%%%%%%%%%%%%

\subsubsection{Bound for Multi-Qubit Measurements}\label{sec:bound for multiqubit measurements}

We are now interested in making statements about the optimal strategy for Bayesian estimation using a sequence of $N$ consecutive single-qubit probes. Unfortunately, the posterior even after one measurement is no longer Gaussian (or symmetric). Therefore, determining the optimal single-qubit measurements and updating the prior becomes problematic for large numbers of measurements. This may not be an issue in an actual measurement, where each qubit gives a single outcome based on which the next measurement is chosen. However, we are interested in the variance of the posterior averaged over all possible sequential measurement outcomes, the set of which grows exponentially. Having $2^{N}$ potentially different posterior distributions makes such an approach computationally infeasible.

We shall therefore refrain from obtaining the exact expression for the optimal expected variance $\overline{V}_{\!\mathrm{post}}$ after $N$ sequential single-qubit measurements with updated directions. Instead, we construct a bound based on the Bayesian Cram{\'e}r-Rao inequality~(\ref{eq:Bayesian cramer rao with QFI}). We note that the updating procedure can be entirely thought of as part of the choice of measurement direction, while the probe state $\ket{+}$ remains the same throughout. Further recall that the QFI entails an optimization over all possible measurements including correlated measurements that can depend on previous outcomes. A lower bound for $\overline{V}_{\!\mathrm{post}}$ in the classical case is hence obtained from the QFI for the state $\ket{+}^{\otimes N}$, which we have previously determined in Eq.~(\ref{eq:pure state unitary encoding Fisher product state}) to be given by $\mathcal{I}(\ket{+}^{\otimes N})=N$. Inserting into Ineq.~(\ref{eq:Bayesian cramer rao with QFI}) we arrive at the bound
\begin{align}
    \overline{V}_{\!\mathrm{post}} &\geq\,\frac{\sigma^{2}}{1\,+\,N\sigma^{2}}\,,
    \label{eq:Bayesian cramer rao bound for Gaussian prior and classical strategy}
\end{align}
where we have used that $I\bigl(p(\theta)\bigr)=1/\sigma^{2}$.
%, and the result is illustrated in Fig.~\ref{fig:variance classical strategies}.
Any classical strategy, may it consist of parallel or sequential measurements, must give an expected variance larger than this bound. This result also extends to the (asymptotic) behaviour of the Holevo phase variance $V_{\!\phi}$ of Eq.~(\ref{eq:holevo phase var}) since $V_{\!\phi}$ reduces to the MSE as $\sigma\rightarrow0$ (see, e.g.,~\cite[p.~7]{Berry:PhD2002}). Consequently, the Holevo phase variance of any successful sequential measurement strategy will approach the behaviour of the MSE. The faster (in terms of the number of measurements) the strategy decreases the phase variance, the sooner one will enter a regime where the bound of Ineq.~(\ref{eq:Bayesian cramer rao appendix 2}) applies. Moreover, the bound in Ineq.~(\ref{eq:Bayesian cramer rao bound for Gaussian prior and classical strategy}) is not tight and might significantly overestimate the performance of classical strategies since the optimization in the QFI also includes entangled measurements. We therefore complement this bound by an investigation into the optimal parallel strategy in Section~\ref{sec:optimal parallel strategy}.

%%%%%%%%%%%%%%%%%%%%%%%%%%%%%%%%%%%%%%%%%%%%%%%%%%%%%%%%%%%%%%%%%%%%%%%%%%%%%%%%%%%%%%%%%%%%%%%%%%%%%%%%%%%%%%%%%%%%%%%%%%%%%

\subsubsection{Optimal Parallel Strategy}\label{sec:optimal parallel strategy}

Having obtained the previous lower bound for $\overline{V}_{\!\mathrm{post}}$ for the optimal classical strategy, one may wonder, how close a practical classical strategy may come to this bound. To address this question, we now consider the optimal classical, parallel strategy for Gaussian priors. That is, we compute $\overline{V}_{\!\mathrm{post}}$ in the case where $N$ qubits are identically prepared and measured (i.e., without intermediate updates) with the optimal single-qubit strategy based on the prior information (see Section~\ref{sec:Single-qubit measurements}). The probe state is hence $\ket{+}^{\otimes N}$ and for each qubit we perform the POVM with elements $\tilde{E}_{\pm}$ as in Eq.~(\ref{eq:single qubit measurement POVM}). Since the state is invariant under the exchange of qubits, it is irrelevant, which of the $N$ qubits give results ``$+$", and which give results ``$-$", we note that there are only $N+1$ qualitatively different measurement outcomes.

\begin{figure}[ht!]
\includegraphics[width=0.44\textwidth]{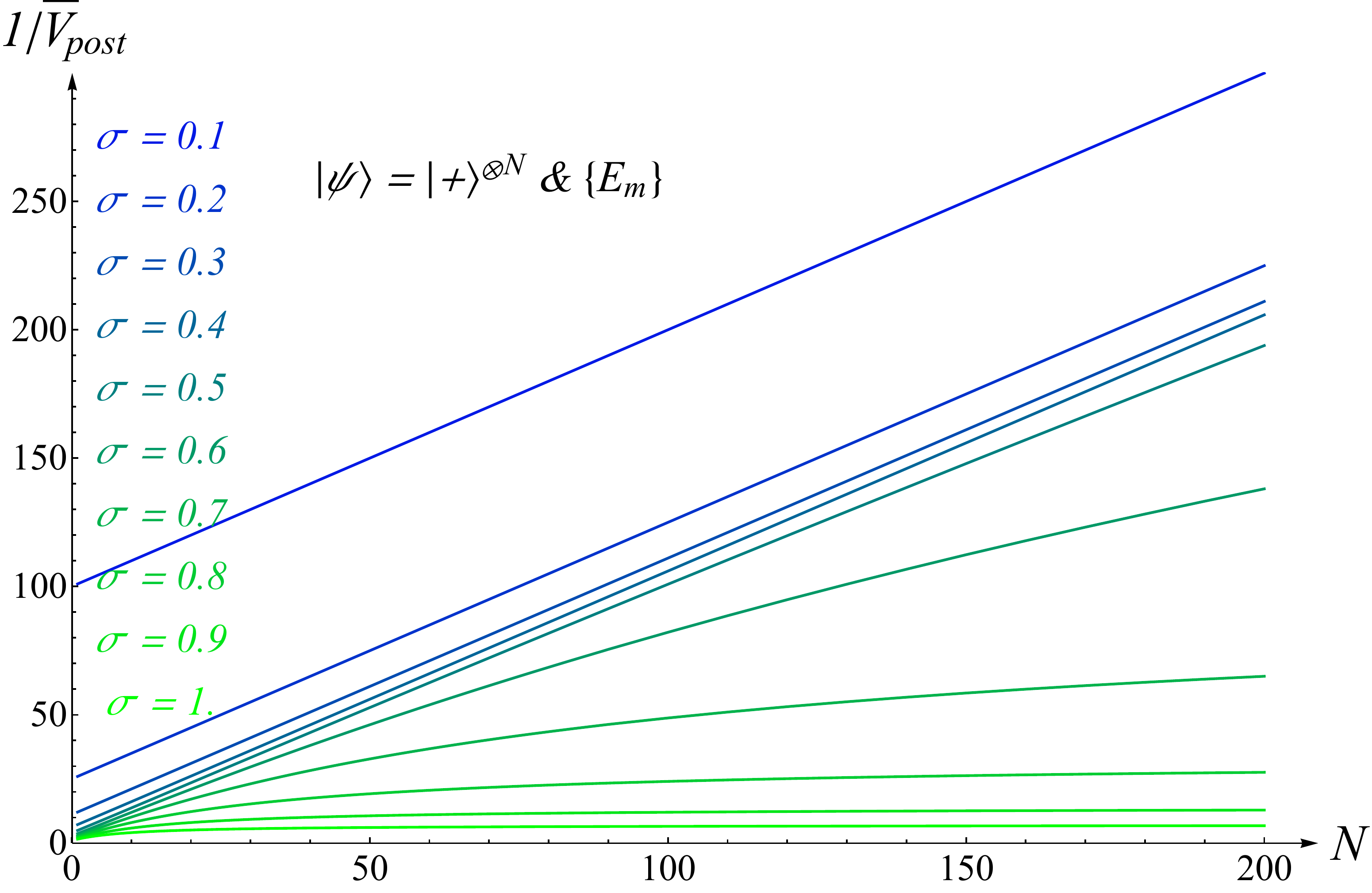}\\
\caption{\label{fig:variance classical parallel}%(Color online)
\textbf{Optimal parallel classical strategy}.
The inverse expected variance of the posterior $1/\overline{V}_{\!\mathrm{post}}$ is shown for the optimal, classical, parallel strategy when starting from a Gaussian prior of width $\sigma$ (shown for $\sigma=0.1,\ldots,1$ in steps of $0.1$) as functions of the qubit number $N$. As indicated by the different curves for $1/\overline{V}_{\!\mathrm{post}}$, the variance decreases as $1/N$ for small widths, but decreases less strongly for larger values of $\sigma$. Note that the decrease in performance may be attributed to the choice of the MSE cost function and the Gaussian priors.
\vspace*{-3mm}}
\end{figure}

We label these outcomes by $m=0,1,\ldots,N$, which we take to be the number of outcomes ``$-$". In other words, for the given state this measurement is equivalent to the POVM with element $E_{m}$ from Eq.~(\ref{eq:single qubit n minus outcome POVM}). The conditional probability to obtain the outcome $m$, given that the parameter takes the value $\theta$ is then
\begin{align}
    p(m|\theta) &=\,p(+|\nr\theta\nr)^{N-m}p(-|\nr\theta\nr)^{m}\binom{N}{m}\,.
\end{align}
We then insert for $p(\pm|\nr\theta\nr)$ from Eq.~(\ref{eq:classical single-qubit cond prob}) and find
\begin{align}
    &\tr\bigl(E_{m}\Gamma\bigr)  \,=\,\binom{N}{m}\int\!d\theta\,p(\theta)\,p(+|\nr\theta\nr)^{N-m}p(-|\nr\theta\nr)^{m}\nonumber\\[1mm]
    &\ \ =\,\frac{1}{2^{N}}\binom{N}{m}\sum\limits_{k=0}^{N-m}\sum\limits_{k\pr=0}^{m}
    \binom{N-m}{k}\binom{m}{k\pr}(-1)^{k\pr}\,I_{k+k\pr}\,,
    \label{eq:class parallel tr Gamma Em}
\end{align}
where the quantity $I_{k+k\pr}$ is given by
\begin{align}
    I_{n}   &=\,\frac{1}{\sqrt{2\pi}\sigma}\int\!d\theta\,e^{-\frac{\theta^{2}}{2\sigma^{2}}}\sin^{n}\nl\theta\,
    \label{eq:inegral In}
\end{align}
and powers of the sine function arise from the binomial expansion of $\bigl(1\pm\sin(\theta-\theta_{o})\bigr)$ followed by a substitution $\theta-\theta_{o}\rightarrow\theta$. Now it is easy to see that the integral $I_{n}$ vanishes for odd $n$. To solve the integral in Eq.~(\ref{eq:inegral In}) for even $n$, we use the trigonometric identity (which holds only for even $n$)
\begin{align}
    \sin^{n}\nl\theta  &=\frac{1}{2^{n}}\binom{n}{\tfrac{n}{2}}
    +\frac{2}{2^{n}}\sum\limits_{l=0}^{\tfrac{n}{2}-1}(-1)^{\tfrac{n}{2}-l}\binom{n}{l}\cos\bigl([n-2l]\theta\bigr),
    \label{eq:superweird trig identity}
\end{align}
as well as the integral formula
\begin{align}
    \frac{1}{\sqrt{2\pi}\sigma}\int\limits_{-\infty}^{\infty}\!d\theta\,e^{-\frac{\theta^{2}}{2\sigma^{2}}}\,\cos(a\theta)    &=\,e^{-a^{2}\sigma^{2}/2}.
    \label{eq:Gaussian integral formula cos}
\end{align}
Combining Eqs.~(\ref{eq:class parallel tr Gamma Em}) to~(\ref{eq:Gaussian integral formula cos}) we obtain
\begin{align}
    &\hspace*{-1mm}\tr\bigl(E_{m}\Gamma\bigr)=
    \sum\limits_{k=0}^{N-m}\sum\limits_{k\pr=0}^{m}
    \frac{\bigl((-1)^{k\pr}+(-1)^{k}\bigr)N!}{2^{N+k+k\pr+1}k!k\pr!(N\nl-\nl m\nl-\nl k)!(m\nl-\nl k\pr)!}\nonumber\\[1mm]
    &\hspace*{-1mm}\times\!\left\{\nl\frac{(k\nl+\nl k\pr)!}{[(\tfrac{k+k\pr}{2})!]^{2}}
    +2\!\!\!\!\!\sum\limits_{l=0}^{\tfrac{k+k\pr-2}{2}}\!\!\!\frac{(-1)^{\tfrac{k+k\pr-2l}{2}}e^{-\tfrac{(k+k\pr-2l)^{2}\sigma^{2}}{2}}(k\nl+\nl k\pr)!}{l!(k+k\pr-l)!}\nl\right\}
\end{align}
For the Gaussian prior we then need to further compute
\begin{align}
    \overline{V}_{\!\mathrm{post}}    &=\,\sigma^{2}+\theta_{o}^{2}
    \,-\,\sum\limits_{m}\frac{\bigl(\tr(E_{m}\nr\eta)\bigr)^{2}}{\tr(E_{m}\nr\Gamma)} %\nonumber\\[1mm]
    %&
    \,=\,\sigma^{2}\,-\,\sum\limits_{m}\frac{\gamma_{m}^{2}}{\tr(E_{m}\nr\Gamma)}\,,
    \label{eq:average final variance gaussian prior simplified}
\end{align}
\vspace*{-1.0mm}
where
\begin{align}
    \gamma_{m}  &=\,\binom{N}{m}\int\!d\theta\,\theta\,p(\theta)\,p(+|\nr\theta\nr)^{N-m}p(-|\nr\theta\nr)^{m}\nonumber\\[1mm]
    &=\,\frac{1}{2^{N}}\binom{N}{m}\sum\limits_{k=0}^{N-m}\sum\limits_{k\pr=0}^{m}
    \binom{N-m}{k}\binom{m}{k\pr}(-1)^{k\pr}\,J_{k+k\pr}\,.
    \label{eq:class parallel tr small gamma Em}
\end{align}
Here we have a different integral, $J_{k+k\pr}$, which is of the form
\begin{align}
    J_{n}   &=\,\frac{1}{\sqrt{2\pi}\sigma}\int\!d\theta\,\theta\,e^{-\frac{\theta^{2}}{2\sigma^{2}}}(\sin\theta)^{n}.
    \label{eq:inegral Jn}
\end{align}
Since $\theta\,\exp(-\tfrac{\theta^{2}}{2\sigma^{2}})=-\sigma^{2}\tfrac{\partial}{\partial\theta}\exp(-\tfrac{\theta^{2}}{2\sigma^{2}})$ we can easily integrate by parts and write
\begin{align}
    J_{n}   &=\,\frac{n\sigma}{\sqrt{2\pi}}\int\!d\theta\,e^{-\frac{\theta^{2}}{2\sigma^{2}}}(\sin\theta)^{n-1}\cos\theta.
    \label{eq:inegral Jn simplified}
\end{align}
When $k+k\pr=n$ is even, the integral vanishes. When $k+k\pr=n$ is odd, on the other hand, then $(n-1)$ is even and we can use the trigonometric identity from Eq.~(\ref{eq:superweird trig identity}) along with the formula
\begin{align}
    \cos\theta\cos\bigl([n-1-2l]\theta\bigr)    &=\,\frac{1}{2}\cos\bigl([n-2-2l]\theta\bigr)\\[1mm]
    &\ +\frac{1}{2}\cos\bigl([n-2l]\theta\bigr).
\end{align}
This, together with the Gaussian integral of Eq.~(\ref{eq:Gaussian integral formula cos}) allows us to rewrite $J_{n}$ as
\begin{align}
    &J_{n}=\,\frac{n\sigma^{2}[1-(-1)^{n}]}{2^{n}}\left\{e^{-\tfrac{\sigma^{2}}{2}}\binom{n-1}{\tfrac{n-1}{2}}\right.\\[1mm]
    &\hspace*{-1mm}+\!\sum\limits_{l=0}^{\tfrac{n-3}{2}}(-1)^{\tfrac{n-2l-1}{2}}\!\binom{\nl n-1}{l\nl}\!
    \!\left.\bigl(e^{-\tfrac{(n-2l-2)^{2}\sigma^{2}}{2}}+e^{-\tfrac{(n-2l)^{2}\sigma^{2}}{2}}\bigr)\!\right\}.\nonumber
\end{align}
With this, the average variance of the posterior of Eq.~(\ref{eq:average final variance gaussian prior simplified}) can be computed, which we have done for up to $N=200$ qubits. The results, depicted in Fig.~\ref{fig:variance classical parallel}, show that $\overline{V}_{\!\mathrm{post}}$ decreases at most as $1/N$, as expected. Moreover, the data suggests that the parallel classical strategy is close the bound of Ineq.~(\ref{eq:Bayesian cramer rao bound for Gaussian prior and classical strategy}) when the width of the prior is much smaller than $\pi/2$. For instance, when $\sigma=0.1$, the relative deviation of the data for $\overline{V}_{\!\mathrm{post}}$ from the bound $\overline{V}_{\!\mathrm{min}}=\sigma^{2}/(1+N\sigma^{2})$ as quantified by $\Delta\overline{V}:=\bigl(\overline{V}_{\!\mathrm{post}}-\overline{V}_{\!\mathrm{min}}\bigr)/\overline{V}_{\!\mathrm{min}}$ is $\Delta\overline{V}<5\times10^{-7}$ for $N=1$ and $\Delta\overline{V}<1.65\times10^{-5}$ for $N=90$. For $\sigma=0.5$ the corresponding deviations are already at $\Delta\overline{V}\approx6.6\times10^{-3}$ and $2.9\times10^{-2}$ for $N=1$ and $N=90$, respectively, and for larger values of $\sigma$ the deviation of $1/\overline{V}_{\!\mathrm{post}}$ from a function increasing linearly with $N$ is already clearly visible in Fig.~\ref{fig:variance classical parallel}.

Having thoroughly investigated the performance of classical estimation strategies in Bayesian scenarios, we will next turn to strategies involving genuine quantum features.

%%%%%%%%%%%%%%%%%%%%%%%%%%%%%%%%%%%%%%%%%%%%%%%%%%%%%%%%%%%%%%%%%%%%%%%%%%%%%%%%%%%%%%%%%%%%%%%%%%%%%%%%%%%%%%%%%%%%%%%%%%%%

\subsection{Quantum Advantage in Bayesian Estimation}\label{sec:Quantum Advantage in Bayesian Estimation}

With respect to the local estimation scenario, Bayesian estimation is made considerably more complicated by the in principle arbitrary shape of the prior. Consequently, results on optimality are scarcely available apart from very special cases such as phase estimation for flat priors~\cite{BerryWiseman2000}, for which an optimal (albeit with respect to a different cost function for the variance) pair of probe state and measurement have been determined. Here, we will discuss a slightly modified version of the scheme of Ref.~\cite{BerryWiseman2000} as an example and show that it can lead to a scaling advantage also for other choices of priors (and cost functions).

The probe state in question is a superposition of $N$-qubit computational basis states, where one representative $\ket{n}\subtiny{-1}{0}{\mathrm{un}}=\ket{1}^{\otimes n}\ket{0}^{\otimes N-n}$ is selected for each Hamming weight, i.e., from each subspace with a fixed number of qubits in the state $\ket{1}$. That is, $\ket{n}\subtiny{-1}{0}{\mathrm{un}}$ is a unary encoding of the integer $n$. For flat priors [and using the Holevo phase variance~\cite{Holevo1984} instead of the MSE of Eq.~(\ref{eq:final variance given m})], the optimal probe state $\ket{\psi\subtiny{0}{0}{\mathrm{sine}}}$ is of the form
\begin{align}
    \ket{\psi\subtiny{0}{0}{\mathrm{sine}}}  &=\,\sum\limits_{n=0}^{N}\,\psi\sub{0}{-1}{n}\,\ket{n}\subtiny{-1}{0}{\mathrm{un}}\,,
    \label{eq:probe_state 2 appendix}
\end{align}
where the coefficients are chosen with a sinusoidal profile (see, e.g., Ref.~\cite{BerryWiseman2000}), i.e.,
\begin{align}
    \psi\sub{0}{-1}{n}  &=\,\sqrt{\frac{2}{N+2}}\,\sin\Bigl(\frac{(n+1)\pi}{N+2}\Bigr)\,.
    \label{eq:Berry wiseman state appendix}
\end{align}
For the sake of illustration, we will study the performance of this particular state that we will refer to as the \emph{sine state} for the MSE and Gaussian priors of finite width. Nonetheless, it is crucial to note that the optimal probe state for phase estimation with any prior (and variance) must be of the form of Eq.~(\ref{eq:probe_state 2 appendix}) for some choice of coefficients. This is due to the fact that $\ket{\psi\subtiny{0}{0}{\mathrm{sine}}}$ already contains one representative eigenvector of $U_{\theta}$ (and $H$) for each of its different eigenvalues. Adding any other components outside of the span of $\{\ket{n}\subtiny{-1}{0}{\mathrm{un}}\}_{n=0,\ldots,N}$ would hence not provide any more information about the phase $\theta$. After the unitary dynamics $U_{\theta}$, the probe state is thus of the form
\begin{align}
    U_{\theta}\ket{\psi\subtiny{0}{0}{\mathrm{sine}}}  &=\,
    e^{-iN\theta/2}\sum\limits_{n=0}^{N}\,\psi\sub{0}{-1}{n}\,e^{i\nr n\nr \theta}\,\ket{n}\subtiny{-1}{0}{\mathrm{un}}\,.
\end{align}
Also note that the probe state we have chosen is not symmetric with respect to the exchange of the different qubits. However, relinquishing this symmetry requirement allows us to operate in an $(N+1)$-dimensional subspace of the total Hilbert space of dimension $2^{N}$, which will prove to be crucial for the efficient implementation of the estimation scheme in MBQC.

As a measurement strategy for our example, we will consider a quantum Fourier transform (QFT) in the subspace spanned by the vectors $\ket{n}\subtiny{-1}{0}{\mathrm{un}}$, followed by computational basis measurements. This measurement can be represented by a POVM with elements $E_{k}=\ket{e_{k}}\!\bra{e_{k}}$ for $k=0,1,2,\ldots,N$ and $E_{N+1}=\mathds{1}-\sum_{k=0,\ldots,N}E_{k}$, where
\begin{align}
    \ket{e_{k}} &=\,\frac{1}{\sqrt{N+1}}\sum\limits_{n=0}^{N}e^{i\nr n\nr\tfrac{2\pi k}{N+1}}\,\ket{n}\subtiny{-1}{0}{\mathrm{un}}\,.
    \label{eq:DFT}
\end{align}
Practically, we can ignore the POVM element $E_{N+1}$, as the corresponding outcome never occurs for the chosen probe state (in the absence of noise). With this, we are now in a position to compute $\overline{V}_{\!\mathrm{post}}$ from Eq.~(\ref{eq:average final variance}) where we again assume a Gaussian prior as in Eq.~(\ref{eq:Gaussian prior}). We hence need to calculate
\begin{align}
    \overline{V}_{\!\mathrm{post}}    &=\,\sigma^{2}\,+\,\theta_{o}^{2}
    \,-\,\sum\limits_{k}\frac{\bigl(\tr(E_{k}\nr\eta)\bigr)^{2}}{\tr(E_{k}\nr\Gamma)}\,.
    \label{eq:average final variance Gaussian prior}
\end{align}
To rewrite this quantity, it is useful to first determine $\tr(E_{k}\nr\rho(\theta))$ where $\rho(\theta)= U_{\theta}\ket{\psi\subtiny{0}{0}{\mathrm{sine}}} \!\!\bra{\psi\subtiny{0}{0}{\mathrm{sine}}}U_{\theta}^{\dagger}$, for which we obtain
\begin{align}
    \tr(E_{k}\nr\rho(\theta))   &=\,\sum\limits_{m,n=0}^{N}\,\frac{\psi_{m}^{*}\psi\sub{0}{-1}{n}}{N+1}\,e^{i(m-n)\bigl(\tfrac{2\pi k}{N+1}-\theta\bigr)}\,.
    \label{eq:trace Ek rho theta BW}
\end{align}
With this, we quickly find
\begin{align}
    \tr(E_{k}\nr\Gamma) &=\,\sum\limits_{m,n=0}^{N}\,\frac{\psi_{m}^{*}\psi\sub{0}{-1}{n}}{N+1}\,e^{i(m-n)\bigl(\tfrac{2\pi k}{N+1}-\theta_{o}\bigr)}\,e^{-\tfrac{(m-n)^{2}\sigma^{2}}{2}},
    \label{eq:tr Ek Gamma general}
\end{align}
\begin{align}
    %\\[1mm]
    \tr(E_{k}\nr\eta)   &=\,\sum\limits_{m,n=0}^{N}\,\frac{\psi_{m}^{*}\psi\sub{0}{-1}{n}}{N+1}\,\bigl(\theta_{o}-i(m-n)\sigma^{2}\bigr)\times
    \nonumber\\
    &\ \ \ \ \ \times\,e^{i(m-n)\bigl(\tfrac{2\pi k}{N+1}-\theta_{o}\bigr)}\,e^{-\tfrac{(m-n)^{2}\sigma^{2}}{2}}\,.
    \label{eq:tr Ek eta general}
\end{align}

\begin{figure}[ht!]
%%%trim={<left> <lower> <right> <upper>}
\includegraphics[width=0.48\textwidth,trim={0cm 0.1cm 0cm 0cm},clip]{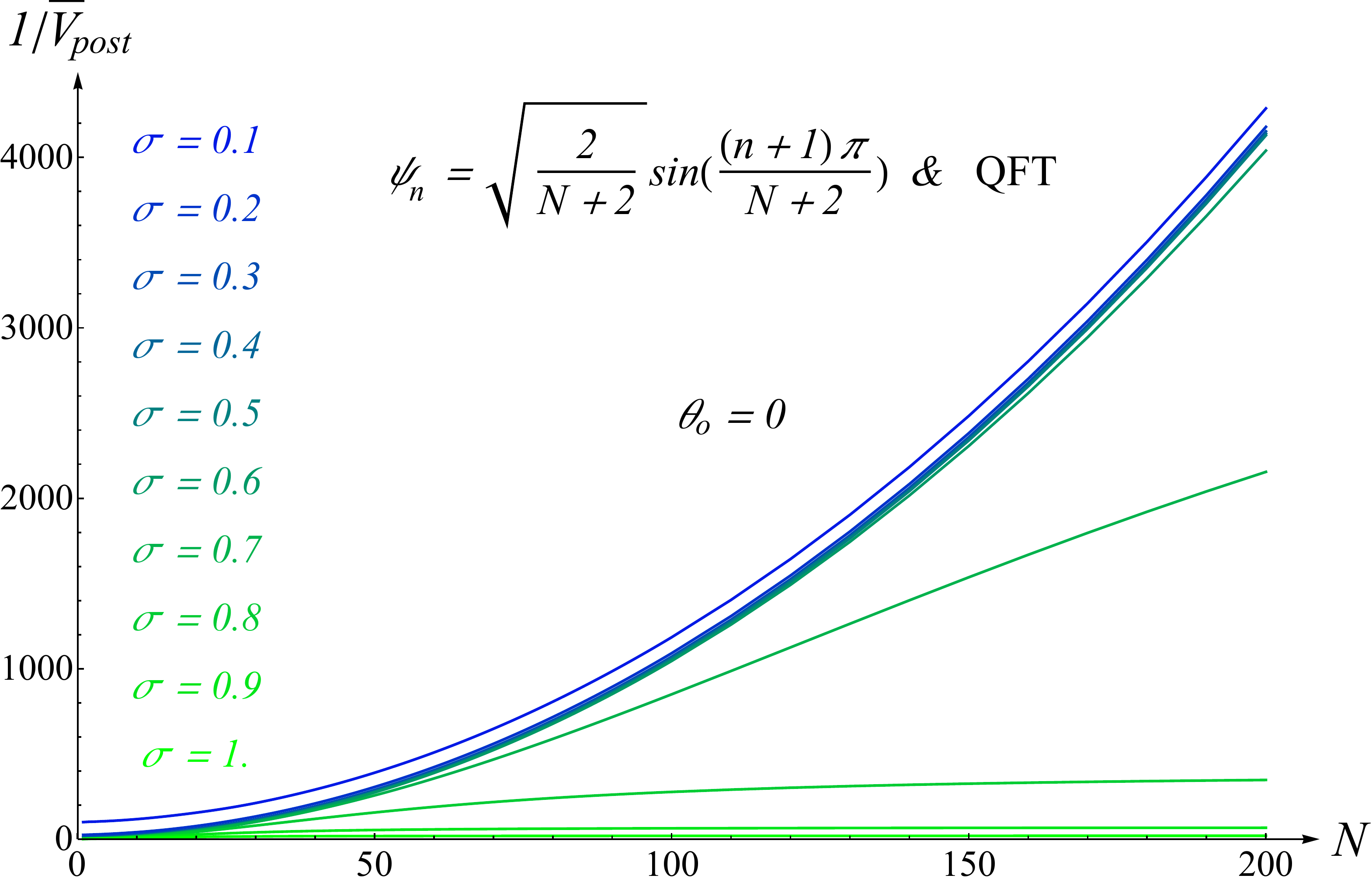}\\
\caption{\label{fig:Berry-Wiseman}%(Color online)
\textbf{Example for quantum strategy:}. The inverse average variance of the posterior $1/\overline{V}_{\!\mathrm{post}}$ is plotted against the qubit number $N$ for the quantum measurement strategy using the sine states from Eq.~(\ref{eq:Berry wiseman state appendix}) with QFT measurements for $\theta_{o}=0$ and for values of $\sigma$ from $0.1$ (blue) to $1$ (green) in steps $0.1$. Additional plots for other values of $\theta_{o}$ and comparisons with classical strategies can be found in Fig.~\ref{fig:Berry-Wiseman comparison}.
%Appendix~\ref{sec:Additional Data for Bayesian Phase Estimation}.
}
\end{figure}

Inserting Eqs.~(\ref{eq:tr Ek Gamma general}) and~(\ref{eq:tr Ek eta general}) back into~(\ref{eq:average final variance Gaussian prior}) and making use of $\sum_{k}E_{k}=\mathds{1}$, and the normalization of $\rho(\theta)$ and $p(\theta)$, as well as the identity $\sum_{k}\exp\bigl(i(m-n)\tfrac{2\pi k}{N+1}\bigr)=(N+1)\delta_{mn}$, one arrives at
\begin{align}
    \overline{V}_{\!\mathrm{post}}    &=\,\sigma^{2}\,-\,\sum\limits_{k}\frac{\gamma_{k}^{2}}{\tr(E_{k}\nr\Gamma)}\,,
    \label{eq:average final variance Gaussian prior simplified}
\end{align}
\vspace*{-1.5mm}
where $\gamma_{k}$ is given by
\begin{align}
    \gamma_{k}  &=\tfrac{i\sigma^{2}}{N+1}\sum\limits_{m,n=0}^{N}\!\psi_{m}^{*}\psi\sub{0}{-1}{n}(m-n)\,e^{i(m-n)\bigl(\tfrac{2\pi k}{N+1}-\theta_{o}\bigr)}
    \,e^{-\tfrac{(m-n)^{2}\sigma^{2}}{2}}.
    \label{eq:small gamma k quantum strategy}
\end{align}
%\vspace*{-1mm}
Finally inserting Eqs.~(\ref{eq:trace Ek rho theta BW}) and~(\ref{eq:small gamma k quantum strategy}) into the formula for $\overline{V}_{\!\mathrm{post}}$ in Eq.~(\ref{eq:average final variance Gaussian prior simplified}), the average variance of the posterior for the sine state and the QFT measurement can be evaluated numerically. The results for up to $N=200$ qubits and for the prior centered at $\theta_{o}=0$ are shown in Fig.~\ref{fig:Berry-Wiseman}.

The plots in Fig.~\ref{fig:Berry-Wiseman} indicate that for narrow priors (e.g. for $\sigma=0.1,\ldots,0.5$) the example quantum strategy exhibits a quadratic scaling gap with respect to all classical measurements schemes,  meaning that the variance in the quantum strategy decreases more strongly with $N$ than classically possible. As discussed in Ref.~\cite{JarzynaDemkowiczDobrzanski15}, this is possible for all priors under certain regularity assumptions, but the explicit form of the optimal states and measurements is generally not known. Indeed, we cannot conclude that the strategy that we discuss here is optimal, but (at least) for narrow Gaussian priors ($\sigma\leq0.5$) we find that it directly outperforms even the (overly optimistic) bound on classical strategies from Ineq.~(\ref{eq:Bayesian cramer rao bound for Gaussian prior and classical strategy}) already for $N=6$ qubits. For broader priors, we can not report a scaling advantage for this example, but this is to be expected using the MSE. However, recall that the
\clearpage
\begin{figure}[ht!]
(a)\includegraphics[width=0.48\textwidth,trim={0cm 0cm 0cm 0cm},clip]{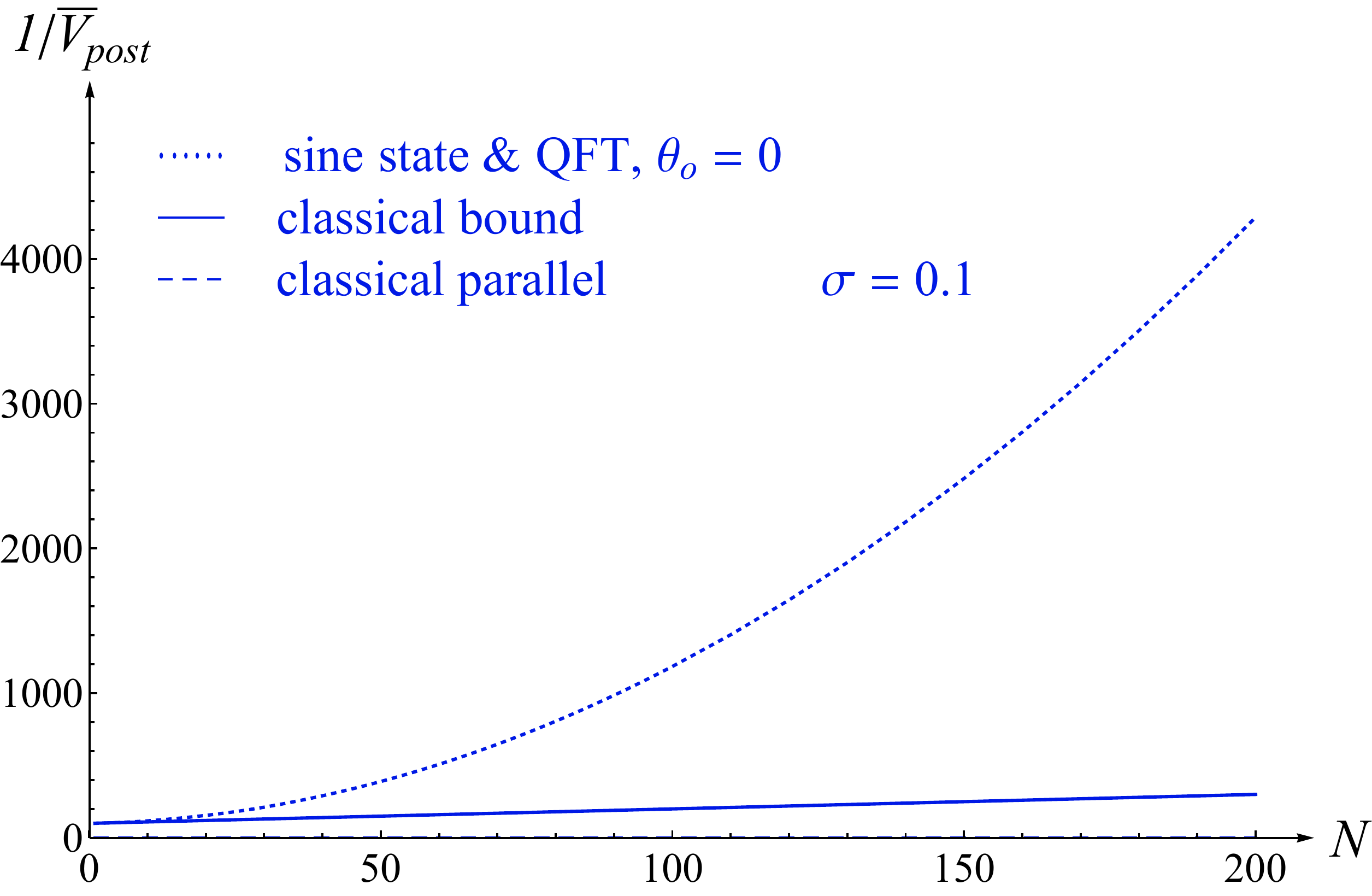}
(b)\includegraphics[width=0.48\textwidth,trim={0cm 0cm 0cm 0cm},clip]{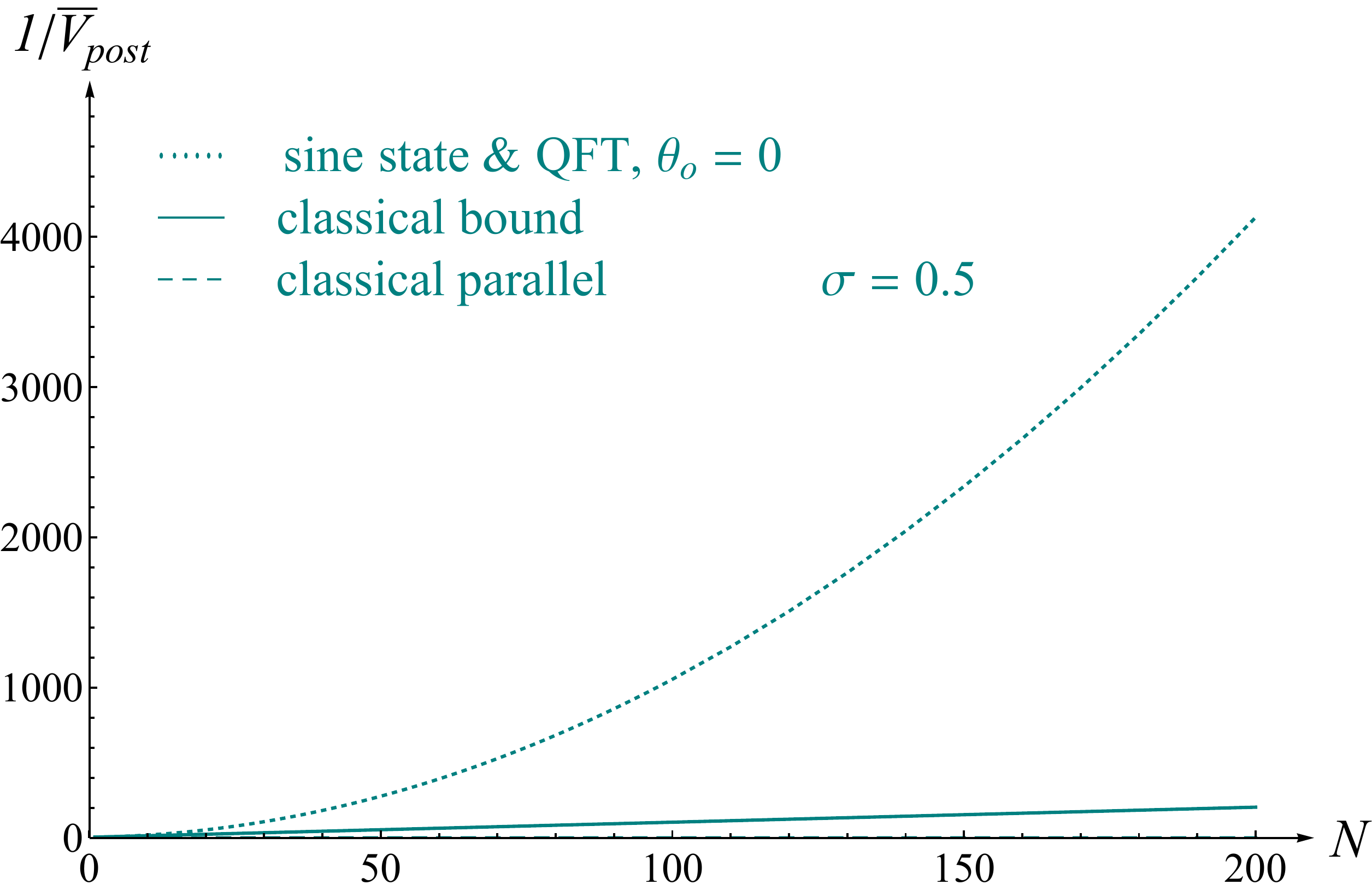}
(c)\includegraphics[width=0.48\textwidth,trim={0cm 0cm 0cm 0cm},clip]{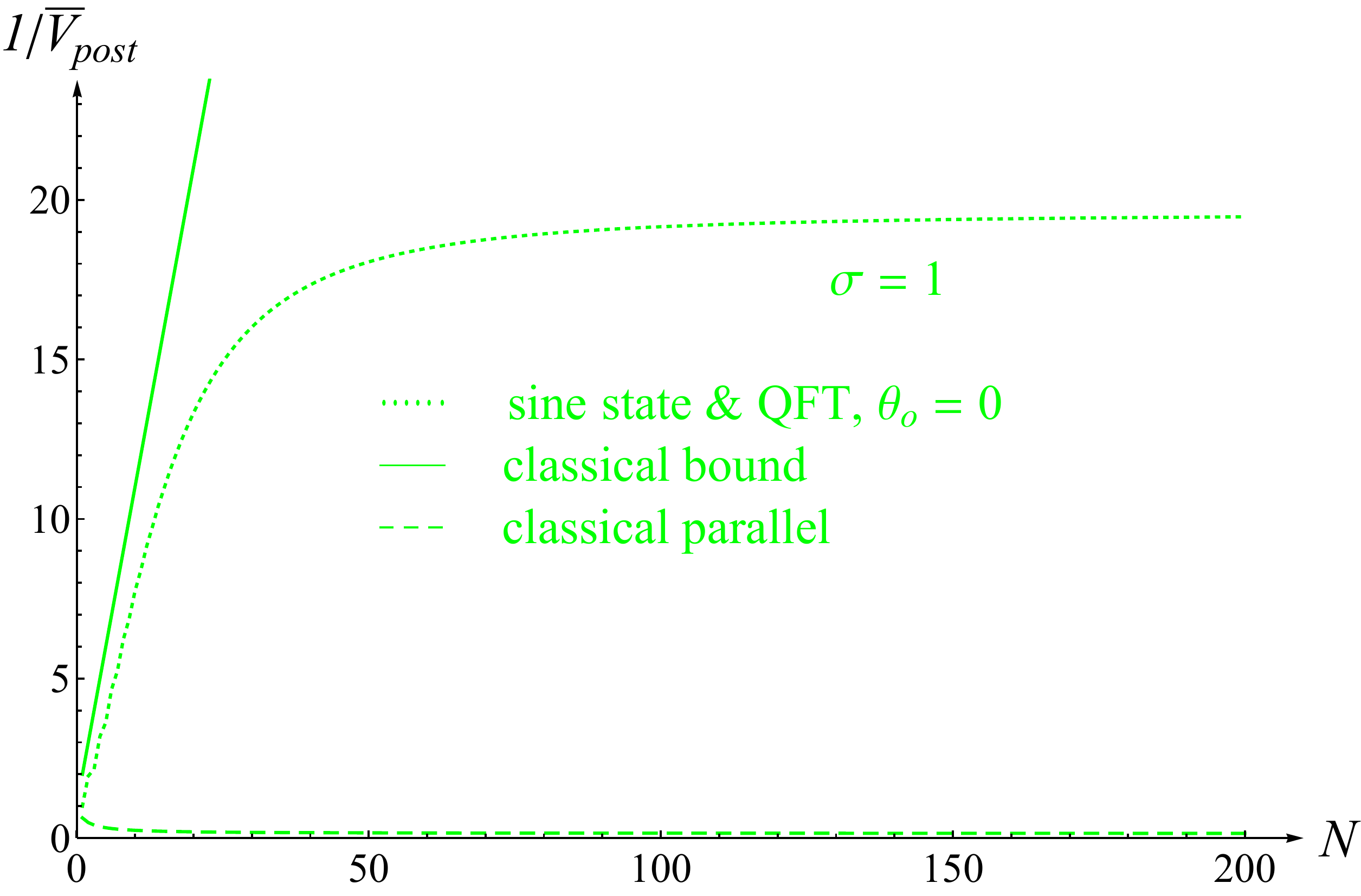}
\caption{\label{fig:Berry-Wiseman comparison}%(Color online)
\textbf{Quantum strategy vs. classical bound}. The inverse average variance of the posterior $1/\overline{V}_{\!\mathrm{post}}$ is plotted (dots) against the qubit number $N$ for the quantum measurement strategy using the sine states from Eq.~(\ref{eq:Berry wiseman state appendix}) with QFT measurements for $\theta_{o}=0$ and values of $\sigma=0.1, 0.5$, and $1$ in (a), (b), and (c), respectively. The solid lines correspond to the Bayesian Cram{\'e}r-Rao bound of Ineq.~(\ref{eq:Bayesian cramer rao bound for Gaussian prior and classical strategy}), which overestimates the best sequential classical strategy. The dashed lines correspond to the optimal classical parallel strategy from Fig.~\ref{fig:variance classical parallel}. As can be seen in (a) and (b), the quantum strategy using the sine states may outperform the best classical strategy for small prior widths $\sigma$, providing a scaling advantage, i.e., $1/\overline{V}_{\!\mathrm{post}}$ increases stronger than linearly with $N$. However, for larger $\sigma$ it performs worse, that is, it still outperforms the optimal classical parallel strategy, but only by a constant improvement, as can be seen in (c).}
\vspace*{-10mm}
\end{figure}
\newpage
\noindent
measurement strategy we discuss here is known to be optimal in the case of flat priors for an appropriately chosen cost function~\cite{BerryWiseman2000}, and our results are hence complimentary in the sense that we provide numerical evidence for optimal scaling in a regime of narrow priors. Additional plots for direct comparison with the classical bounds can be found in Fig.~\ref{fig:Berry-Wiseman comparison}.

%%%%%%%%%%%%%%%%%%%%%%%%%%%%%%%%%%%%%%%%%%%%%%%%%%%%%%%%%%%%%%%%%%%%%%%%%%%%%%%%%%%%%%%%%%%%%%%%%%%%%%%%%%%%%%%%%%%%

\subsection{Bayesian Frequency Estimation}\label{sec:Bayesian frequency estimation in MBQC}

In this appendix we investigate on Bayesian frequency estimation, i.e., the case where the parameter to be estimated is the angular frequency, $\omega$, rather than the phase
$\theta$, i.e., such that $\theta=\omega t$.  The key difference of frequency estimation compared to phase estimation is that in the
former we have the freedom to optimize over the interrogation time $t$.  We shall do this for some of the states and measurement previously considered for phase estimation. Specifically, for the optimal classical parallel measurement strategy and for the quantum strategy using the sine states and QFT measurements from Eqs.~(\ref{eq:Berry wiseman state appendix}) and~(\ref{eq:DFT}), respectively.

More precisely, the dynamical evolution of each qubit is described by the unitary transformation $U(\omega t)=e^{-i\omega t Z/2}$, and our prior information about $\omega$ is given by the normal distribution
\begin{equation}
p(\omega)=\sqrt{\frac{1}{2\pi\Delta^2}}e^{\frac{-\omega^2}{2\Delta^2}},
\label{gaussianfrequency}
\end{equation}
where we have assumed without loss of generality that the mean frequency is centered at $\omega_{0}=0$. With the initial state as in Eq.~(\ref{eq:probe_state 2 appendix}) the matrix elements of the operators $\Gamma$ and $\eta$ of Eqs.~(\ref{eq:Gamma}) and~(\ref{eq:eta}), respectively,
\vspace*{-3mm}
\begin{figure}[hb!]
\includegraphics[keepaspectratio,width=0.49\textwidth]{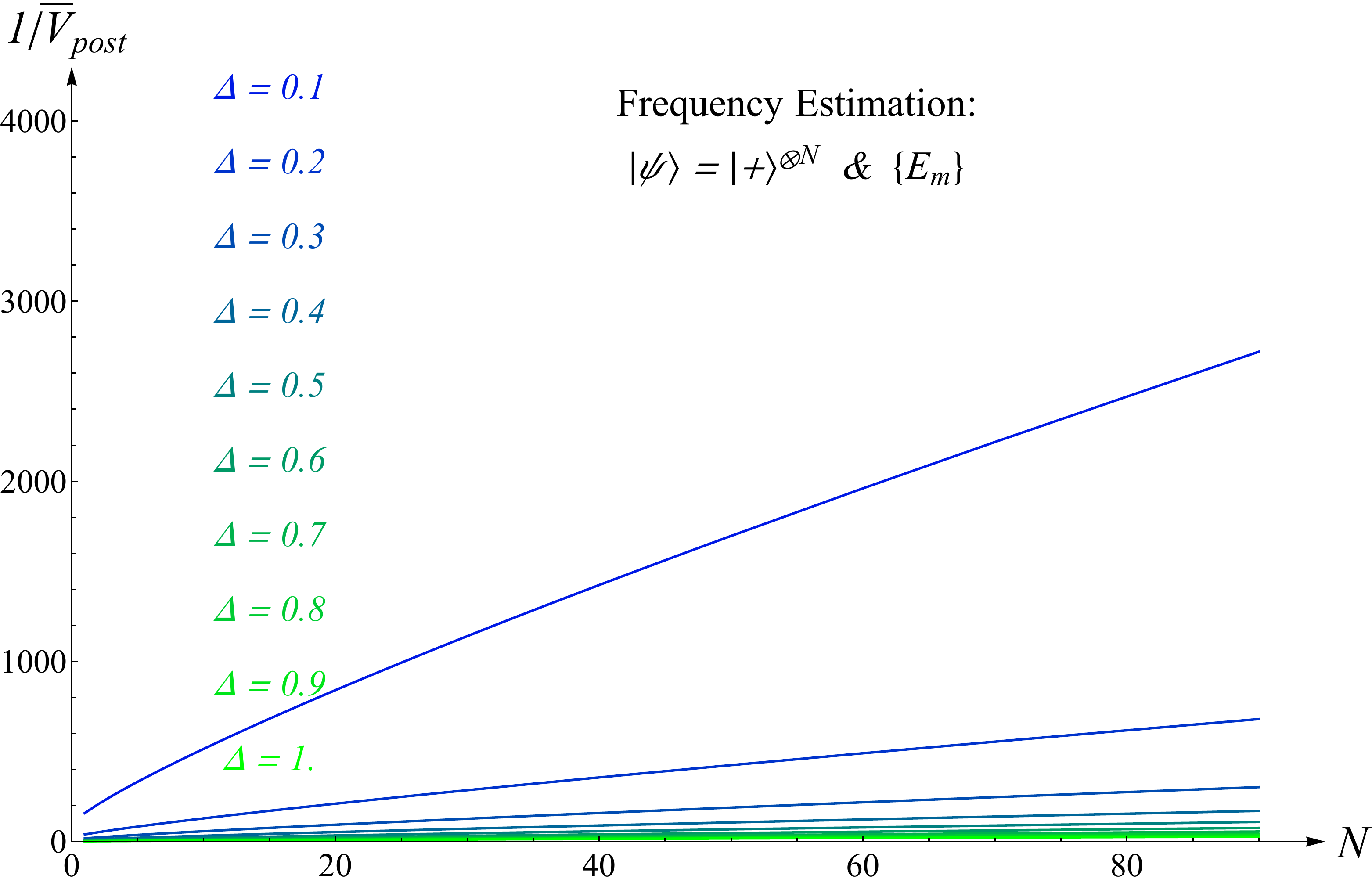}
\vspace*{-2mm}
\caption{\textbf{Optimal classical parallel frequency estimation}. The inverse expected variance of the posterior $1/\overline{V}_{\!\mathrm{post}}$, optimized over the interrogation time, is shown for the optimal, classical, parallel strategy [with probe state $\ket{+}^{\otimes N}$ and POVM with elements $E_{m}$ as in Eq.~(\ref{eq:single qubit n minus outcome POVM})] when starting from a Gaussian prior (in frequency space) of width $\Delta$ in units of $\Delta^{2}$. The horizontal axis shows the qubit number $N$. %Although $N$ will take on the values of integers larger or equal $1$, the curves have been plotted for continuous values of $N$ for the purpose of illustration.
\label{fig:frequency classical parallel}}
\end{figure}
\clearpage

\begin{figure}[ht!]
\includegraphics[keepaspectratio,width=0.48\textwidth]{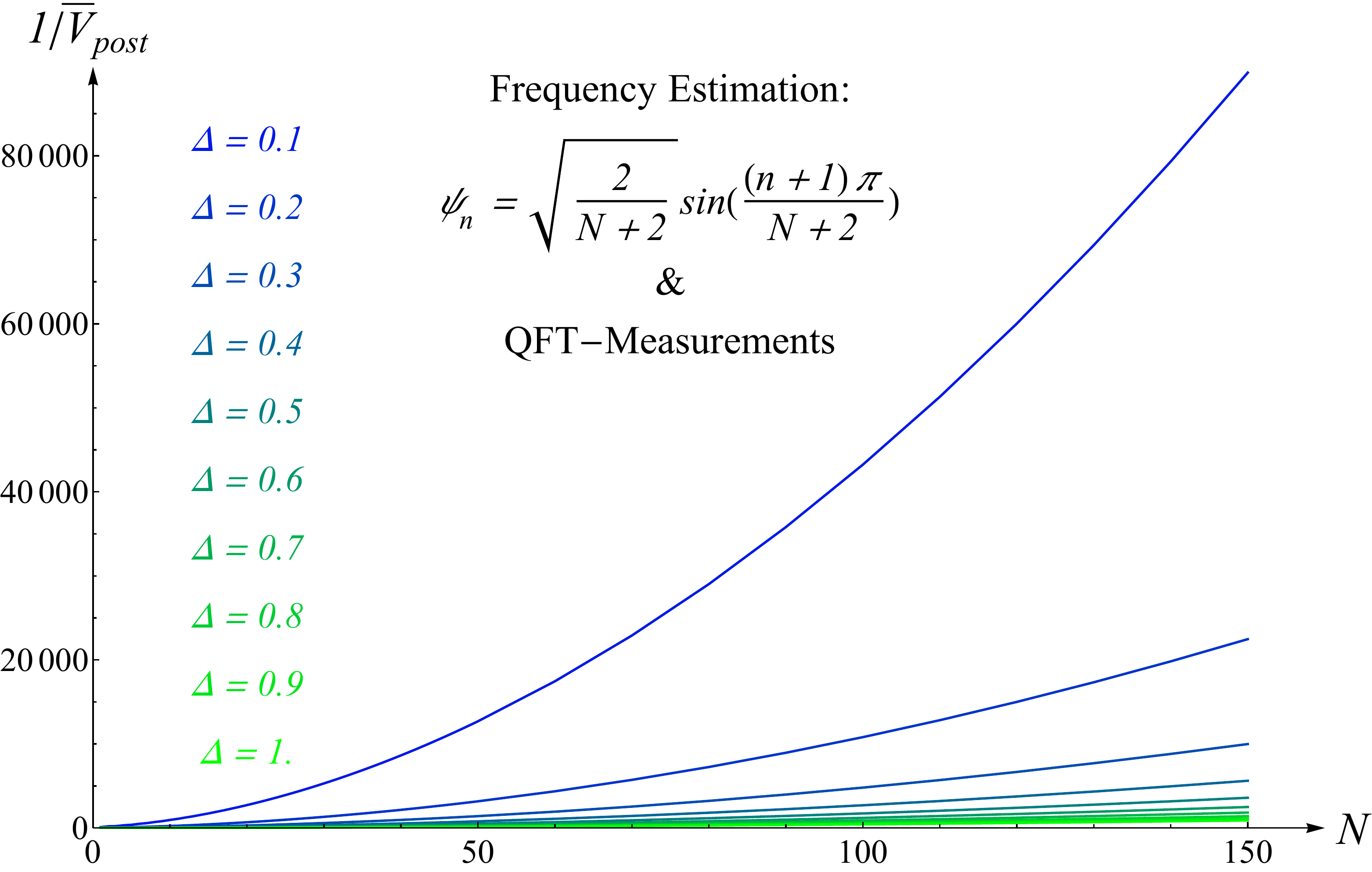}
\vspace*{-2mm}
\caption{\textbf{Frequency estimation: sine states \& QFT measurements}. The inverse average variance of the posterior $1/\overline{V}_{\!\mathrm{post}}$, optimized over the interrogation time~$t$, is plotted against the qubit number $N$ for the quantum measurement strategy using the sine states from Eq.~(\ref{eq:Berry wiseman state appendix}) with QFT measurements for $\omega_{o}=0$ and for values of $\Delta$ from $0.1$ (blue) to $1$ (green) in steps $0.1$, in units of $\Delta^{2}$.
\label{fig:frequency BW DFT}}
\end{figure}
\begin{figure}[ht!]
\includegraphics[keepaspectratio,width=0.48\textwidth]{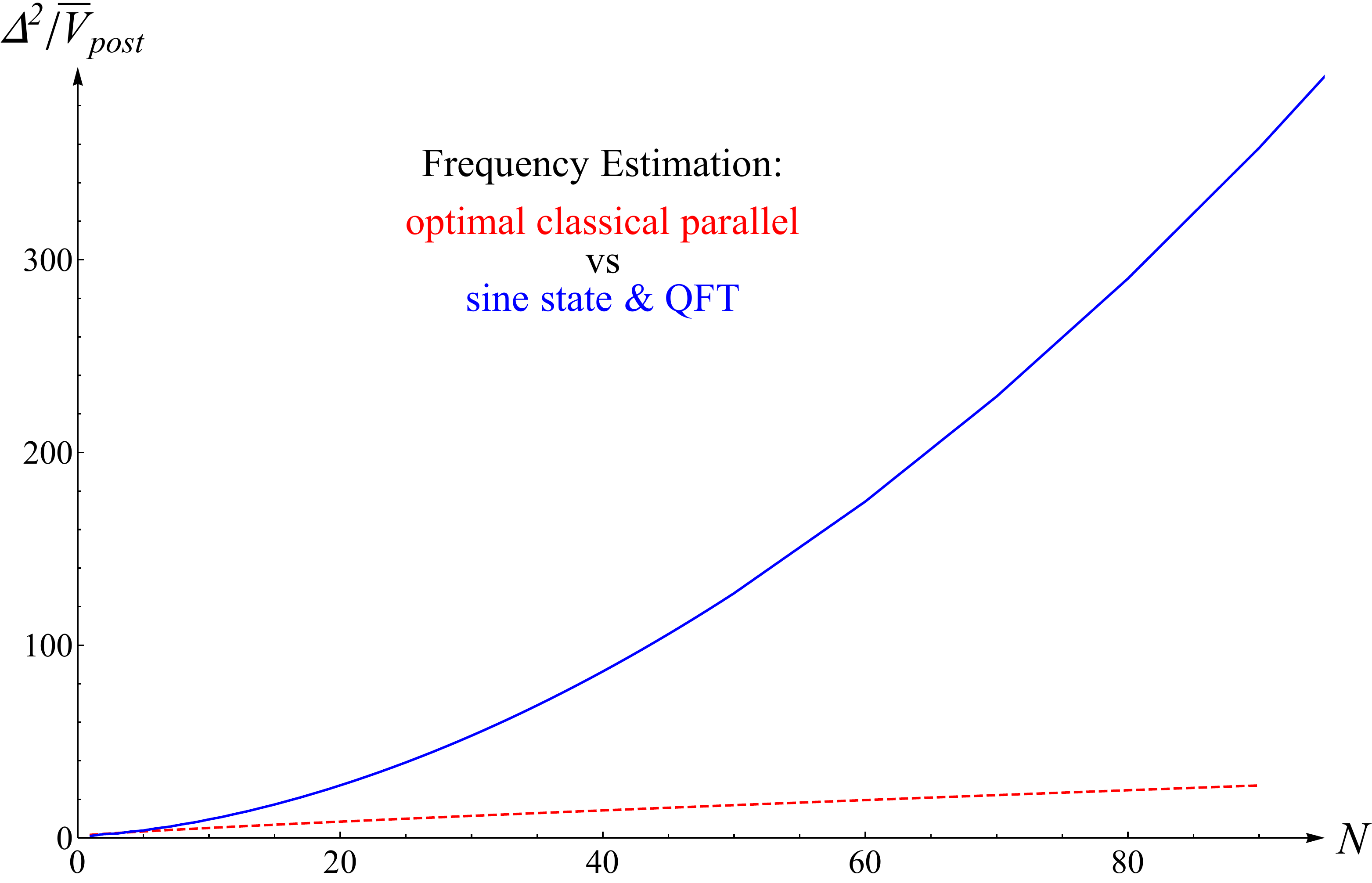}
\caption{\textbf{Frequency estimation comparison}. The inverse average variance of the posterior $\Delta^{2}/\overline{V}_{\!\mathrm{post}}$, optimized over the interrogation time~$t$ and plotted against the qubit number $N$, is compared for the optimal classical parallel strategy (red, dashed) and the quantum strategy using the sine states and QFT measurements (blue, solid). For the plotted range one can clearly see that the quantum strategy provides a scaling advantage with respect to the best parallel classical measurements, that is, the solid blue curve for $\Delta^{2}/\overline{V}_{\!\mathrm{post}}$ increases quadratically with $N$, while the the dashed, red curve only increases linearly with $N$.
\label{fig:frequency comparison}}
\end{figure}

\noindent
read
\begin{align}
\label{Gammafreq}
\Gamma&=\sum_{n,m=0}^N\psi_n\psi_m^*e^{\frac{-(m-n)^2\tau^2}{2}}\ketbra{n}{m}\,,\\
\eta&=-i\tau\Delta\sum_{n,m=0}^N(m-n)\psi_n\psi_m^*e^{\frac{-(m-n)^2\tau^2}{2}}\ketbra{n}{m}\,,
\label{etafrequency}
\end{align}
where we have defined the dimensionless parameter $\tau\equiv t\Delta$.  The final average variance is again given by
Eq.~\eqref{eq:average final variance Gaussian prior}. However, due to the dependence of the Fisher information on $\tau$, we need to optimize the
average final variance over all $\tau$. The results for the optimal, classical parallel strategy (see Appendix~\ref{sec:optimal parallel strategy}) and for the quantum strategy using the sine states and QFT measurements (see Appendix~\ref{sec:Quantum Advantage in Bayesian Estimation}) are plotted in Figs.~\ref{fig:frequency classical parallel} and~\ref{fig:frequency BW DFT}, respectively, and a comparison is shown in Fig.~\ref{fig:frequency comparison}.

\subsection{Measurement-Based Quantum Computation}\label{sec:Measurement-Based Quantum Computation}

\subsubsection{Basics of MBQC}\label{sec:basics of MBQC}

In this appendix, we will briefly review the basic concepts of MBQC, but we direct the interested reader to more detailed reviews in Refs.~\cite{BrowneBriegel2006, BriegelBrowneDuerRaussendorfVanDenNest2009}. In this computational paradigm, established in Refs.~\cite{RaussendorfBriegel2001, RaussendorfBrowneBriegel2003}, a specific entangled state (e.g., a cluster state) is prepared in an array of qubits. Using the entanglement present in the system along with local measurements on a subset of the qubits, (arbitrary) unitary transformation may be implemented on the remaining qubits (if the cluster is large enough). Here, we will focus on MBQC based on 1D and 2D cluster states, i.e., graph states~\cite{HeinDuerEisertRaussendorfVanDenNestBriegel2006} based on regular, linear or rectangular lattices. Each vertex of the graph
\newpage

\begin{figure}[hb!]
\includegraphics[width=0.41\textwidth]{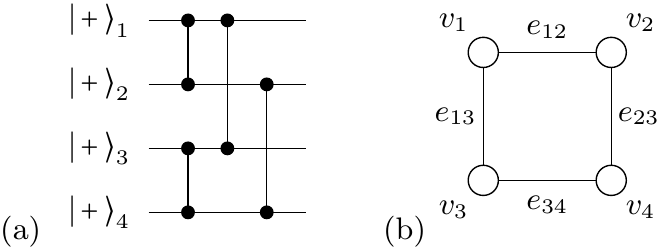}
\caption{\label{fig:square cluster}%(Color online)
\textbf{Two-dimensional cluster states}. In (a) the circuit representation of a two-dimensional (2D) cluster state with $4$ qubits is shown. Each horizontal line represents a qubit, initialized in $\ket{+}_{i}$ ($i=1,2,3,4$), and time goes from left to right. The vertical lines
(\nr\protect\raisebox{-3pt}{\protect\begin{tikzpicture}
\protect\draw[fill=black,line width=1pt] (0,0.135) circle (1.35pt);
\protect\draw[black,line width=0.3pt] (0,0.135) -- (0,-0.135);
\protect\draw[fill=black,line width=1pt] (0,-0.135) circle (1.35pt);
\end{tikzpicture}}\nr) represent controlled phase gates $C\nl Z_{ij}$ applied to the respective qubit pairs $(i,j)$. Fig.~\ref{fig:square cluster} shows the graph structure of the 2D cluster of (a), with vertices $v_{i}$ on a square lattice connected by edges $v_{ij}$.}
\end{figure}

\noindent
corresponds to a qubit initialized in the state $\ket{+}$, while edges connecting the vertices indicate that controlled phase gates $C\nl Z$, given by
\begin{align}
    C\nl Z_{ij}  &=\,\ket{0}\!\!\bra{0}_{i}\otimes\mathds{1}_{j}\,+\,\ket{1}\!\!\bra{1}_{i}\otimes Z_{j}\,=\,C\nl Z_{ji}\,,
    \label{eq:controlled phase gate appendix}
\end{align}
have been applied to these pairs of qubits. A simple example for a cluster state is shown in Fig.~\ref{fig:square cluster}.

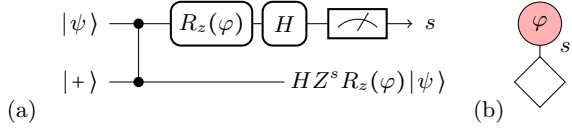
\begin{figure}[ht!]
\includegraphics[width=0.42\textwidth]{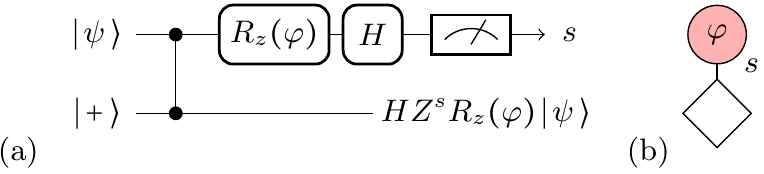}
\vspace*{-1mm}
\caption{\label{fig:single_qubit gate_teleportation}%(Color online)
\textbf{Single-qubit gate teleportation}. (a) After the entangling $C\nl Z$ operation on a pair of qubits prepared in the states $\ket{\psi}$ and $\ket{+}$, respectively, the first qubit is measured in the basis $\{R_{z}(\varphi)H\ket{s}|s=0,1\}$, where $\varphi$ specifies an angle in the $x$\textendash$y$ plane. Here, the rounded boxes correspond to applications of single-qubit gates, where $H=(X+Z)/\sqrt{2}$ is the Hadamard gate, and the symbol~\protect\raisebox{-3pt}{\protect\begin{tikzpicture}[Measurement/.style={rectangle,draw=black,fill=white,thick,minimum height=4mm, minimum width=8mm}]
    \protect\node at (4.0,0) [Measurement]{};
    \protect\draw[color=black] (4.27,-0.05) arc (45:135:0.38);
    \protect\draw[color=black] (4,-0.1) -- (4.15,0.15);
\end{tikzpicture}}
indicates a measurement in the computational basis $\{\ket{s}|s=0,1\}$ with outcome $s$. The remaining qubit is then left in the state $HZ^{s}R_{z}(\varphi)\ket{\psi}$. Up to the outcome-dependent local correction $HZ^{s}$ (and an irrelevant global phase) the output qubit hence carries the result of the computation, $R_{z}(\varphi)\ket{\psi}$. (b) In a graphical notation (see, e.g., Ref.~\protect\cite{BrowneBriegel2006}) for the circuit in (a), measured qubits are represented by circles inscribed with the corresponding measurement angle $\varphi$, while output qubits are indicated by diamonds~(\protect\raisebox{-3pt}{\protect\begin{tikzpicture}[Diamond/.style={diamond,draw=black,thick,fill=white,minimum size=3mm}]
    \protect\node at (4.0,0) [Diamond]{};
\end{tikzpicture}}). The connecting lines between qubits indicate the initial application of $C\nl Z$ gates, and the symbols for input qubits, which may be prepared in arbitrary states are coloured in red, whereas all other qubits are assumed to have been initialized in the state $\ket{+}$.
}
\end{figure}

The essence of the working principle of a measurement-based computation is captured by single-qubit gate teleportation~\cite{GottesmanChuang1999}. That is, by measuring one of the qubits of an entangled pair in a suitable local basis and applying local correction operators dependent on the outcome on the other qubit, a desired quantum gate can be effectively implemented on the remaining qubit, as illustrated in Fig.~\ref{fig:single_qubit gate_teleportation}. Concatenating this procedure for a chain of

\clearpage

\noindent
qubits in a 1D cluster state, arbitrary single-qubit gates may be performed in such a way that only local corrections on the final qubit are required.

\begin{figure}[hb!]
\includegraphics[width=0.35\textwidth]{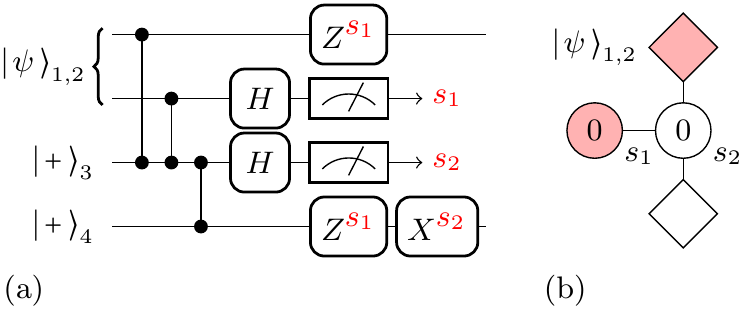}
\vspace*{-1mm}
\caption{\label{fig:CNOT in MBQC}%(Color online)
\textbf{CNOT gate in MBQC}. The four-qubit circuit in (a), and the corresponding measurement pattern in (b) illustrate how the measurement of two of the qubits in a 2D cluster, followed by local Pauli corrections on the two remaining qubits dependent on the measurement outcomes $s_{i}$ ($i=1,2$) can realize an effective CNOT gate in an MBQC architecture. The notation is as in Fig.~\ref{fig:single_qubit gate_teleportation}. Note that of the two input qubits marked red in (b) one is measured, but the other is also an output qubit.}
\end{figure}

Although the measurement-based implementation of the CNOT gate [$C\nl X$] in the notation of Eq.~(\ref{eq:controlled phase gate appendix})] is not possible in a 1D cluster, it can be achieved in two dimensions~\cite{VanDenNestMiyakeDuerBriegel2006}, as is demonstrated by a simple example in Fig.~\ref{fig:CNOT in MBQC}. Since the combination of arbitrary single-qubit gates with the CNOT gate is computationally universal, one may hence prepare an arbitrary quantum state (e.g., for performing parameter estimation) from a 2D cluster.

\begin{figure}[ht!]
\includegraphics[width=0.45\textwidth]{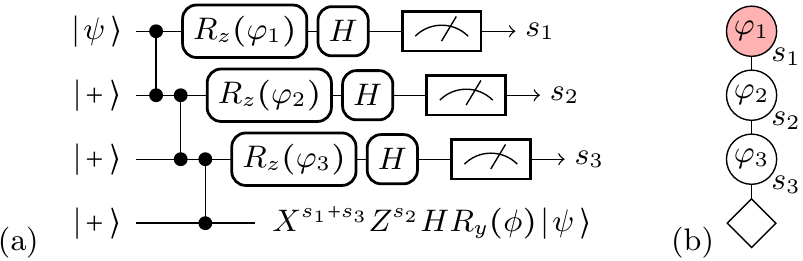}
\vspace*{-1mm}
\caption{\label{fig:Y-rotation in MBQC}%(Color online)
\textbf{Pauli-$Y$ rotation in MBQC}. (a) The circuit representation of the MBQC realization of a Pauli-$Y$ rotation is shown. Measuring the first three qubits in bases in the $x$-$y$ plane rotated with respect to the $X$-basis by $\varphi_{1}=\tfrac{\pi}{2}$, $\varphi_{2}=(-1)^{s_{1}}\phi$, and $\varphi_{3}=(-1)^{s_{2}+1}\tfrac{\pi}{2}$, respectively, and applying the local Pauli corrections $X^{s_{1}+s_{3}} Z^{s_{2}} H$ dependent on the measurement outcomes $s_{i}=0,1$ $(i=1,2,3)$ leaves the fourth qubit in the desired state. (b) Graphical representation of the circuit in (a) following the notation of Fig.~\ref{fig:single_qubit gate_teleportation}.
}
\end{figure}

\subsubsection{Probe State Preparation in MBQC}\label{sec:Probe State Preparation in MBQC}

In this last appendix, we present details on the conversion of the circuit for generating probe states (shown in Fig.~\ref{fig:circuit for BW} of the main text) to an MBQC measurement pattern. To do this, let us first see how a $Y$-rotation can be performed in MBQC, and consider the concatenation of three steps of single-qubit gate teleportation (see Fig.~\ref{fig:single_qubit gate_teleportation}) as shown in Fig.~\ref{fig:Y-rotation in MBQC}. That is, we prepare a one-dimensional four-qubit cluster state, where the first qubit is initialized in an arbitrary state $\ket{\psi}$. The first three qubits are then measured with angles $\varphi_{1}$, $\varphi_{2}$, and $\varphi_{3}$, respectively, leaving the fourth qubit in the state (up to a global phase)
\begin{align}
    X^{s_{1}+s_{3}} Z^{s_{2}} H
    R_{z}\bigl((-1)^{s_{2}}\varphi_{3}\bigr)
    R_{x}\bigl((-1)^{s_{1}}\varphi_{2}\bigr)
    R_{z}(\varphi_{1}) \ket{\psi}\,.
    \label{eq:y rotation gate teleported state}
\end{align}
Noting that a $Y$-rotation about an arbitrary angle $\phi$ can be written as $R_{y}(\phi)=R_{z}(-\tfrac{\pi}{2})R_{x}(\phi)R_{z}(\tfrac{\pi}{2})$, selecting measurement angles $\varphi_{1}=\tfrac{\pi}{2}$, $\varphi_{3}=(-1)^{s_{2}+1}\tfrac{\pi}{2}$, and $\varphi_{2}=(-1)^{s_{1}}\phi$ in Fig.~\ref{fig:Y-rotation in MBQC} realizes $R_{y}(\phi)$ up to appropriate local corrections on the last qubit.

\begin{figure}[hb!]
\includegraphics[width=0.45\textwidth]{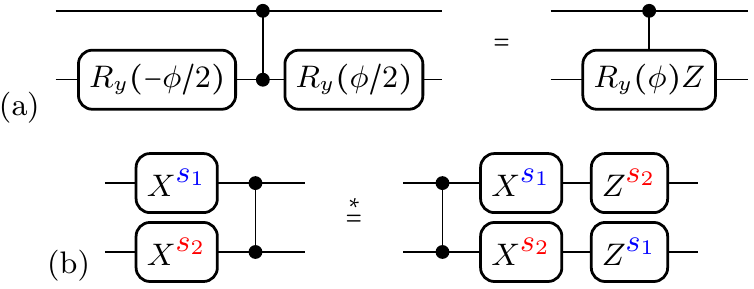}
\caption{\label{fig:controlled_rotation MBQC}%(Color online)
\textbf{Circuit identities}. The circuit in (a) implements the controlled $R_{y}(\phi)Z$ rotation by incorporation of a controlled phase gate $C\nl Z$. In (b) a circuit identity for commuting single-qubit $X$-operations past the controlled phase gates (featuring, e.g., in cluster states) is shown. The symbol $\stackrel{*}{=}$ indicates equality up to a possible global phase.}
\end{figure}

\newpage
With this strategy, we are able to implement $R_{y}(\phi_{1})$. One may even commute the Hadamard correction with the $Y$-rotation to switch the initial state of the qubit from $\ket{+}$ to $\ket{0}$, as required in Fig.~\ref{fig:circuit for BW} of the main text. For the remaining controlled rotations, we make use of the simple identity $R_{y}(\phi)Z=ZR_{y}(-\phi)$, which allows us to utilize the $C\nl Z$-gates naturally appearing in the cluster state to perform the operation $C\nl R_{y}(\phi)$, as shown in the circuit in Fig.~\ref{fig:controlled_rotation MBQC}~(a). The spurious application of the operator $Z$ before the rotation can be disregarded, since all qubits in the circuit in Fig.~\ref{fig:circuit for BW} are assumed to be in the state $\ket{0}$ in the beginning. This initialization step can be included as for $R_{y}(\phi_{1})$ before.

Since we already know from the circuit in Fig.~\ref{fig:Y-rotation in MBQC} how to implement rotations $R_{y}(\phi)$ for arbitrary angles, all that is left to do to translate the preparation circuit in Fig.~\ref{fig:circuit for BW} to MBQC is to commute the local $X$-corrections past the $C\nl Z$-gate appearing on the left-hand side of Fig.~\ref{fig:controlled_rotation MBQC}~(a), as shown in Fig.~\ref{fig:controlled_rotation MBQC}~(b), such that all local corrections can be applied in the final step of the state preparation. We hence arrive at the MBQC measurement pattern generating the sine state $\ket{\psi\subtiny{0}{0}{\mathrm{sine}}}$, which is shown in Fig.~\ref{fig:MBQC_pattern for BW state} of the main text.

\end{document}